
\input phyzzx
%
%
\catcode`\@=11 
\def\papers{\papersize\headline=\paperheadline\footline=\paperfootline}
\def\papersize{\hsize=40pc \vsize=53pc \hoffset=0pc \voffset=1pc
   \advance\hoffset by\HOFFSET \advance\voffset by\VOFFSET
   \pagebottomfiller=0pc
   \skip\footins=\bigskipamount \normalspace }
\catcode`\@=12 
\papers
%

\def\square{\kern1pt\vbox{\hrule height 1.2pt\hbox{\vrule width 1.2pt\hskip 3pt
   \vbox{\vskip 6pt}\hskip 3pt\vrule width 0.6pt}\hrule height 0.6pt}\kern1pt}

\def\bra#1{\langle #1 |}
\def\ket#1{| #1 \rangle}
\def\vev#1{\langle #1 \rangle}
\def\ov{{\overline}}
\def\hep{{hepth@xxx.lanl.gov. \#}}
\def\A{{\cal A}}

\def\C{{\cal C}}
\def\D{{\cal D}}
\def\H{{\cal H}}
\def\F{{\cal F}}

\def\M{{\cal M}}
\def\N{{\cal N}}
\def\O{{\cal O}}
\def\P{{\cal P}}

\def\R{{\cal R}}
\def\V{{\cal V}}
\def\bz{{\overline z}}
\def\da{{\downarrow}}
\def\ua{{\uparrow}}
\def\e{{\epsilon}}
\def\g{{\gamma}}
\def\k{{\kappa}}
\def\l{{\bigl[}}
\def\r{{\bigr]}}
\def\ov{\overline}
\def\spr{\mathop{{\sum}'}}
\def\mapdown#1{\Big\downarrow
   \rlap{$\vcenter{\hbox{$\scriptstyle#1$}}$}}
\def\mapup#1{\Big\uparrow
   \rlap{$\vcenter{\hbox{$\scriptstyle#1$}}$}}
\def\wt{\widetilde}
\def\wh{\widehat}
\overfullrule=0pt
\baselineskip 13pt plus 1pt minus 1pt
\pubnum{IASSNS-HEP-92/41 \cr
MIT-CTP-2102\cr
hep-th/9206084}
\date{June 1992}
\titlepage
\title{CLOSED STRING FIELD THEORY: QUANTUM ACTION \break
\break AND THE {}B-V MASTER EQUATION}
\author{Barton Zwiebach \foot{Permanent address: Center for Theoretical
Physics, MIT, Cambridge, Mass. 02139. Supported in part by D.O.E.
contract DE-AC02-76ER03069 and NSF grant PHY91-06210.}}
\address{School of Natural Sciences \break
Institute for Advanced Study \break
Olden Lane \break
Princeton, NJ 08540}
\abstract{The complete quantum theory of covariant closed strings
is constructed in detail. The nonpolynomial action is defined by
elementary vertices satisfying recursion relations that give rise to
Jacobi-like identities for an infinite chain of string field products.
The genus zero string field algebra is the homotopy Lie algebra $L_\infty$
encoding the gauge symmetry of the classical theory. The higher
genus algebraic structure implies the Batalin-Vilkovisky (BV) master
equation and thus consistent BRST quantization of the quantum action. From
the $L_\infty$ algebra, and the BV equation on the off-shell state
space we derive the $L_\infty$ algebra, and the BV equation on physical
states that were recently constructed in $d=2$ string theory.
The string diagrams are surfaces with minimal area metrics, foliated by
closed geodesics of length $2\pi$. These metrics generalize quadratic
differentials in that foliation bands can cross. The string vertices
are succinctly characterized; they include the surfaces whose foliation bands
are all of height smaller than $2\pi$.}
\endpage

\singlespace

\REF\saadizwiebach{M. Saadi and B. Zwiebach, `Closed string
field theory from polyhedra', Ann. Phys. {\bf 192} (1989) 213.}
\REF\kks{T. Kugo, H. Kunitomo and K. Suehiro, `Non-polynomial closed string
field theory', Phys. Lett. {\bf 226B} (1989) 48.}
\REF\kugosuehiro{T. Kugo and K. Suehiro,  `Nonpolynomial closed
string field theory: action and gauge invariance', Nucl.
Phys. {\bf B337} (1990) 434.}
\REF\kaku{M. Kaku, `Geometrical derivarion of string field theory from
first principles: closed strings and modular invariance. Phys. Rev.
{\bf D38} (1988) 3052;\hfill\break
M. Kaku and J. Lykken, `Modular Invariant closed string field theory',
Phys. Rev. {\bf D38} (1988) 3067.}
\REF\zwiebachqcs{B. Zwiebach, `Quantum closed strings from minimal
area', Mod. Phys. Lett. {\bf A5} (1990) 2753.}
\REF\zwiebachtalk{B. Zwiebach, `Recursion Relations in Closed String
Field Theory', Proceedings of the ``Strings 90'' Superstring Workshop.
Eds. R. Arnowitt, et.al. (World Scientific, 1991) pp 266-275.}
\REF\sonodazwiebach{H. Sonoda and B. Zwiebach, `Closed string field theory
loops with symmetric factorizable quadratic differentials',
Nucl. Phys. {\bf B331} (1990) 592.}
\REF\zwiebachma{B. Zwiebach,  `How covariant closed string
theory solves a minimal area problem',
Comm. Math. Phys. {\bf 136} (1991) 83 ; `Consistency of closed string
polyhedra from minimal area', Phys. Lett. {\bf B241} (1990) 343.}
\REF\siegel{W. Siegel, `Introduction to String Field Theory' (World Scientific,
Singapore, 1988).}
\REF\witten{E. Witten, `Noncommutative geometry and
string field theory', Nucl. Phys. {\bf B268} (1986) 253.}
\REF\batalinvilkovisky{I. A. Batalin and G. A. Vilkovisky,
`Quantization of gauge theories with linearly dependent generators', Phys.
Rev. {\bf D28} (1983) 2567.}
\REF\henneaux{M. Henneaux, `Lectures on the antifield-BRST formalism for
gauge theories', Proc. of the XXI GIFT Meeting.}
\REF\teitelboim{M. Henneaux and C. Teitelboim, `Quantization of gauge
systems', Princeton University Press, Princeton, New Jersey, 1992, to
be published.}
\REF\sen{A. Sen, Nucl. Phys. {\bf B345} (1990) 551; {\bf B347} (1990) 270.}
\REF\ranganathan{K. Ranganathan, `A Criterion for flatness in minimal
area metrics', MIT-preprint MIT-CTP-1945, to appear in Comm. Math. Phys.}
\REF\wolfzwiebach{M. Wolf and B. Zwiebach, `The plumbing of minimal
area surfaces', IASSNS-92/11, submitted to Comm. Math. Phys. \hep 9202062.}
\REF\wittenzwiebach{E. Witten and B. Zwiebach, `Algebraic Structures
and Differential Geometry in 2D String Theory', IASSNS-HEP-92/4,
to appear in Nucl. Phys. B. \hep 9201056}
\REF\everlinde{E. Verlinde, `The Master Equation of 2D String Theory'
IASSNS-HEP-92/5, to appear in Nucl. Phys. B. \hep 9202021.}
\REF\sentalk{A. Sen, `Some applications of string field theory',
TIFR/TH/91-39, Talk at the `Strings and Symmetries, 1991', conference
held at Stony Brook, May 25-30, 1991. \hep 9109022.}
\REF\stasheffi{J. Stasheff, `Towards a closed string field
theory: topology and convolution algebra', University of North
Carolina preprint, UNC-MATH-90/1, to appear.}
\REF\stasheff{J. Stasheff, `Homotopy associativity of H-spaces, II.',
Trans. Amer. Math. Soc., {\bf 108} (1963) 293; `H-Spaces from a homotopy
point of view', Lecture Notes in Mathematics
{\bf 161} Springer Verlag, 1970.}
\REF\wittennew{E. Witten, `Chern-Simons gauge theory as a string theory',
IASSNS-92/38, June 1992.}
\REF\kontsevich{M. Kontsevich, `The Cohomology of Graphs', to appear.}
\REF\thorn{C. B. Thorn, `Perturbation Theory for Quantized
String Fields', Nucl. Phys. B287 (1987) 61.}
\REF\thornpr{C. B. Thorn, `String field theory', Phys. Rep.
{\bf 174} (1989) 1.}
\REF\bochicchio{M. Bochicchio, `Gauge fixing for the field
theory of the bosonic string', Phys. Lett. {\bf B193} (1987) 31.}
\REF\seneqm{A. Sen, `Equations of motion in non-polynomial closed
string field theory and conformal invariance of two dimensional field
theories', Phys. Lett. {\bf B241} (1990) 350.}
\REF\kugoojima{T. Kugo and I. Ojima, `Local Covariant Operator Formalism
of Non-Abelian Gauge Theories and Quark Confinement Problem', Suppl.
Prog. Theor. Phys. {\bf 66} (1979) 1.}
\REF\strebel{K. Strebel, ``{\it Quadratic Differentials}'', Springer-Verlag
(1984)}
\REF\ranganew{K. Ranganathan, `On the background independence of
overlap string interactions', MIT preprint, to appear.}
\REF\kugozwiebach{T. Kugo and B. Zwiebach, `Target space duality
as a symmetry of string field theory', YITP/K-961, IASSNS-HEP-92/3,
January 1992, to appear in Prog. Theor. Phys. \hep 9201040.}
\REF\sarojasen{R. Saroja and A. Sen, `Picture changing operators
in closed fermionic string field theory', TIFR-TH-92/14, February
1992. \hep 9202087.}
\REF\shenker{S. Shenker, unpublished. See also S. Shenker, `The strength
of nonperturbative effects in string theory'. Proc. Cargese Workshop
on Random Surfaces, Quantum Gravity and Strings, 1990.}
\REF\bpz{A. A. Belavin, A. M. Polyakov, and A. B. Zamolodchikov,
`Infinite conformal symmetry in two-dimensional quantum field
theory', Nucl. Phys. {\bf B241} (1984) 333.}
\REF\schubert{C. Schubert, `The finite gauge transformations in closed
string field theory', MIT preprint, MIT-CTP-1977, May 1991, to appear
in Mod. Phys. Lett. {\bf A}.}
\REF\alvarez{L. Alvarez-Gaume, C. Gomez, G. Moore and C. Vafa, Nucl.
Phys. {\bf B303} (1988) 455;\hfill\break
C. Vafa, Phys. Lett. {\bf B190} (1987) 47.}
\REF\nelson{P. Nelson, Phys. Rev. Lett. {\bf 62} (1989) 993;\hfill\break
H. S. La and P. Nelson, Nucl. Phys. {\bf B332} (1990) 83;\hfill\break
J. Distler and P. Nelson, Comm. Math. Phys. {\bf 138} (1991) 273.}

\chapter{Introduction and Summary of Results}

One can hardly overstate the need for a complete formulation
of string theory. This necessity arises both at a conceptual level and at
a computational level. At the conceptual level we do not yet know
the underlying principles of string theory. Such understanding appears
to be a prerequisite for addressing deep issues of quantum gravity
in the context of string theory. At the computational level we do not have yet
the ability to calculate nonperturbative effects. Such effects seem to be
needed in order to obtain realistic string models.

The subject of this paper is the covariant theory of closed string fields
[\saadizwiebach--\zwiebachma ] This theory, I believe, represents concrete
progress in the construction of a complete formulation of string theory.
The current formulation of closed string field theory
is based in the BRST approach, which originated in the work of Siegel
[\siegel ]. The theory appears to be the natural closed string analog of the
interacting open string field theory constructed by Witten [\witten ].
As we will discuss, it succeeds
in generating the perturbative definition of the theory starting from an
action based on a gauge principle. It also makes conceptually manifest the
factorization and unitarity of the string amplitudes. The closed string field
theory is apparently the first field theory for which the most sophisticated
machinery for quantization, the Batalin-Vilkovisky (BV) field-antifield
formalism [\batalinvilkovisky--\teitelboim ],
is necessary and useful in its full form. At present, however, the
closed string field theory is not yet a complete formulation of string
theory. Its most glaring shortcoming is that its formulation requires
making a choice of a conformal background. In a complete formulation, this
background should arise as a classical solution. Therefore, the
question whether string field theory is the `right' approach to string
theory, is at present the question whether string field theory can be
formulated without having to choose a conformal background.
We may be encouraged by the fact that despite initial difficulties the
construction of covariant closed string field theory was possible.
The resulting theory, as will be elaborated here, has an
interesting and novel algebraic structure arising from subtle properties
of moduli spaces of Riemann surfaces. Moreover, as recent work of Sen
indicates, the string field theory, while not {\it manifestly} independent
of the conformal background chosen to formulate it, does incorporate
such background independence to some degree [\sen ].
There is therefore little evidence that we face difficulties that cannot
altogether be solved in the framework of string field theory by extensions
or modifications of the current formulations.
\medskip
The real obstruction to the construction of
closed string field theory was, all along, the absence of a decomposition
of the moduli spaces of Riemann surfaces compatible with covariant Feynman
rules. A decomposition suitable for the classical theory (genus zero surfaces)
was found in Refs.[\saadizwiebach ,\kks ] by incorporating and generalizing
the closed string extension of the open string vertex [\witten ] and the
tetrahedron of Ref. [\kaku]. The mathematically precise proof that this was
a correct decomposition
revealed that the string diagrams were given by minimal area metrics
[\zwiebachma ], and this suggested a natural way to generalize the
construction for higher genus surfaces.
The decomposition of moduli space for the quantum theory was addressed in
Ref.[\zwiebachqcs ] using minimal area metrics and the ideas of
[\sonodazwiebach ], and the algebraic structure of the resulting theory
was sketched in Refs.[\zwiebachqcs ,\zwiebachtalk ].
The assumptions used in these works regarding plumbing properties of
minimal area metrics have now been established to a large degree in
recent works by Ranganathan, Wolf, and the author
[\ranganathan ,\wolfzwiebach ].

Even though the main difficulty was geometrical, working out the
algebraic implications of the geometrical constructions is both
instructive and necessary. This amounts to giving the explicit construction
of the string field vertices, with due care of technical issues like
antighost insertions, and proving the set of identities they satisfy. These
properties must also be shown to guarantee the consistency of the
quantum theory constructed with the string field vertices.
This is one of the main purposes of the present paper, it amounts
to giving the detailed derivation and construction of the quantum closed
string field theory sketched in Refs.[\zwiebachqcs ,\zwiebachtalk ].
Essentially we do for the full quantum theory what Kugo and Suehiro, in
their important work, did for the classical theory [\kugosuehiro ].  We will
also present new results involving an improved understanding of the algebraic
structure of the theory, as well as new results on the concrete specification
of
the string vertices and the characteristics of minimal area metrics.
Furthermore, we will show how to derive the algebraic structures on
physical states recently discovered [\wittenzwiebach ,\everlinde ]
in the context of two-dimensional string theory from the corresponding
generalized structures of closed string field theory that are defined on the
full off-shell state space.
Let us consider some of the main points in the present paper.
\medskip
\noindent
$\underline{\hbox{Algebraic Structure of Closed String Field Theory}}.\,$
The algebraic ingredients of the classical closed string field theory,
as emphasized in [\sentalk ], are a set of multilinear, graded-commutative
string products $m_n$, with $n\geq 2$, ($n$ is the number of string fields
to be multiplied) satisfying a set of Jacobi-type identities involving the BRST
operator $Q$. This structure was seen to correspond to a homotopy Lie algebra
[\stasheffi ]. This homotopy Lie algebra will be denoted as $L_\infty$,
in analogy to the homotopy associative $A_\infty$ algebras [\stasheff ],
which are relavant to open string field theory and Chern-Simons theory
[\wittennew ,\kontsevich ].
We identify the closed string $L_\infty$ structure precisely by choosing a set
of conventions that eliminates the need for ghost insertions in the identities
relating the string products.\foot{I thank E. Witten for suggesting
such change of conventions.} This allows to interpret the BRST operator
as generating the first product $m_1$ of the homotopy Lie algebra
($\l B \r \equiv QB$ where $B$ is a string field), which streamlines the proof
of gauge invariance. We compute the algebra of gauge transformations, and
determine the field dependent structure constants and the extra terms
that vanish upon use of the equations of motion. At the quantum level we
have string products labeled by genus. The identity
satisfied by all the products is given. This structure is not anymore
an $L_\infty$ structure due to the appearance of new terms involving
a `trace' over two of the entries of the string products. This new structure
is shown to guarantee that the master action satisfies the quantum BV master
equation. To this end, following the earlier work in open strings
[\thorn--\bochicchio ], we give a convenient assignment of fields and
antifields in closed string field theory.
Extending a result of Sen [\seneqm ], we show that upon
shifting the string field by a term that {\it does not} satisfy the classical
field equations, the $L_\infty$ algebra of the classical theory involving
products $m_n$ with $m\geq 1$, acquires the zeroth order product $m_0$, this
product being just a special state vector. Despite the indirect construction,
this is the algebraic structure of closed string field theory formulated around
a background that is {\it not} conformal. If we denote the special state by
$F$,
then the lowest identities of the resulting $L_\infty$ algebra imply that the
generalized $Q$ satisfies $QF=0$, and $Q^2 B = -[F,B]$ for all string fields
$B$. The new $Q$ also fails to be a derivation with respect to the product,
the failure being related to $F$ being nonzero. We suggest the possible
relevance of these structures to the problem of formulating string field
theory without use of conformal backgrounds.
\medskip
\noindent
$\underline{\hbox{Algebraic Structures on Physical States}}.\,$
The $L_\infty$ structure and the BV structure of the closed string field theory
is defined on the complete off-shell state space of the theory. On the other
hand in recent works similar structures were obtained on the physical states
of two-dimensional string theory [\wittenzwiebach ,\everlinde ]. The homotopy
Lie algebra on physical states has $m_0 = m_1 = 0$, and the first
nonvanishing product is $m_2$. Here we show that, as conjectured in these
works, the algebraic structures on physical states can be derived from those
on the off-shell state space. This is {\it general}, and not peculiar to
$D=2$ string theory. We do this by showing how to truncate the basic
identity of string field theory to physical states. This cannot be done
naively with the original string products because of two reasons. First,
when the closed string field products $m_n$ (for $m\geq 3$) are used
to multiply physical fields, the resulting product is
a physical field plus a BRST-trivial field, plus an {\it unphysical}
field, which is problematic since it cannot be absorbed as part of
the physical
fields. Thus the original products do not close on physical fields.
A second difficulty is that the original products, since they involve
subsets of moduli spaces, would not agree with the physical products,
which were defined by correlators over the full moduli spaces.
These two difficulties have the same solution. We take a one parameter
family of string field products, as the parameter grows the
subsets defining the products grow, until they eventually cover the
full moduli spaces. In this limit we can see that BRST decoupling
of trivial states implies that the product of physical states cannot
give an unphysical state, and we are able to derive the identities
on physical states from the main identity of string field theory.
For the full master equation this requires the use of the
Kugo-Ojima quartet mechanism [\kugoojima ].
\medskip
\noindent
$\underline{\hbox{From Riemann surfaces to Feynman diagrams}}.\,$
The main difficulty in constructing string field theory was due
to the lack of a suitable decomposition of the moduli spaces of
closed Riemann surfaces. Any naive set of
Feynman rules led to multiple or infinite overcounting of surfaces.
Given a Riemann surface one must to be able to associate to
it a {\it unique} Feynman diagram. This is achieved with a generalized
minimal area problem [\zwiebachma ] which asks for the
(conformal) metric of least possible area under the condition that
all nontrivial closed curves on the surface be longer than or equal
to some fixed length, conventionally chosen to be $2\pi$.
Staring at the surface with
its minimal area metric one can reconstruct the Feynman diagram from
which it arises (\S6.4). The idea is that a minimal area metric must have
closed geodesics of length $2\pi$ that foliate the surface. These
geodesics form foliation bands, that is, annuli foliated by
homotopic geodesics. The height of a band is defined to be the
shortest distance (on the band) between its boundaries.
In contrast with metrics that arise from Jenkins-Strebel quadratic
differentials [\strebel ], where the geodesics (horizontal trajectories)
intersect in zero measure sets (critical graphs), the minimal
area metrics can have bands of geodesics that cross. \foot{One such
example was given in Ref.[\wolfzwiebach], Fig.5.} A band of height
greater or equal to $2\pi$ cannot be crossed by any other band of
geodesics and must be isometric to a flat cylinder of circumference
$2\pi$ [\ranganathan ,\wolfzwiebach ]. These are the propagators.
The bands of infinite height are the external legs of the diagram.
If the surface has no finite height foliation
of height bigger than $2\pi$, then the whole string
diagram corresponds to an elementary interaction $\V_{g,n}$, where
$n\geq 0$ is the number of infinite height foliations, and $g$ is the
genus of the surface. On each propagator, which is
a flat cylinder, we mark two closed geodesics, each a distance
$\pi$ from each boundary of the cylinder. We also mark one curve,
a distance $\pi$ from the boundary on the semi-infinite cylinders.
The string diagram is cut open along all these curves and the surface
breaks up into a number of semi-infinite cylinders,
the external legs, a number of finite cylinders, the
propagators, and a number of surfaces with boundaries, that correspond
to the elementary interactions. Those surfaces have `stubs' of height
$\pi$ on each of their boundaries, arising from the height $\pi$
foliation that was left attached from the bands that gave rise to
the propagators. It is interesting that a foliation of height
less than $2\pi$, even though it may look like a propagator
(if it happens not to be crossed by another foliation), must not be
considered as such.
It is as if, due to the minimum length condition of $2\pi$ on the
string diagram, propagation by a distance smaller than $2\pi$ does not
qualify as free propagation. The closed geodesics correspond to the
closed strings, and the fact that the bands of foliations cross, for
genus one and above, may be interpreted as saying that it is not
clear where the string is. This crossing is necessary for modular
invariance, and only happens within the elementary vertices. The
difficulty with modular invariance in closed string theory
was due to our inability to tell in which direction the strings propagate.
This ambiguity does not exist for the minimal
area string diagrams; propagators are
unambiguously determined.
\medskip
\noindent
$\underline{\hbox{The uses of Stubs and the BV Equation}}.\,$
The role of the stubs is to prevent the appearance of short curves
upon sewing. Since we leave a stub of length $\pi$ on every boundary,
upon gluing two boundaries and creating new closed curves, these
will be longer than or equal to $2\pi$. The stubs allowed us to
characterize the elementary interactions of the complete quantum theory of
closed strings as the surfaces with no finite height foliation of length
greater than $2\pi$. The stubs may not be
essential for the classical theory since sewing in tree configurations
does not introduce short curves. That is why, at the classical level only,
we can specify the elementary interactions as the surfaces with
no finite height foliations; these are the restricted polyhedra
of [\saadizwiebach ,\kks ]. Such interactions, which are contact-type
overlaps, would be expected to be background independent [\ranganew ]
in the sense discussed recently in [\kugozwiebach ]. Then, as in open string
field theory, the only background dependent ingredient of the classical
theory would be the BRST operator.
On the other hand stubs may be relevant in the classical theory, for
using stubs string products are well defined for a set of string
fields larger than that for which overlap-type string products exist.
Furthermore, stubs are helpful in closed superstring field theory to prevent
the collission of picture changing operators [\sarojasen ].
At the quantum level the stubs are essential and have the great
virtue of making the full quantum BV equation true without
need of regularization (in usual formulations of particle field theory, or in
string theory with contact interactions, the `delta' term in the BV equation
needs regularization.)
The closed string field theory master action amounts to a perturbative
solution of the BV equation. It is of great interest to understand if this
series solution has a finite radius of convergence or it is an asymptotic
expansion. The case for the latter possibility has been emphasized by
Shenker, who also indicated the possible relation with nonperturbative
effects in string theory [\shenker ].
\medskip
\noindent
$\underline{\hbox{On Vacuum graphs}}.\,$ The BV master equation does not
give us any constraint on the vacuum graphs of the string field theory.
This is so because in the BV equation the master action always appears
differentiated with respect to fields (or antifields). The master equation
only guarantees the gauge independence of the S-matrix, and therefore
it does not ensure that the string field theory will give the values of
vacuum graphs that one would expect. Our geometrical recursion relations
(\S5 ) determine the `vacuum vertices' of the string field theory for
genus greater or equal to two. In this paper we will include in the master
action the $\hbar$ dependent constants that arise from the vacuum vertices.
This is done, even though the constants drop out of the master
equation, because it seems likely that such constants would actually
arise from a fundamental underlying action (once we find it!) upon
expansion around a classical solution. The case of genus one is quite
puzzling. The geometrical recursion relations do not determine this
vacuum vertex. If we would compute this vacuum graph using just the
propagator, we would get an infinite cover of the moduli space. If we insist
on getting the genus one vacuum graph to work we must use a cutoff propagator
instead of stubs on the vertices. The cutoff-propagator would include tubes
with length in the interval $[ 2\pi ,\infty]$ rather than in the interval
$[0,\infty]$. If we do this there is no need for stubs; the price we
pay is a somewhat less natural kinetic term, which would contain,
in addition to the BRST operator, an insertion of $\exp (2\pi L_0^+)$.
It seems clear that we need further understanding of background
independence in order to clarify the issues above.
\bigskip
Let us now indicate briefly the contents of this work. Sections
\S2, \S3 and \S4 are essentially algebraic. In \S2 we
discuss all the conventions that we will use and derive basic
relations necessary for the later sections. We give a detailed
discussion of Hermitian conjugation, BPZ conjugation [\bpz ],
and inner products. We discuss sewing kets and bras (the so-called
reflector states), their BRST properties and their exchange symmetry
properties. We elaborate on the properties of basis states and
their conjugate states. In \S3 we study the string field and the
kinetic term. We explain in detail the subsidiary conditions on
the string field, the ghosts number conditions, and Grassmanality.
We discuss target space ghost number. The kinetic term is given
and its basic properties are established. We also specify bases
for fields and antifields, give the conventions for the BV formalism
and verify that the kinetic term satisfies the master equation.
In \S4 the subject is the algebraic structure of the string field
theory. We state the properties of the string products and give
the main identity satisfied by the products, along with examples.
We also introduce the multilinear functions and give their properties.
We then give the classical action, the field equation,
prove gauge invariance, and compute the commutator of two gauge
transformations.\foot{This commutator has been independently
obtained by Schubert [\schubert ].} Finally we give the quantum master action,
show it satisfies the quantum master equation, derive the master
transformations
and the BRST transformations. This section
concludes with a mathematical definition for the $L_\infty$ algebra
and an analysis of the algebraic structure that would arise if closed string
field theory were constructed around a non-conformal background.

Sections \S5 and \S6 deal with the geometrical issues that underlie
the formulation of the string field theory. In \S5 we define precisely
the notion of string vertices $\V_{g,n}$ (which are sets of Riemann surfaces),
and state the basic conditions they must satisfy. We then discuss and
elaborate on the fundamental geometrical relation [\sonodazwiebach ],
satisfied if the string vertices, via the Feynman rules, generate
a decomposition of the relevant moduli spaces. In \S6 we show how the
minimal area problem determines the subsets $\V_{g,n}$. Our discussion
is based fully on the properties of the geodesics in minimal area
metrics. We state and prove a cutting theorem, which together with
the results of [\wolfzwiebach ] completes the study of the sewing
properties of minimal area metrics. We also state what remains to be proven
in order to have complete mathematical treatment of minimal area metrics.

Section \S7 combines the geometrical and the algebraic structures in order to
construct the string products and multilinear functions and prove their
basic properties. This requires a detailed analysis of Schiffer variations
(deformations of punctured Riemann surfaces with local coordinates) and the
construction of a set of differential forms on the space $\wh\P_{g,n}$
(the space of surfaces of genus $g$ and $n$ punctures, with local coordinates
at the punctures specified up to phases). Using heavily
results of the operator formalism of Alvarez-Gaume, {\it et.al.} [\alvarez ],
further developed and clarified by Nelson and others [\nelson ],
we show that the differential forms are well defined if the off-shell states
that label the forms satisfy the subsidiary conditions  $b_0^-=L_0^-=0$.
The BRST operator is shown to act as a total derivative on the
forms defined in $\wh\P_{g,n}$. We define the string products and the
string multilinear functions and establish some of their simple properties.
Finally we prove the hermiticity of the string field vertices.

In section \S8 we establish the main identity satisfied by the string
field products, this involves a detailed discussion of sewing of
string field vertices. Section \S9 is the final section. It shows how to
obtain the algebraic structures on physical states from the algebraic
structures on the whole state space.
\medskip
The present paper is long because of the large amount of material covered
and because I wished to present a complete and useful set of conventions
for closed string field theory along with detailed proofs of the most
important properties. The material presented here is about 40$\%$ completely
new; about $40\%$ is proofs of results only sketched or announced before,
and $20\%$ is background material included for completeness.

\REF\leclair {A. LeClair, M. E. Peskin, and C. R. Preitschopf, Nucl.
Phys. {\bf B317} (1989) 411.}

\chapter{Conformal Field Theory, Conjugation and Inner Products}

In this section we will give the conventions with which we will
formulate the string field theory and some of the basic ingredients of
the formalism. We begin with a description of the relevant ideas
of conformal field theory, and a study of the ghost sector and
its vacuum structure. Then we turn to defining hermitian conjugation
and the associated antilinear inner product. Next we consider
BPZ conjugation and introduce the associated linear inner product.
These are essential to be able to write actions. We discuss bases
of states for conformal field theory and derive a series of basic
properties that will be essential to discuss the string field theory.
The first work emphasizing the use of conformal field theory
to define string field theory is that of LeClair, Peskin and Preitschopf
[\leclair ].

\section{Conformal Field Theory}

Bosonic string field theory is formulated using a conformal field theory.
This conformal field theory has a matter sector, which is some
conformal field theory of total central charge equal to $(+26)$
and a ghost sector, which is the conformal field theory of the
reparametrization ghosts and has central charge equal to $(-26)$.
The total conformal field theory has therefore total central charge
equal to zero.

Consider the conformal field theory formulated in the $z$-plane
with $z= \exp (\tau + i\sigma )$. While the whole construction of
the string field theory can be carried out abstractly, we can
describe very concretely the ghost sector of the conformal field
theory. Consider the case when the conformal field theory has
holomorphic and antiholomorphic fields
$\phi (z), \ov \phi (\bz )$ and stress energy tensors
$T(z), \ov T (\bz)$. The field $\phi(z)$ is said to be a primary
field of conformal dimension $d$ if
$$T(z) \phi(w) \sim {d\over (z-w)^2} \phi (w) + {1\over z-w}
\partial_w \phi (w) + \cdots ,\eqn\opematt$$
corresponding to the transformation law
$$\phi'(z') (dz')^d = \phi (z) (dz)^d.\eqn\transflaw$$
The field is expanded in a basis of operators (oscillators)
in the form
$$\phi(z) = \sum_n {\phi_n \over z^{n+d}}, \quad
\phi_n = \int {dz \over 2 \pi i} z^{n+d-1} \phi(z) . \eqn\oscexpmat$$
The operator $\phi_n$ is of dimension $(-n)$ (recall $z$ counts as
dimension minus one). The total stress tensor, has central charge zero
and is a dimension two field (it transforms as a quadratic differential
on a surface). Its oscillator expansion is given by
$$T(z) = \sum_n {L_n \over z^{n+2}}, \quad
\ov T(z) = \sum_n {\ov L_n \over \ov z^{n+2}}.\eqn\oevo$$

The SL(2,C) vacuum, denoted by $\ket{{\bf 1}}$, is the vacuum in the
asymptotic past and it is associated with $z=0$, with no operator
inserted there. It follows from \oscexpmat\ and regularity that
$$ \phi_n \ket{{\bf 1}} = 0 \quad \hbox{for} \quad n\geq -d + 1.\eqn\annihvac$$
One introduces a state in the dual space, the SL(2,C) vacuum in the
asymptotic future $z=\infty$, denoted by $\bra{{\bf 1}}$. It also
follows from \oscexpmat\ (and \transflaw\ for $z'=1/z$) that
$$\bra{{\bf 1}} \phi_n = 0 \quad \hbox{for} \quad n\leq d-1 .\eqn\annihavac$$

The canonical example of matter field theory is the theory of $26$
free bosons $i\partial X^\mu (z)$, $i\ov \partial\ov X^\mu (\bz)$,
of dimensions $(1,0)$ and $(0,1)$ respectively (the first entry in
$(\cdot , \cdot )$ refers to the dimension with respect to the
holomorphic part of the stress tensor, and the second gives the
dimension with respect to the antiholomorphic part of the stress
tensor). The mode expansions are:
$$i\partial X^\mu (z)= \sum_n {\alpha_n^\mu \over z^{n+1}},
\quad  i\ov \partial \,\ov X^\mu (\bz)=
\sum_n {\ov \alpha_n^\mu \over \bz^{n+1}}.
\eqn\bosonicosc$$
Here the oscillators satisfy the algebra
$$[\alpha_n^\mu , \alpha_n^\nu ]
= [\ov \alpha_n^\mu , \ov \alpha_n^\nu ] = n g^{\mu\nu} \delta_{n+m,0},
\eqn\commutx$$
where $g^{\mu\nu}$ is the Minkowskian metric $(-,+\cdots +)$ (the case
of compactified coordinates involves a few more subtleties, see for
example the recent treatment in [\kugozwiebach ]). Moreover
$\alpha_0^\mu = \ov \alpha_0^\nu = \wh p^\mu$. We then have
$$\alpha_n \ket{{\bf 1}}
= \ov \alpha_n \ket{{\bf 1}} = 0 \quad \hbox{for} \quad n\geq 0,
\quad\hbox{and} \quad
 \bra{{\bf 1}}\alpha_n =
\bra{{\bf 1}}\ov\alpha_n = 0 \quad\hbox{for}\quad n\leq 0. \eqn\actvacm$$
One actually defines a family of vacua $\ket{{\bf 1}, p}$, and dual
vacua $\bra{{\bf 1}, p}$
labelled by the eigenvalue of the momentum operator $\wh p$ via
$$\wh p \,\ket{{\bf 1}, p} = p \ket{{\bf 1}, p}, \quad\hbox{and}\quad
\bra{{\bf 1}, p}\, \wh p = \bra{{\bf 1}, p} p.\eqn\momvac$$
For these vacua the oscillators that annihilate it are given by
$$\alpha_n \ket{{\bf 1},p}
= \ov \alpha_n \ket{{\bf 1},p} = 0 \quad \hbox{for} \quad n> 0,
\quad\hbox{and} \quad
 \bra{{\bf 1},p}\alpha_n =
\bra{{\bf 1},p}\ov\alpha_n = 0 \quad\hbox{for}\quad n< 0, \eqn\actvacmat$$
as follows from the two previous equations.

Let us now consider the ghost conformal field theory.
We have ghost fields $c(z)$ and $ \ov c (\bz )$
of dimensions $(-1,0)$ and $(0,-1)$ respectively, and
antighost fields $b(z)$ and $\ov b (\bz)$ of dimensions
$(2,0)$ and $(0,2)$ respectively. We therefore have
$$c(z) = \sum_n {c_n\over z^{n-1}},\quad
\ov {c}(\bz ) = \sum_n {\bar{c}_n\over\bar{z}^{n-1}},\eqn\first$$
$$b(z ) = \sum_n {b_n\over z^{n+2}},\quad
\ov {b}(\bz ) = \sum_n {\bar{b}_n\over \bar{z}^{n+2}}.\eqn\second$$
The modes satisfy the anticommutation relations
$$\{b_n , c_m\} = \{ \bar{b}_n , \bar{c}_m \} = \delta_{m+n,0},\eqn\anticom$$
with all other anticommutators equal to zero.
It is convenient to define new zero modes by linear combinations
of the old ones
$$c_0^\pm={1\over 2}( c_0 \pm \bar{c}_0),
\quad b_0^\pm = b_0\pm\bar{b}_0.\eqn\newzm$$
It follows from equations \annihvac\ and \annihavac\ that
$$\eqalign{
b_n \ket{{\bf 1},p} &= \bar{b}_n \ket{{\bf 1},p} = 0 \quad\hbox{for}\quad
n\geq -1 \cr
c_n \ket{{\bf 1},p} &= \bar{c}_n \ket{{\bf 1},p} = 0,\quad\hbox{for}
\quad n\geq 2, \cr} \eqn\ghostvac$$
and
$$\eqalign{
\bra{{\bf 1},p} b_n &= \bra{{\bf 1},p}\bar{b}_n = 0
\quad\hbox{for}\quad n\leq 1, \cr
\bra{{\bf 1},p}c_n &= \bra{{\bf 1},p} \bar{c}_n =0,\quad
\hbox{for}\quad  n\leq -2.
\cr}\eqn\ghostavac$$

We define the first quantized ghost number operator $G$ by
$$ G = 3 + \biggl[ {1\over 2} (c_0 b_0 - b_0c_0) +
\sum_{n=1}^\infty (c_{-n}b_n -b_{-n}c_n) + \hbox{a.h.} \biggr] ,
\eqn\ghostdef$$
where a.h. denotes an identical contribution from the antiholomorphic
sector. We included the constant shift of $(+3)$ in order to have
$G\ket{{\bf 1},p} = 0$, that is, the ghost number of the
SL(2,C) vacuum is zero.

Due to the presence of two zero modes the state space
space breaks into four sectors built on top of four
different vacua forming a representation of the algebra
of zero modes. These are defined as follows:
$$\eqalign{
\ket{\da \da,p } &\equiv c_1\bar{c}_1 \ket{{\bf 1},p} , \cr
\ket{\ua  \da ,p }&\equiv c_0^+ \ket{\da  \da ,p} , \cr
\ket {\da  \ua ,p} &\equiv c_0^- \ket {\da  \da ,p} , \cr
\ket {\ua  \ua ,p } &\equiv c_0^+ c_0^- \ket{\da \da ,p} .\cr}
\eqn\zerovac$$
The state $\ket{\da\da ,p}$ is a state of ghost number $(+2)$ and it
is annihilated by $b_0^{\pm}$. This state is annihilated by all
$b_n$ and $c_n$ with $n\geq 1$ (similarly for the antiholomorphic
oscillators). From left to right, the first entry in the kets refers to the
$(+)$ zero modes and the second entry refers to the $(-)$ zero modes.

The conformal field theory stress tensor is the sum of the matter stress
tensor $(T_m(z),\ov T_m(\ov z))$, and the ghost stress tensor
$(T_g(z),\ov T_g(\ov z))$. The latter is given by
$$T_g(z) = -2b(z)\cdot \partial c(z) -\partial b(z) \cdot c(z),\eqn\ghset$$
and the basic operator product expansion is
$$b(z)c(w) \sim {1\over z-w} . \eqn\opebc$$
The BRST operator of the theory is given by
$$Q = \int {dz\over 2\pi i} c(z) \bigl( T_m(z)+{1\over 2}T_g(z)\bigr)
+\int {d\ov z\over 2\pi i} \ov c(\ov z) \bigl( \ov T_m(\ov z)+ {1\over 2}
\ov T_g(\ov z)\bigr),\eqn\brstdef$$
where the operators in the integrand are normal ordered. Using the
above three equations, one can verify that
$$\{ Q , b(z) \} = T_m(z)+T_g(z) = T(z),\quad
\{ Q , \ov b(z) \} = \ov T_m(\ov z)+\ov T_g(\ov z) = \ov T(\ov z).
\eqn\frbb$$
This equation implies that $Q$ is of the form
$$Q = c_0 L_0 + \ov c_0 \ov L_0 + \cdots ,\eqn\espqb$$
where the dots indicate terms that do not involve the zero
modes of the ghost fields. If we define
$$L_0^\pm = L_0 \pm \ov L_0, \eqn\lpmst$$
this together with \newzm\ imply that \espqb\ can be
written as
$$Q = c_0^+ L_0^+ + c_0^- L_0^- + \cdots .\eqn\espqb$$

The contribution of the ghosts to $L_0$ is given by
$$L_{0(g)} = -1+\sum_{n=1}^\infty n(b_{-n}c_n + c_{-n}b_n),\eqn\emghost$$
with a similar expression for the antiholomorphic part. We then have that
$L_{0(g)}$ and $\ov L_{0(g)}$ annihilate the vacua
$$L_{0(g)} \ket{{\bf 1},p} = \overline L_{0(g)} \ket{{\bf 1},p} = 0.
\eqn\virvac$$
The $\ket{\da\da ,p}$ vacua are tachyonic as far as the ghost energy
momentum is concerned, one has
$$L_{0(g)} \ket{\da\da ,p} = \overline L_{0(g)} \ket{\da\da ,p} = -1\cdot
\ket{\da\da ,p}.\eqn\lzeroonvac$$
Since, for the usual bosonic string theory $L_{0(m)}+\ov L_{0(m)}=p^2+\cdots$,
where the dots correspond to terms that annihilate the $\ket{\da\da ,p}$
vacua, one has that the total $L_0^+$ operator on these vacua gives
$$L_0^+ \ket{\da\da ,p} = (p^2-2)\ket{\da\da , p}.\eqn\lasitc$$

\section{Hermitian Conjugation, BPZ conjugation and Inner Products}

Given a conformal field theory we can define an
antilinear inner product, arising from Hermitian conjugation
and a bilinear inner product, arising from BPZ conjugation.
Both types of conjugations relate states built on the
in-vacuum to states built on the out-vacuum. In order to
be able to define these products we need to define a basic overlap.
For the conformal field theory giving rise to bosonic string theory, we take
$$\bra{{\bf 1},p} c_{-1}\bar{c}_{-1}c_0^+ c_0^- c_1 \bar{c}_1
\ket{{\bf 1}, p'} =(2\pi)^d \cdot \delta^d (p-p'), \eqn\begininner$$
where $\delta^d$ is the $d$-dimensional delta function.
The absence of an $i$ in the right hand side, while very
convenient to avoid unnecessary factors of $i$, will result in a
slightly unusual formula for hermitian conjugation. For other
conformal field theories one may take different definitions for
the basic overlap.

\noindent
$\underline{\hbox{Hermitian conjugation and}\, (\cdot , \cdot )}.\,$
Hermitian conjugation will be relevant to us since there is a need to impose
a reality condition on the string field, as we will see in \S3.1.
Let us consider for concreteness the bosonic string. The zero mode
operators for the matter, namely the $\wh x$'s and $\wh p$'s, are both
hermitian operators: $\wh x^\dagger = \wh x$ and
$\wh p^\dagger = \wh p$. One also has
$$ \ket{{\bf 1},p} = e^{ip\wh x} \ket{{\bf 1}},\quad
\hbox{and}\quad \bra{{\bf 1},p} = (\ket{{\bf 1},p})^\dagger =
\bra{{\bf 1}}e^{-ip\wh x}, \eqn\hermconj$$
which, in agreement with our earlier definition, implies
$\bra{{\bf 1},p} \wh p = \bra{{\bf 1},p} p$. One defines for all
the oscillators (all $n$)
$$\alpha_n^\dagger = \alpha_{-n}, \quad
c_n^\dagger = c_{-n},\quad
b_n^\dagger = b_{-n},\eqn\hermosc$$
and similarly for the antiholomorphic oscillators. Finally the hermitian
conjugate of a general state is defined via
$$ \biggl( A_1 \cdots A_n \ket{{\bf 1},p} c \biggr)^\dagger
\equiv \ov c \bra{{\bf 1},p} A_n^\dagger \cdots A_1^\dagger , \eqn\genherm$$
where the $A$'s represent oscillators, $c$ is a number, and the bar denotes
complex conjugation.
We will denote the hermitian conjugate state corresponding to
$\ket{A}$ as $\bra{A_{hc}}$. The antilinear inner product is defined
as follows: given two ket states $\ket{A}$ and $\ket{B}$ one defines
$$(A , B ) \equiv \bra{A_{hc}} B \rangle .\quad
\hbox{Antilinear Inner Product}
\eqn\alinear$$
The label `$hc$' is supressed for momentum eigenstrates; $\bra{p}$ will
always denote the hermitian conjugate of $\ket{p}$.
As desired, the inner product is antilinear in the first
argument and linear in the second
$$\eqalign{
( A\alpha + A' \alpha ' , B) &= \alpha^* (A,B) + {\alpha'}^* (A', B),\cr
( A, B \beta + B' \beta ') &=  (A,B)\beta  + (A, B')\beta '.
\cr}\eqn\bsicprop$$
It follows from the definition of the basic overlap in \begininner\ that
$$ (A,B)^\dagger = - (B,A), \eqn\conjherm$$
or in other words $(\bra{A_{hc}}B\rangle )^\dagger = -\bra{B_{hc}}A\rangle$.
The unusual minus sign arises due to the absence of an $i$ factor in the right
hand side of \begininner , and because the hermitian conjugate of the
$(c_{-1}\ov c_{-1} c_0^+ c_0^- c_1 \ov c_1)$ is minus itself.
We have used the dagger in the left hand side
of \conjherm\ instead of complex conjugation in order to take into
account the fact that the c-numbers that multiply the states may have
nontrivial statistics.

Hermitian conjugation of an operator is defined in analogy to \genherm\
and clearly gives a normal ordered operator if we start with one.
We then have, by construction, that
$$ (\O A , B ) = (  A ,\O^\dagger B). \eqn\adjointeqn$$
The BRST operator can be shown to be hermitian, and we therefore have
$$ (QA, B) = (A, QB),\eqn\qhermitian$$
for any two states $A, B$ (of whatever statistics).

Note finally that we can use the inner product to define a
norm in the usual way. Equation \begininner\ implies that only
states of ghost number three have nonvanishing norm. The inner
product we have defined pairs states of ghost numbers that add
up to six. In terms of the first quantized ghost number operator
$G$ introduced earlier, the ghost number of $\bra{{\bf 1}}$ is six,
\foot{Note that $G$ is of the form $G= 3+{\cal G}$, where
${\cal G}$ is an antihermitian operator.}
any ghost oscillator acting on it subtracts one unit of ghost
number and any antighost oscillator adds one unit.
It is actually more convenient to think of the ghost number of
$\bra{{\bf 1}}$ as zero, then any ghost (antighost) oscillator adds
(subtracts) one unit of ghost number, just as for states. In this
convention, that we will follow throughout, the ghost number of
a state does not change under hermitian conjugation.

\noindent
$\underline{\hbox{BPZ conjugation and}\, \langle \cdot , \cdot \rangle }.\,$
There is another type of conjugation which will be most useful in
our writing of actions. It is BPZ conjugation [\bpz ], and relates
in-states to out-states.
Given a state $\ket{A}= A(0)\ket{{\bf 1}}$, with $A$ a normal ordered
operator, one defines the associated BPZ conjugate state
$$\bra{A} \equiv \bra{{\bf 1}} I\circ A(0), \eqn\defbpzconj$$
where $I$ denotes the conformal mapping $I(z) = 1/z$\foot{The
choice of $1/z$ instead of the standard one of $-1/z$ in BPZ is more
convenient for closed strings.}. The
conformal mapping acting on the operator is defined as
$I \circ A(z,\ov z) \equiv A'(z'(z), \ov z '(\ov z ) )$
where $z'=I(z), \ov z ' = I (\ov z) $ and $A'(z',\ov z ')$ is related
to $A(z,\ov z )$ via the conformal transformation properties of the operator.

For the mode expansion \oscexpmat\ one takes
$$\phi_n^T \equiv I\circ \phi_n
=\int {dz'\over 2\pi i} {z'}^{n+d-1} \phi ' (z')
=\int {dz'\over 2\pi i}{z'}^{n+d-1}\bigl( {-1\over {z'}^2}\bigr)
\phi\bigl( {1\over z'}\bigr)
= (-1)^d\phi_{-n},\eqn\defbpzosc$$
where we have used the superscript $(T)$ to denote this type of
conjugation. The above formula enables us to calculate the BPZ conjugate
of a string of oscillators:
$$ (c \cdot a_n b_m c_p \cdots \ket{{\bf 1}})^T
=\bra{{\bf 1}} \, c \cdot (a_n)^T (b_m)^T (c_p)^T \cdots\eqn\bconjosc$$
Note that the order of the operators does not change
and that the c-number is not complex conjugated. The ghost number
of the BPZ conjugate state is the same as that of the original state.
It follows from \defbpzosc\ that
$\wh p^T = - \wh p $, and therefore we must take $(\ket{{\bf 1},p})^T$
$=\bra{{\bf 1}, -p}$
The reader can check that the BPZ conjugates of the various
ghost vacua (\zerovac ) are given by
$$\eqalign{
\bra{\da \da ,p } &\equiv \bra{{\bf 1},-p} c_{-1}\bar{c}_{-1} , \cr
\bra{\ua  \da ,p }&\equiv  -\bra{{\bf 1},-p} c_0^+ c_{-1}\bar{c}_{-1} , \cr
\bra {\da  \ua ,p} &\equiv -\bra{{\bf 1} ,-p}c_0^-c_{-1}\bar{c}_{-1} , \cr
\bra {\ua  \ua ,p} &\equiv \bra{{\bf 1},-p}
c_0^+ c_0^-c_{-1}\bar{c}_{-1} .\cr}
\eqn\zerovacbra$$
Using BPZ conjugation one defines an inner product which is
linear on both arguments
$$\langle A, B \rangle \equiv \bra{A} c_0^- \ket{B} ,
\quad\hbox{Linear Inner Product}. \eqn\linearip$$
The utility of including the factor of $c_0^-$ in the inner product
arises because our string fields will be defined to be annihilated
by $b_0^-$, and the naive inner product of such states would vanish.
In order to elucidate the properties of this inner product let us
discuss the so-called conjugate states.

Given a complete basis of states in a conformal field theory
$\{ \ket{\Phi_r} \}$, where $r$ is a label, which may include indices
and/or continuous parameters, one defines the {\it conjugate states}
$\{ \bra{\Phi_r^c} \}$ (not hermitian conjugates nor BPZ conjugates)
via the condition
$$\vev{\Phi_r^c | \Phi_s}= \delta_{rs}.\eqn\conjstates$$
The conjugate states form a basis for all the states built on
the out-vacuum (bra-states). Let us understand more explicitly
this conjugation. We can construct a basis by using
states built with oscillators. For example, the state conjugate to
the vacuum is
$$\bra{{\bf 1}^c}=(2\pi)^{-d}\bra{{\bf 1}}\,c_{-1}
\ov c_{-1}c_0^+c_0^- c_1\ov c_1, \eqn\conjvacuum$$
It then follows by BPZ conjugation that
$\ket{{\bf 1}^c} = (2\pi)^{-d}c_{-1}
\ov c_{-1}c_0^+c_0^-c_1\ov c_1\ket{{\bf 1}}$ and that
$\bra{{\bf 1}} {\bf 1}^c\rangle = 1$. Doing the same operations
with an arbitrary state, we have
$$\eqalign{
{}&\ket{\Phi_s} = (c_{-n_1} \cdots c_{-n_i})(b_{-m_1} \cdots b_{-m_j})
(a_{-p_1}\cdots a_{-p_k})\ket{{\bf 1}} , \cr
{}&\bra{\Phi_s^c} = \bra{{\bf 1}^c} (a_{p_k}\cdots a_{p_1})
(c_{m_j} \cdots c_{m_1})(b_{n_i} \cdots b_{n_1}) , \cr
{}&\bra{\Phi_s} = (-)^{i+k} \bra{{\bf 1}} (c_{n_1} \cdots c_{n_i})
(b_{m_1} \cdots b_{m_j})
(a_{p_1}\cdots a_{p_k}) , \cr
{}&\ket{\Phi_s^c} = (-)^{k+j}
(a_{-p_k}\cdots a_{-p_1})
(c_{-m_j} \cdots c_{-m_1})(b_{-n_i} \cdots b_{-n_1})
\ket{{\bf 1}^c} . \cr}\eqn\exampleconj$$
It follows from the above that $\bra{\Phi_s}\Phi_s^c \rangle =(-1)^{(i+j)}$,
but $(i+j)$ simply denotes the statistics of the state $\ket{\Phi_s}$.
We therefore infer that corresponding to \conjstates\ we have
\foot{Whenever a field appears
in the exponent of a sign factor, as in $(-)^\Phi$, it denotes what
is usually called the Grassmanality $\epsilon (\Phi )$ of the field
$\Phi$. This Grassmanality $\epsilon (\Phi )$ is an even integer, if
$\Phi$ is Grassmann even, or an odd integer, if $\Phi$ is Grassmann odd.}
$$\bra{\Phi_r}\Phi_s^c\rangle = (-)^{\Phi_r}\delta_{rs}.\eqn\otherorder$$
Completeness of the states
implies that the identity operator can be written as
$${\bf 1} = \sum_r \ket{\Phi_r} \bra{\Phi_r^c}
= \sum_r \ket{\Phi_r^c} \bra{\Phi_r} (-)^{\Phi_r} ,\eqn\completeness$$
as can be checked by multiplying arbitrary states from the left and from
the right, and making use of \conjstates\ and \otherorder .
We claim now that for arbitrary states $\ket{A}$ and $\ket{B}$ of
definite statistics one has
$$\bra{A} B\rangle = \bra{B} A \rangle \, (-1)^{AB}.\eqn\symmbpz$$
This is established by writing $\bra{A} = \sum a_s \bra{\Phi_s^c}$
and $\ket{B} = \sum_s \ket{\Phi_s}b_s$, where $a_s$ and $b_s$ are
c-numbers, possibly anticommuting. It then follows that
$\bra{A} B\rangle = \sum a_sb_s$, and a short computation shows that
$\bra{B} A\rangle = \sum a_sb_s (-)^{(a_s+\Phi_s)(b_s+\Phi_s)}$
where the sign factor is simply $(-)^{AB}$, since the states were
assumed to be of definite statistics.
\medskip
\noindent
$\underline{\hbox{The Reflector States}}.\,$ We can use the conjugate states
to give a simple expression for a state that implements BPZ conjugation.
One defines
$$\bra{R_{12}} = \sum_r {}_2\bra{\Phi_r}\, {}_1\bra{\Phi_r^c},
\eqn\reflector$$
where following our notation $\bra{\Phi_r}$ denotes the BPZ conjugate of
$\ket{\Phi_r}$, and the subscripts on the bras are the labels for the Hilbert
spaces. The bra $\bra{R}$ may be thought as an object in the tensor product of
two Hilbert spaces (it defines a multilinear function on a pair of
states by the operation of contraction). It was constructed so that
$$\bra{R_{12}} A\rangle_1 = {}_2\bra{A},\quad\hbox{and}\quad
\bra{R_{12}} A\rangle_2 = {}_1\bra{A}, \eqn\magiccon$$
the first of which follows directly from expanding the state $A$
along the basis $\ket{\Phi_s}$ and using \conjstates , and the second
by expanding $A$ along the basis $\ket{\Phi_s^c}$ and
using \otherorder . This suggests that $\bra{R_{12}}$ is actually
symmetric in the labels $1$ and $2$:
$$\bra{R_{12}} = \bra{R_{21}}.\eqn\symreflexx$$
Rather than prove this, it is useful to derive slightly more general
rearrangement properties:
$$\eqalign{
{}&\sum_{G(\Phi_s )=G} \ket{\Phi_s}_1 \ket{\Phi_s^c}_2 =
\sum_{G(\Phi_s )=6-G} \ket{\Phi_s}_2 \ket{\Phi_s^c}_1 , \cr
{}&\sum_{G(\Phi_s )=G} {}_1\bra{\Phi_s}\, {}_2\bra{\Phi_s^c} =
\sum_{G(\Phi_s )=6-G}{}_2\bra{\Phi_s}\, {}_1\bra{\Phi_s^c}.\cr}\eqn\rprop$$
To establish the first relation we begin with the left hand side and
introduce two complete sums over states
$$\eqalign{
{}&\quad\sum_{G(\Phi_s )=G} \sum_{p,q} (-)^{\Phi_p} \ket{\Phi_p^c}_1
\, \bra{\Phi_p} \Phi_s\rangle \,
\ket{\Phi_q}_2 \,
\bra{\Phi_q^c}\Phi_s^c\rangle ,\cr
{}&=\sum_{G(\Phi_s )=G}
\sum\limits_{{G(\Phi_p ) = 6-G}\atop{G(\Phi_q ) = 6-G}}
(-)^{\Phi_p} (-)^{\Phi_p\Phi_q} \ket{\Phi_q}_2
\ket{\Phi_p^c}_1 \,
\bra{\Phi_p} \Phi_s\rangle \,
\bra{\Phi_q^c}\Phi_s^c\rangle ,\cr}\eqn\dvance$$
where we used the fact that the overlap only couples states whose ghost
numbers add up to six to restrict the sums to a space of fixed ghost
number, and we moved the state $\ket{\Phi_q}$ all the way to the left.
Note that since $\bra{\Phi_p}\Phi_s\rangle$ is an ordinary number,
the statistics of $\Phi_p$ is the same as that of $\Phi_s$ for all
nonvanishing terms in the above sum. We can therefore replace
$(-)^{\Phi_p\Phi_q} \bra{\Phi_q^c} \Phi_s^c\rangle$ by
$ (-)^{\Phi_s\Phi_q} \bra{\Phi_q^c} \Phi_s^c\rangle$
which equals $\bra{\Phi_s^c} \Phi_q^c\rangle$, to find
$$\sum\limits_{G(\Phi_p ) = 6-G \atop G(\Phi_q ) = 6-G} \sum_s
(-)^{\Phi_p} \ket{\Phi_q}_2 \ket{\Phi_p^c}_1 \,
\bra{\Phi_p} \Phi_s\rangle \,
\bra{\Phi_s^c}\Phi_q^c\rangle , \eqn\dvancei$$
where we can now freely sum over all possible $\Phi_s$ since
in any case the ones with ghost number different from $G$ would
vanish by ghost number conservation. Since the sum over
$\Phi_s$ is simply the identity operator we find that the above
reduces to
$$\sum_{G(\Phi_q) = 6-G} \sum_p
\ket{\Phi_q}_2 \, (-)^{\Phi_p} \ket{\Phi_p^c}_1 \,
\bra{\Phi_p} \Phi_q^c\rangle  =
\sum_{G(\Phi_q) = 6-G}
\ket{\Phi_q}_2 \ket{\Phi_q^c}_1 , \eqn\dvanceii$$
where again we made the sum over states $\Phi_p$ unconstrained,
since terms with other ghost numbers vanish, and then
recognized the identity operator to obtain the result we were
after. The second identity in \rprop\ is established in an
exactly analogous way. The symmetry $\bra{R_{12}} = \bra{R_{21}}$
of the reflector is an immediate consequence of this second identity.

We also have the useful relations
$$\eqalign{
{}&\bra{R_{12}} (c_n^{(1)} + c_{-n}^{(2)} ) = 0,\cr
{}&\bra{R_{12}} (b_{n}^{(1)} - b_{-n}^{(2)}) = 0,\cr
{}&\bra{R_{12}} (Q_1 + Q_2) = 0.
\cr}\eqn\qonref$$
The first two identities, which hold for all $n$, can be obtained as
special cases of \magiccon . The last equation can be checked directly
in particular cases, but follows from basic operator formalism ideas
[\alvarez ]; the reflector simply represents a two punctured sphere and
contour deformation implies that the two BRST operators cancel each other.
We can use the reflector to write the basic overlap as
$$\bra{A}B\rangle = \bra{R_{12}} A\rangle_1 \ket{B}_2,\eqn\innerccc$$
and the inner product we defined before as
$$\langle A ,  B\rangle = \bra{R_{12}} A\rangle_1 c_0^{-(2)} \ket{B}_2.
\eqn\innerprccc$$
Note that the overlap is nondegenerate, that is $\bra{B}A\rangle =0$ for all
$B$
implies $\ket{A}= 0$. Furthermore, the {\it inner product is nondegenerate}
too.
First note that in $\langle A,B\rangle$ the terms in $A$ and $B$ proportional
to
$c_0^-$ drop out, and therefore $A$ and $B$ can be assumed to be
annihilated by $b_0^-$. The statement of nondegeneracy is therefore

$$\hbox{If}\quad\langle B , A\rangle = 0 , \,
\forall B ,\, \hbox{and}\,\, b_0^- \ket{A} = 0
,\quad\hbox{then}\,\,\ket{A} = 0, \eqn\nondeg$$

\noindent
since the nondegeneracy of the overlap
$\langle B , A \rangle = \bra{B}c_0^-\ket{A}= 0$ for all $B$ implies
$c_0^-\ket{A}=0$, but then multiplying by $b_0^-$ and using
$b_0^-\ket{A}=0$ one readily sees that $\ket{A}=0$.
The inner product satisfies the two following properties

$$\langle A , B \rangle = (-)^{(A+1)(B+1)} \langle B , A \rangle,\eqn\skd$$

\noindent
and, for $A$ and $B$ annihilated by $b_0^-$ and $L_0^-$, we have

$$\langle QA , B \rangle = (-)^A \langle A , QB \rangle . \eqn\brstinp$$

\noindent
Equation \skd\ follows directly from \innerprccc , the symmetry of
$R_{12}$ and the first relation in \qonref\ (used for $c_0^-$). Equation
\brstinp\ is less
straightforward. First note that the left hand side can be written as
$$(-)^{A+1} \bra{R_{12}} A \rangle_1 Q^{(2)} c_0^{-(2)} \ket{B}_2, \eqn\stt$$
using \qonref . We now introduce the factor $1=\{ b_0^{-(2)},c_0^{-(2)}\}$
into the above to find
$$(-)^{A+1} \bra{R_{12}} A \rangle_1\,
( b_0^{-(2)} c_0^{-(2)} +  c_0^{-(2)} b_0^{-(2)})
\, Q^{(2)} c_0^{-(2)} \ket{B}_2. \eqn\stti$$
The first term in the parenthesis vanishes because $b_0^{-(2)}$ can
be pushed all the way to the left until it hits the reflector,
becomes $b_0^{-(1)}$ and kills $\ket{A}_1$. The second term gives
$$(-)^{A+1} \bra{R_{12}} A \rangle_1\,
\biggl( c_0^{-(2)} \{ b_0^{-(2)}, Q^{(2)} \} c_0^{-(2)}
- c_0^{-(2)} Q^{(2)}  \{ b_0^{-(2)}, c_0^{-(2)} \} \biggr)
\ket{B}_2, \eqn\sttii$$
where we used $b_0^- \ket{B} = 0$. The first anticommutator gives
$L_0^-$, which commutes with $c_0^-$ and annihilates $\ket{B}$,
therefore the only remaining term is
$$(-)^{A} \bra{R_{12}} A \rangle_1\,
c_0^{-(2)} Q^{(2)} \ket{B}_2 = (-)^A \langle A , QB \rangle , \eqn\sttiii$$
as we wanted to show.

Related to the reflector $\bra{R}$, we introduce a ket reflector
$\ket{R}$ defined by
$$\ket{R_{12}}\equiv \sum_s \ket{\Phi_s}_1 \ket{\Phi_s^c}_2
= \sum_s \ket{\Phi_s}_2 \ket{\Phi_s^c}_1 ,\eqn\conjreflector$$
where the equality, implying $\ket{R_{12}} = \ket{R_{21}}$,
is an immediate consequence of the first equation in \rprop . It follows
that
$$\bra{R_{ij}}R_{jk}\rangle = \sum_{r,s} {}_i\bra{\Phi_r}
{}_j\bra{\Phi_r^c}\Phi_s \rangle_j \ket{\Phi_s^c}_k =
\sum_s {}_i\bra{\Phi_s} \cdot \ket{\Phi_s^c}_k =
\sum_s (-)^{\Phi_s} \ket{\Phi_s^c}_k \cdot {}_i\bra{\Phi_s}, \eqn\contract$$
where the last expression to the right is simply the identity
operator (see Eqn. \completeness ) except for the fact that the labels
of the Hilbert spaces do not coincide. This operator simply relabels
states (or equivalently, it maps $\H_i$ to $\H_k$)
$$\bra{R_{ij}}R_{jk}\rangle \ket{A}_i = \ket{A}_k . \eqn\refcollapse$$
In analogy with the properties of the reflector, this conjugate reflector
satisfies
$$\eqalign{
(c_n^{(1)} + c_{-n}^{(2)} ) &\ket{R_{12}} = 0,\cr
(b_n^{(1)} - b_{-n}^{(2)} ) &\ket{R_{12}} = 0,\cr
(Q_1 + Q_2) &\ket{R_{12}} = 0.
\cr}\eqn\qonconjref$$

It is useful for our future applications to introduce some
further notation. We define the `tilde' states which are
simply related to the conjugate states by multiplication
by the antighost zero mode $b_0^-$
$$\ket{\wt \Phi_s} \equiv b_0^- \ket{\Phi_s^c}. \eqn\tildestates$$
We also introduce the tilde reflector
$$\ket{\wt R_{12}} \equiv  b_0^{-(1)}\ket{R_{12}}=
 b_0^{-(2)}\ket{R_{12}},\eqn\reflectilde$$
where the equality follows from \qonconjref . This reflector
can therefore be written as
$$\ket{\wt R_{12}} =  \sum_s b_0^{-(2)}\ket{\Phi_s}_1 \ket{\Phi_s^c}_2
= \sum_s (-)^{\Phi_s} \ket{\Phi_s}_1 \ket{\wt \Phi_s}_2 .\eqn\ijnji$$
It should be noted that this reflector does not satisfy a simple
BRST property; we have $(Q_1 + Q_2 ) \ket{\wt R_{12}} \not= 0$.
It is useful to introduce the `primed' reflectors, $\ket{{R'}_{12}}$
and $\ket{{\wt R'}_{12}}$ which are simply the reflectors but with
the sum over states restricted to states annihilated by $L_0^-$
$$\ket{{R'}_{12}}= \P_1\P_2 \ket{R_{12}} , \quad
\ket{{\wt R'}_{12}} = \P_1\P_2 \ket{{\wt R}_{12}}, \eqn\primrefs$$
where $\P_i$, given by
$$\P_i=\int{d\theta \over 2\pi}e^{i\theta (L_0^{(i)}-\ov L_0^{(i)})},\eqn\prj$$
is the projector, in the $i$-th Hilbert space, to states annihilated
by $(L_0-\ov L_0)$. It satisfies $\P L_0^- = L_0^- \P = 0$. Since the BRST
operator satisfies $\{ Q, b_0^- \} = L_0^-$, and commutes with $\P$, we have
that $\{ Q , b_0^- \P \} = 0$. As a consequence the primed reflectors
satisfy simple BRST properties
$$(Q_1+Q_2)\ket{{R'}_{12}}= 0,\quad (Q_1+Q_2)\ket{{\wt R'}_{12}}=0.\eqn\qprim$$
We can write the primed reflectors as
$$\ket{{R'}_{12}} =  \spr_s \ket{\Phi_s}_1 \ket{\Phi_s^c}_2, \quad
\ket{\wt{R'}_{12}}=\spr_s(-)^{\Phi_s}\ket{\Phi_s}_1\ket{\wt\Phi_s}_2,\eqn\prf$$
where the primed sums remind us that we are only summing over states
annihilated by $L_0^-$.
Note that since the conjugate of a state annihilated by $L_0^-$ is also
annihilated by $L_0^-$, it is actually enough to have one projector
rather than two in each of the expressions in \primrefs . Thus we write
$$\ket{{R'}_{12}}= \P_1\ket{R_{12}} , \quad
\ket{{\wt R'}_{12}} = b_0^{-(1)} \P_1\ket{R_{12}}. \eqn\prfi$$

We give, for illustration, the oscillator form of
the reflector state for the bosonic string:
$$\bra{R_{12}} = \int {dp\over (2\pi )^d}\, {}_1\bra{\da\da ,p}
{}_2\bra{\da\da ,-p}\, \exp(E_{12}) (c_0^{+(1)} + c_0^{+(2)})
(c_0^{-(1)} + c_0^{-(2)}),\eqn\explicitr$$
where
$$E_{12} = \sum_{n\geq 1} {1\over n} \biggl( \alpha_n^{(1)}
\cdot \alpha_n^{(2)} + c_n^{(1)}b_n^{(2)} - b_n^{(1)}c_n^{(2)}
\biggr) + \hbox{a.h.},\eqn\defexpli$$
with a.h. denoting the antiholomorphic part.
This state corresponds to a two punctured sphere with
local coordinates $z_1(z) = z$ around $z=0$, and
$z_2(z)= 1/z$ around $z=\infty$.
\medskip
\noindent
$\underline{\hbox{A Rearrangement Property}}.\,$ We conclude this
section by establishing a result that will be needed in the following
sections. We first show that
$$\ket{\Phi_s} = \alpha c_0^- \ket{\Phi_{s'}}, \,\, \hbox{with}
\,\, b_0^-\ket{\Phi_{s'}}=0,\quad\hbox{implies}\quad
\ket{\Phi_s^c} = (\alpha )^{-1} (-)^{\Phi_{s'}} \ket{\wt\Phi_{s'}}.
\eqn\conjwithb$$
This is proven by first realizing that under the assumptions
$\ket{\Phi_s^c}$ must be of the form
$\ket{\Phi_s^c} = b_0^- \ket{\chi}$. Then one requires that
$\bra{\Phi_s^c}\Phi_r\rangle = \delta_{rs}$ to find that
$\ket{\chi} = (\alpha )^{-1}(-)^{\Phi_{s'}}\ket{\Phi_{s'}^c}$.
The rearrangement property that we wish to establish reads

$$\sum_{G(\Phi_s ) = G} \hskip-8pt f(G) \ket{\Phi_s}_1 \O \ket{\wt\Phi_s}_2
=-\hskip-10pt\sum_{G(\Phi_s ) = 5-G} \hskip-8pt f(G) \ket{\wt\Phi_s}_1 \O
\ket{\Phi_s}_2. \eqn\mrp$$

Here $\O$ stands for some arbitrary operator, and $f(G)$ is some
arbitrary function of the ghost number. By construction, the sum of
states on the left hand side runs over all states $\Phi_s$ of the
specified ghost number, and annihilated by $b_0^-$ (since otherwise
they would be annihilated). The equation also holds if the sum is
restricted to states annihilated by $L_0^-$. The left hand side of this
equation can be written as
$$
=\sum_{G(\Phi_s ) = G} \hskip-8pt f(G) (-)^{(\O + 1 ) \Phi_s} \O b_0^{-(2)}
\ket{\Phi_s}_1 \ket{\Phi_s^c}_2
=\hskip-8pt\sum_{G(\Phi_s ) = 6-G} \hskip-8pt f(G) (-)^{(\O + 1 ) \Phi_s} \O
b_0^{-(2)} \ket{\Phi_s}_2 \ket{\Phi_s^c}_1, \eqn\asdf$$
where use was made of \rprop .  Notice now that $\ket{\Phi_s}_2$
must be of the form $c_0^{-(2)}\ket{\Phi_{s'}}_2$ and therefore we can
use \conjwithb\ to replace
$$ b_0^{-(2)} \ket{\Phi_s}_2 \ket{\Phi_s^c}_1 \,
= b_0^{-(2)} c_0^{-(2)} \ket{\Phi_{s'}}_2 (-)^{\Phi_{s'}} \ket{\wt\Phi_{s'}}_1
= \ket{\Phi_{s'}}_2 \ket{\wt\Phi_{s'}}_1 (-)^{\Phi_{s'}},\eqn\hjfd$$
and therefore back in \asdf\ we have
$$=\hskip-10pt\sum_{G(\Phi_{s'})=6-G-1}\hskip-8pt f(G)(-)^{(\O +1)(\Phi_{s'}
+1)}\O \ket{\Phi_{s'}}_2 \ket{\wt\Phi_{s'}}_1 (-)^{\Phi_{s'}},\eqn\ufjrh$$
where we shifted the ghost number and wrote the sign factors in terms
of the new set of states we are summing over. Since a state and its tilde
conjugate commute we write
$$= -\hskip-8pt\sum_{G(\Phi_{s'})=5-G}\hskip-8pt f(G)(-)^{\O \wt\Phi_{s'}}
\O \ket{\wt\Phi_{s'}}_1 \ket{\Phi_{s'}}_2 =-\hskip-8pt\sum_{G(\Phi_{s'})=5-G}
\hskip-8pt f(G)\ket{\wt\Phi_{s'}}_1 \O \ket{\Phi_{s'}}_2,\eqn\trww$$
which is the result we wanted to establish.

\chapter{The String Field and Its Kinetic Term}

In this section we begin our discussion of string field theory
by describing the string field, and discussing in detail the
subsidiary conditions it must satisfy, its reality condition,
statistics, and ghost number. It is useful to distinguish between
first quantized ghost number and target space ghost number. We
show that the ghost number anomalies of the first quantized
path integrals precisely imply that a suitably defined target
space ghost number is conserved.
We are then able to write the kinetic term for the closed string
field theory and to show its hermiticity and its gauge invariance.
The semirelative cohomology is seen to be the type of cohomology
that describes the physical spectrum. Finally, we review the basics
of the Batalin-Vilkovisky formalism, give the master equation,
specify a basis for fields and antifields in closed string field
theory and prove that the kinetic term satisfies the master equation.

\section{The Dynamical String Field}

An arbitrary string field is a vector in the Hilbert space of the
conformal field theory. \foot{Technically speaking, the state
space of the matter plus ghosts conformal field theory form
an indefinite metric Hilbert space, or simply an inner-product space.}
Given a basis of states $\{ \ket{\Phi_s}\}$, an arbitrary string field
is simply an arbitrary linear superposition of the basis states:
$$\ket{\Psi} = \sum_s \ket{\Phi_s} \psi_s ,\eqn\setupsf$$
Here each {\it target space field} $\psi_s$ is the component
of the vector $\ket{\Psi}$ along the basis vector $\ket{\Phi_s}$.
The target space fields are in general complex numbers, and may be
Grassmann even or odd. The {\it dynamical}
string field is the dynamical variable of the string field theory,
that is, the string field that enters into the action. This string
field is not totally arbitrary. There are conditions limiting the
possibilities in the general expansion above. These conditions are
the subject of the present subsection.
\medskip
\noindent
$\underline{\hbox{Subsidiary Conditions}}.\,$ We will demand that
the closed string field be annihilated both by $b_0^-$ and
$L_0^-$, namely
$$(b_0 -\ov b_0 ) \ket{\Psi} = (L_0 - \ov L_0) \ket{\Psi} = 0.\eqn\bsubcon$$
The necessity of such conditions is well understood by now; if we do
not impose them we do not know how to write a kinetic term for the
closed string field (see \S3.2) nor how to write interactions
(see \S7.3). The same conditions
apply for the string field $\ket{\Lambda}$ which corresponds to the
gauge parameter. Given the $b_0^-$ condition, the closed string field
can be expanded as
$$\ket{\Psi} = \ket{\phi_s, \da\da }\psi_s + \ket{\phi_s, \ua \da}
\wt \psi_s , \eqn\expcsf$$
where $\ket{\phi_s,\da\da}$ denotes the subset of the basis states
$\ket{\Phi_s}$ which are built on the vacuum $\ket{\da\da}$ (and
similarly for the second term).
\medskip
\noindent
$\underline{\hbox{Ghost Number}\,\, G}.\,$ The ghost number of a
component of a string field is conventionally
defined to be given by the ghost number of the (first quantized) state.
Thus if we have $G \ket{\Phi_s} = G_s \ket{\Phi_s}$ we define
$$G ( \ket{\Phi_s} \psi_s ) \equiv  (G\ket{\Phi_s})\psi_s =
G_s (\ket{\Phi_s} \psi_s).\eqn\defghonstat$$
For the classical theory the dynamical closed string field
$\ket{\Psi}$ entering in the string action is taken to be of
ghost number $G=+2$.
For the general master action the string field $\ket{\Psi}$
contains fields of all integer ghost numbers.
\medskip
\noindent
$\underline{\hbox{Target Space Ghost Number}\, \, g^t}.\,$
It is possible to assign to the target space fields
a conserved ghost number $g^t$. For a target space field $\psi_s$
entering the string field as $\ket{\Phi_s}\psi_s$, we simply define
$$g^t(\psi_s)= 2- G_s. \eqn\eeiiee$$
Thus in the classical
action all spacetime fields have ghost number $g^t=0$, and in the
master action we have target space fields of all ghost numbers. As
a consequence of ghost number selection rules in the
first quantized formalism, the interactions of target space fields
will conserve target space ghost number. We can explain this now.
When one defines the interaction of $n$ string fields via
Riemann surfaces of genus $g$ one does it by use of a correlator
on those surfaces. One must do an integral over a subset of
the corresponding moduli space $\M_{g,n}$, whose dimensionality
is $6g-6+2n$. Such dimensionality requires precisely the same
number of antighost insertions in the correlator. Moreover,
in order to have a nonvanishing answer, the total
ghost number inside the correlator must be $6-6g\, (=-\hbox{dim}\M_g)$.
Thus we need that the sum of ghost numbers $G_i$ of the $n$ string
fields to be inserted be equal to $2n$, namely, $\sum_{i=1}^n G_i = 2n$.
This implies that $\sum g^t_i = \sum (2-G_i) = 2n -\sum G_i = 0$, which
shows that the ghost numbers of the target space fields add up to zero
for every interaction in the master action.\foot{This
counting applies to all interactions ($n\geq 1$).
There is no interaction for a one punctured
sphere. For the two puncture sphere, which corresponds to the kinetic
term, the above requires $(-2)$
antighost insertions. As we will see, one inserts the operator
$c_0^- Q$ of ghost number two.}
\medskip
\noindent
$\underline{\hbox{Grassmanality}}.\,$
The string field $\ket{\Psi}$ is declared to be
grassman even. This will be true both for the classical action
and for the master action. We also declare the vacua
$\ket{{\bf 1},p}$ to be even
(and similarly for the dual vacua).
The grassmanality of the state $\ket{\Phi_s}$ is defined to be
the grassmanality of the conformal field theory operator
$\Phi_s$ that creates the state by acting on the vacuum. This defines
the grassmanality of the target space field $\psi_s$ as that which
together with the grassmanality of $\ket{\Phi_s}$ makes up an even
object.
It is important to note that statistics and ghost number are correlated.
If the ghost number $G_s$ of a ket $\ket{\Phi_s}$ is even, the
ket must be Grassmann even, and then the corresponding target space field
has even target space ghost number and is Grassman even. Similarly if
the ghost number $G_s$ of a field $\ket{\Phi_s}$ is odd, the
ket must be odd, and then the corresponding target space field has
odd target space ghost number and is Grassmann odd. This implies that
the interactions for target space fields in the master action have
overall even statistics. In other words, the master action is a boson.
\medskip
\noindent
$\underline{\hbox{Reality Condition on the String Field}}.\,$
A priori our definition of the dynamical string field allows for a
complex spacetime field associated to every state of the
conformal field theory. This is clearly twice as much
as we have in string theory, for example, the closed string tachyon is
a real scalar field. It is convenient to impose the following reality
condition, we demand that hermitian conjugation and BPZ conjugation
give the same state up to a sign:
$$ (\ket{\Psi})^\dagger \equiv \bra{\Psi_{hc}} =-
\bra{\Psi}.\eqn\realitycond$$
In formulating closed string field theory there is no necessity
to investigate the implications of the reality conditions on the
expansion of the string field in terms of component fields. In
order to fix the normalization of the action in section \S3.2, we will demand
that the tachyon field that appears conventionally in the string field as:
$$\ket{T} = \int {dp\over (2\pi)^d}\,  \phi(p)\, c_1\ov c_1 \ket{{\bf 1},p},
\eqn\tachyon$$
appear in the closed string field theory action with standard normalization.
The hermitian and BPZ conjugates of the tachyon are given by
$$\eqalign{
{}&\bra{T_{hc}} = \int {dp\over (2\pi)^d} \phi^*(p)\,
\bra{{\bf 1},p}\ov c_{-1} c_{-1}, \cr
{}&\bra{T} = \int {dp\over (2\pi)^d}\, \phi(p)\, \bra{{\bf 1},-p}
c_{-1} \ov c_{-1},\cr}\eqn\realtach$$
and Eqn.\realitycond\ requires
the standard reality condition $\phi^* (p) = \phi (-p)$.
\medskip
\noindent
Summarizing all the conditions on the dynamical string field we have
$$\eqalign{
{}&(b_0 - \ov b_0 ) \ket{\Psi} = (L_0 -\ov L_0) \ket{\Psi} = 0,\cr
{}& (\ket{\Psi})^\dagger = -\bra{\Psi}, \quad \hbox{reality,}\cr
{}& \ket{\Psi} \quad \hbox{is Grassmann even}, \cr
{}&G \ket{\Psi} = 2 \ket{\Psi}, \quad\hbox{for the classical action.}\cr
}\eqn\allcond$$

\section{The Kinetic Term}

Having discussed in detail the relevant issues of
conjugation and the conditions that must be imposed
on the closed string field
we now turn to the closed string field kinetic
term
$$S_{0,2} = {1\over 2} \bra{\Psi} c_0^- Q \ket{\Psi}.\eqn\kineticterm$$
Since the $c_0^-Q$ operator has ghost number two, and the
string field has ghost number two, the total ghost number adds up to
six, as it should.\foot{Most recent works in
closed string field theory use the convention where the
string field $\ket{\Psi}$ is of ghost number $+3$, and
is annihilated by $c_0^-$. The convention chosen in the present
work is useful to make explicit factors of $c_0^-$ dissapear
from the identities satisfied by the string products.}
This shows that the BRST operator alone could not have been a suitable
kinetic operator for any choice of ghost number for the string
field. Thus the necessity of an extra insertion. In order to preserve
gauge invariance in the presence of the insertion of $c_0^-$ the dynamical
string field must satisfy the subsidiary conditions
$b_0^- \ket{\Psi} = L_0^- \ket{\Psi} = 0$.
Using the reality condition on the string field and the
definition of the antilinear inner product, the kinetic term can also
be written as
$$S_{0,2} = -{1\over 2} (\Psi  , c_0^- Q \Psi ) ,\eqn\kinetici$$
and using the definition of the linear inner product, it
can be written as
$$S_{0,2} = {1\over 2} \langle \Psi , Q \Psi \rangle ,\eqn\kineticii$$
which is the nicest expression for the kinetic term and justifies
our introduction of the linear inner product with a ghost zero mode
insertion.
\medskip
\noindent
$\underline{\hbox{Hermiticity of}\, S_{0,2}}.\,$ We check this
beginning with \kinetici\
$$\eqalign{
(\Psi , c_0^-Q\Psi )^\dagger
&= - (c_0^- Q \Psi\,  , \, \Psi ) ,\cr
{}&=-( \Psi\, , \, Qc_0^- \Psi ),\cr}\eqn\provereal$$
where we made use of \conjherm\ and of the hermiticity of $c_0^-$ and
$Q$. Since $\{ Q,c_0^- \}$ does not vanish, we are not yet
through. One must use the fact that $\Psi$
is annihilated by $b_0^-$, and we do this by inserting the
factor $\{ b_0^- , c_0^- \} = 1$ in the above expression
$$\eqalign{
{}&= -(\Psi \, , \, [b_0^- c_0^- + c_0^- b_0^-] Qc_0^- \Psi ) \cr
{}&= -(\Psi \, , \,  c_0^- b_0^-\, Qc_0^- \Psi ) \cr
{}&= +(\Psi \, , \, c_0^- \, Q \{ b_0^- , c_0^- \} \Psi ) \cr
{}&= +(\Psi \, , \, c_0^- Q \Psi ) \cr}\eqn\itisreal$$
proving the hermiticity of the kinetic term.
In the first step we used the hermiticity of $b_0^-$
and $b_0^-\ket{\Psi} = 0$. In the second step we moved the remaining $b_0^-$
to the right until it annihilates the string field. On the way
one picks up the commutator with $Q$ giving $L_0^-$ which annihilates
$c_0^- \Psi$, and the commutator with $c_0^-$, as shown in the
equation.
\medskip
\noindent
$\underline{\hbox{Field Equation}}.\,$ This time we begin with
\kineticii\ and vary the string field to find
$$
2\cdot \delta S_{0,2} =  \langle \delta \Psi ,  Q \Psi \rangle
+ \langle \Psi ,   Q \delta \Psi \rangle ,\eqn\varykin$$
using \skd\ the second term can be written as
$\langle Q\delta \Psi , \Psi \rangle$
and using \brstinp\ we write it as $\langle \delta\Psi , Q \Psi \rangle$.
Therefore the total variation is
$\delta S_{0,2} =  \langle \delta \Psi ,  Q\Psi \rangle$ and the
nondegeneracy of the inner product (see \nondeg ) together with
the fact that $b_0^-Q\ket{\Psi} \equiv 0$ implies that the
equation of motion is simply
$$Q \ket{\Psi} = 0. \eqn\firsteqmii$$
This equation is simply what one would expect naively from
\kineticii .
\medskip
\noindent
$\underline{\hbox{Gauge Invariance}}.\,$ The gauge transformations
of the classical theory are given by
$$\delta \ket{\Psi} = Q \ket{\Lambda},\eqn\gaugetransf$$
where the gauge parameter must satisfy the following conditions
$$\eqalign{
{}&b_0^-\ket{\Lambda} = L_0^- \ket{\Lambda} = 0,\cr
{}& (\ket{\Lambda})^\dagger = -\bra{\Lambda}, \quad \hbox{reality,}\cr
{}& \ket{\Lambda} \quad \hbox{Grassmann odd}, \cr
{}&G \ket{\Lambda} =  \ket{\Lambda}, \quad\hbox{ghost number,}\cr}
\eqn\condgauge$$
which follow from the corresponding conditions on the string field
in Eqn. \allcond\ (the second condition is verified by showing
that $\bra{Q\Lambda} = \bra{\Lambda}Q$ and
$\bra{(Q\Lambda)_{hc}}= \bra{\Lambda_{hc}}Q$). The gauge parameter
can also be expanded in terms of target space gauge parameters
as $\ket{\Lambda} = \sum \ket{\Phi_r} \lambda_r$. Since the
ket $\ket{\Phi_r}$ must have ghost number one, it must be
Grassmann odd, and as a consequence (Eqn.\condgauge ) the target
space gauge parameters are Grassmann even, as expected.

The equation of motion in \firsteqmii\ is clearly gauge invariant under this
gauge transformation. Let us check the gauge invariance of the action
$$\eqalign{
2\cdot\delta S_{0,2} &= \langle Q \Lambda \, , \,  Q \Psi \rangle
+ \langle \Psi , QQ\Lambda \rangle , \cr
{}&= - \langle \Lambda \, , \, QQ \Psi ) = 0, \cr}\eqn\itisgaugeinv$$
where use was made of Eqn. \brstinp\ and of the nilpotency of the BRST
operator.
\medskip
\noindent
$\underline{\hbox{Physical Spectrum}}.\,$ As a consequence of the
field equations and the gauge transformations the physical spectrum
of the theory is defined by the {\it semirelative cohomology}
$$\eqalign{
{}&Q\ket{\Psi} = 0, \quad \ket{\Psi} \equiv \ket{\Psi}+Q\ket{\Lambda}, \cr
{}& b_0^- \ket{\Psi} = b_0^- \ket{\Lambda} = 0.\cr}\eqn\srcoh$$
It is possible to show that all cohomology is always concentrated
on the subspace $L_0^-= 0$, so this condition need not be imposed
explicitly in the cohomology problem (it must be imposed explicitly
on the string field in order to have a gauge invariant kinetic term).
The cohomology computed with no subsidiary conditions on the
string field nor gauge parameters is called {\it absolute} cohomology.
The closed string {\it relative} cohomology is computed imposing
the conditions that both $b_0$ and $\ov b_0$ annihilate the string
field and the gauge parameters.
\medskip
Let us conclude by writing out the string field kinetic term for the
case of the tachyon. Using \tachyon , \realtach , and \espqb\ in the
kinetic term \kineticterm\ we find
$$S_{kin}^{tach} = {1\over 2}\int {dp\over (2\pi )^d} \int {dp'\over (2\pi )^d}
\phi (p')\,\bra{-p',{\bf 1}}c_{-1}\ov c_{-1} c_0^- c_0^+ (L_0+\ov L_0)
c_1\ov c_1 \ket{p,{\bf 1}}\, \phi (p), \eqn\tkterm$$
and using $L_0+\ov L_0 = p^2-2$ (Eqn. \lasitc ) and \begininner\ we find
$$S_{kin}^{tach} = -{1\over 2}\int{dp\over (2\pi )^d}\phi(-p)\,(p^2-2)
\,\phi(p),\eqn\tktwn$$
which is indeed the correctly normalized tachyon kinetic term.

\section{Batalin Vilkovisky Formalism and the Kinetic Term}

In the Batalin Vilkovisky formalism one works with a set
of fields $\psi^s$ and a corresponding set of antifields
$\psi_s^*$. The index $s$ may possibly run over many indices,
discrete and/or continuous. One encodes the grassmanality
of $\psi^s$ in $\epsilon_s$; if $\psi^s$ is Grassmann even (odd),
$\epsilon_s$ is an even (odd) number. The grassmanality of
$\psi_s^*$ is similarly encoded in $\epsilon_{s^*}$.

The fields and antifields are paired, with $\psi_s^*$ denoting the antifield
corresponding to the field $\psi^s$. The elements of a pair have opposite
statistics and their ghost numbers add up to $(-1)$:
$$\epsilon_{s^*} = \epsilon_s + 1, \quad
g^t(\psi^s) + g^t(\psi_s^*) = -1, \eqn\basicpair$$

The master action is built using fields and antifields with
the ``boundary condition'' that when the antifields are set
to zero (and $\hbar$ is set to zero) we should recover the
classical action. The full master action $S$ must satisfy the
complete Batalin-Vilkovisky master equation
$${\partial_r S \over \partial \psi^s}
{\partial_l S \over \partial \psi_s^*}= -\hbar
{\partial_r \over \partial\psi^s}
{\partial_l S \over  \partial\psi_s^*},
\eqn\bvequation$$
where repeated indices are summed, and $\partial_r , \partial_l$
denote derivatives from the
right and from the left respectively.\foot{For $\eta$ a
c-number one has: $\partial_r /\partial \psi^s (\eta\psi^r)
= \eta \delta_s^r$ and $\partial_l /\partial \psi^s (\eta\psi^r)
= (-)^{\epsilon_s \epsilon_\eta} \eta \delta_s^r$.} If this
equation is satisfied it follows that the S-matrix
of the quantum theory is independent of the gauge fixing
conditions [\batalinvilkovisky ].
This equation is sometimes written more briefly as
$${1\over 2} \{ S,S\} = -\hbar \Delta S , \eqn\bveqnshort$$
where the antibracket $\{ \cdot , \cdot \}$ and the $\Delta$
operator are given by
$$\{ G , H \} =  {\partial_r G \over \partial \psi^s}
{\partial_l H \over \partial \psi_s^*}-
{\partial_r G \over \partial \psi_s^*}
{\partial_l H \over \partial \psi^s}, \quad
\Delta ={\partial_r  \over \partial \psi^s}
{\partial_l \over \partial \psi_s^*}.\eqn\defbracket$$
The antibracket satisfies a Jacobi identity of the form
$$(-)^{(\epsilon_F + 1)(\epsilon_H +1)} \{ F , \{ G , H \} \}
+ \hbox{cyclic} = 0.\eqn\jacobi$$

We now define the ``master transformations'', which are the precursors
of the BRST transformations of the gauge fixed theory and of the
gauge transformations.  We denote the master trasformation of an object
$\O$ by $s\O$. They are defined by
$$s \O = \{ \O , S \} \cdot \mu ,\eqn\mastertrans$$
where $\O$ is an arbitrary function of the fields and antifields,
and $\mu$ is an anticommuting $c$-number.
It follows from this definition and that of the antibracket
that
$$s\psi^s = {\partial_l S \over \partial \psi_s^*} \cdot \mu, \quad
\hbox{and} \quad
s\psi_s^* = - {\partial_l S \over \partial \psi_s} \cdot \mu.\eqn\mtfields$$
Note that these transformations do not change the statistics of the
object that is varied.
Let us now perform two transformations in succession, one with parameter
$\mu$ and the other with parameter $\mu'$:
$$\eqalign{
s ( s \O ) &= \{ s\O ,S\} \cdot \mu , \cr
{}&= \{ \{ \O , S \} \cdot \mu' , S \} \cdot \mu , \cr
{}&= \{ \{ \O , S \}  , S \} \cdot \mu'\mu , \cr
{}&\sim \{ \O , \{ S,S\} \} \cdot \mu' \mu,\cr} \eqn\nilpmast$$
where in the last step we used the Jacobi identity. If the
action $S$ satisfies the classical master equation
$\{ S, S \} = 0$ (the $\hbar = 0$ reduction of \bvequation )
then the master transformations are nilpotent. Moreover, in
this case, the action $S$ itself is invariant under master
transformations, since $sS = \{ S, S\} \cdot \mu = 0$.
Thus:
$$\{ S, S\} = 0, \quad \hbox{implies}\quad  s^2 = 0, \quad
\hbox{and} \quad  sS=0.\eqn\nilpinv$$

In the remainder of this subsection we will describe a
basis for fields and antifields in closed string field theory,
show that the kinetic term satisfies the classical
master equation, and conclude by giving two selection
rules showing what type of terms cannot appear in the master
equation.

The basic idea on how to decompose the string field into fields
and antifields is due to Thorn who studied this problem for the
case of open string fields [\thorn ]. What follows is simply
an extension to closed strings. One decomposes the string field
$\ket{\Psi}$ as
$$\ket{\Psi} = \ket{\Psi_-} + \ket{\Psi_+}, \eqn\fafdec$$
where $\ket{\Psi_-}$ contains all the fields and $\ket{\Psi_+}$ will
contain all the antifields. Both $\ket{\Psi_-}$ and $\ket{\Psi_+}$ {\it must be
annihilated by} $b_0^-$ {\it and} $L_0^-$ . The fields appear as follows
$$\ket{\Psi_-} = \spr_{G(\Phi_s)\leq 2} \ket{\Phi_s} \, \psi^s. \eqn\field$$
The sum extends over the basis of states $\ket{\Phi_s}$ with ghost
numbers less than or equal to two, and the prime reminds us that the
states are annihilated by $L_0^-$. The target space fields $\psi^s$
have target space ghost number $g^t(\psi^s) = 2-G(\Phi_s)$.
It follows that the fields include all possible values of $g^t \geq 0$.
Half of the fields appear along states built on the $\ket{\da\da}$
vacuum and the other half along states built on the $\ket{\ua\da}$.
Note that the complete set of fields is much larger than the set of fields
in the classical action, we now include all possible positive ghost numbers.
The antifields appear as
$$\ket{\Psi_+} = \spr_{G(\Phi_s) \leq 2} \ket{\wt\Phi_s} \,
\psi_s^*, \eqn\antifield$$
where $\psi_s^*$ is the antifield corresponding to the
field $\psi^s$ and  $\ket{\wt\Phi_s}= b_0^-\ket{\Phi_s^c}$,
as defined earlier. Since $G(\wt\Phi_s) = 5-G(\Phi_s)$, it follows
that the ghost number of $\psi_s^*$ is
$$g^t(\psi_s^*) =2-G(\wt\Phi_s) = 2-(5-G(\Phi_s)) = -3 + (2-g^t(\psi^s))
=-1 - g^t(\psi^s),\eqn\aishould$$
and therefore, the target space ghost numbers of a field and its
antifield add up to $(-1)$, as it should be. Moreover, the statistics
of the antifield is indeed opposite to that of the field, since
the statistics of the corresponding first quantized states are opposite.
A field corresponding to a state built on the $\ket{\da\da}$ vacuum
is always paired with an antifield corresponding to a state built on the
$\ket{\ua\da}$ vacuum, and viceversa. Note that the ghost number of
the antifields takes all possible negative values.

It is rather interesting that the closed string master action
(with $\hbar = 0$), satisfying the classical master equation,
is simply given by the {\it same expression} that
gives the classical action, but without restriction on the ghost
number of the string field.  Indeed
upon setting all the antifields to zero in such an
expression one recovers the classical action. This happens because
all fields have ghost number greater or equal to zero, and the
only ghost number conserving terms are the ones involving the
ghost zero fields, that is, the classical fields.

Let us now verify that the above ansatz, for the case of the
kinetic term, does indeed satisfy the master equation.
The master kinetic term is given by
$$S_{0,2} = {1\over 2}
\bra{R_{12}} \Psi_1 \rangle \O_{2} \ket{\Psi_2},\eqn\mkin$$
where, for brevity, we introduced the grassmann even operator $\O$ given by
$\O = c_0^-Q$, which satisfies $\bra{R_{12}} (\O_1 - \O_2) = 0$, and
now the string field contains all ghost numbers.
It follows by taking derivatives that
$${\partial_r S_2^0 \over \partial \psi^s} = \bra{R_{12}} \Psi_1\rangle
\O_2 \ket{\Phi_s}_2, \quad\hbox{and}\quad
{\partial_l S_2^0 \over \partial \psi_s^*} = (-)^{\Phi_s + 1}
\bra{R_{34}} \Psi_3\rangle
\O_4 \ket{\wt\Phi_s}_4, \eqn\begcom$$
and therefore back into the classical master equation we have
$$\eqalign{
{\partial_r S_2^0 \over \partial \psi^s}
{\partial_l S_2^0 \over \partial \psi_s^*}
&= -\bra{R_{12}}\Psi_1\rangle \O_2 \bra{R_{34}}\Psi_3\rangle\O_4
\hskip-8pt\spr_{G(\Phi_s)\leq 2} \hskip-8pt
\ket{\Phi_s}_2\ket{\wt\Phi_s}_4 (-)^{\Phi_s} , \cr
&=-{1\over 2}\bra{R_{12}}\Psi_1\rangle \O_2 \bra{R_{34}} \Psi_3\rangle\O_4
\hskip-8pt\spr_{G(\Phi_s)\leq 2}\hskip-8pt (-)^{\Phi_s}
\bigl( \ket{\Phi_s}_2\ket{\wt\Phi_s}_4
+\ket{\Phi_s}_4\ket{\wt\Phi_s}_2 \bigr) , \cr }\eqn\firstpart$$
where we used the symmetry of the terms to the left of the sum
under the exchange of the labels 2 and 4. We now use \mrp\ to replace
the second term in the sum
$$\spr_{G(\Phi_s)\leq 2}(-)^{\Phi_s}
\ket{\Phi_s}_4\ket{\wt\Phi_s}_2
= -\hskip-8pt\spr_{G(\Phi_s)\geq  3}(-)^{\Phi_s+1}
\ket{\wt\Phi_s}_4\ket{\Phi_s}_2
= \hskip-8pt\spr_{G(\Phi_s)\geq  3}(-)^{\Phi_s}
\ket{\Phi_s}_2 \ket{\wt\Phi_s}_4,\eqn\usingmrp$$
and back into \firstpart\ we now obtain a sum
extending over all states
$$=-{1\over 2}\,\bra{R_{12}}\Psi_1\rangle \O_2 \bra{R_{34}}\Psi_3\rangle\O_4
\spr_s(-)^{\Phi_s}\ket{\Phi_s}_2\ket{\wt\Phi_s}_4,\eqn\huerta$$
and the sum is recognized to be simply a primed gluing ket
(see Eqns. \prf\ and \prfi ). We therefore have
$$\eqalign{2\cdot {\partial_r S_2^0 \over \partial \psi^s}
{\partial_l S_2^0 \over \partial \psi_s^*}
&=-\bra{R_{12}}\Psi_1\rangle \O_2 \bra{R_{34}} \O_4\ket{\Psi_3}
b_0^{-(4)} \ket{{R'}_{24}},\cr
&=-\bra{\Psi_2}\O_2 \bra{R_{34}}b_0^{-(3)}\O_3\ket{\Psi_3}\P_2\ket{R_{24}},\cr
&=-\bra{\Psi_2}\O_2 \bra{R_{34}}b_0^{-(3)}\O_3\ket{\Psi_3}\ket{R_{24}},\cr
&=-\bra{\Psi_2} \O_2 b_0^{-(2)}\O_2\ket{\Psi_2},\cr
&=-\bra{\Psi}\O b_0^- \O \ket{\Psi} = 0 ,\cr}\eqn\checkmmast$$
where the projector $\P_2$ was moved all the way until it becomes one
acting on $\bra{\Psi_2}$, use was made of  Eqn.\refcollapse , and of the fact
that $\O b_0^-\O = c_0^-Q b_0^- c_0^- Q$ vanishes between states annihilated by
$b_0^-$. This concludes our verification that
the kinetic term satisfies the master equation.
\medskip
\noindent
$\underline{\hbox{Selection rules}}.\,$ It is clear from the master
equation that field/antifield independent terms in the master action do not
appear in the master equation. It seems possible, at first sight, that
there are field independent terms in the master equation; for example,
tadpole terms (interactions with one field) could give constants in
the left hand side of the equation, and bilinear interactions could
give rise to constants in the right hand side. We now show this is not
possible, we claim that
$${\partial_r S_1^g \over \partial \psi^s}
{\partial_l S_1^g \over \partial \psi_s^*}= 0, \quad\hbox{and}\quad
{\partial_r \partial_l S_2^g \over \partial\psi^s \, \partial\psi_s^*}
=0,\eqn\selrules$$
where $S_n^g$ ($g \geq 0$) denotes
the part of the master action $S$ proportional
to $\hbar^g$ and with $n$ string fields. The reason these terms
vanish is simple. When we have only one string field in the interaction
term (as in $S_1^g$) only a target space field of ghost number zero
can appear, thus a field. Then, the derivative with respect to the
antifield must vanish. For the second case (that of $S_2^g$) one
is dealing with terms with two string fields, the corresponding target
space fields must therefore have opposite ghost numbers, thus they
can never correspond to a field and its antifield.

\chapter{Algebraic Structure of Closed String Field Theory}

In this section we will begin by introducing the string field
products. These products are required to be multilinear,
graded commutative, and to satisfy a generalized type of Jacobi
identity involving the BRST operator. Closely related to the
string products are the string multilinear functions. While the products
are necessary to write the field equations and the gauge transformations,
the multilinear functions are used to write the string field action.
In this section we do not prove the requisite properties of the
string products, this is done in sections \S7 and \S8. Here we simply show,
that given those requisite properties one can build a string field
action that is consistent. In particular we show the gauge invariance
of the classical action and give the algebra of gauge transformations.
For the full quantum theory we show that the action satisfies
the quantum master equation and give the BRST transformations.
In the last subsection we observe and comment on the fact that
the algebraic structure of the classical closed string field
theory corresponds to a homotopy lie algebra $L_\infty$, a cousin
of the homotopy associative $A_\infty$ of Stasheff [\stasheff ].
This $L_\infty$ algebra is defined on the full Fock complex of the
theory, which includes physical, unphysical and trivial states. In
this algebra even the ordinary Jacobi identity is violated. We also
study what happens to the algebraic structure of the theory when
we shift the string field by a term that does not satisfy the
classical equations of motion. The resulting theory simulates closed
string field theory formulated around non-conformal field theories and
may be of interest in the problem of conformal background independence.

\section{The String Products and Their Identities}

Let us denote by ${\cal H}$ the Hilbert space of the combined conformal
field theory of matter and ghosts, and let $B_1,B_2, \cdots$ be
{\it arbitrary} string fields (thought as kets).
By arbitrary we mean that while the states $B_i$ are annihilated
by $b_0^-$ and $L_0^-$, they can have arbitrary ghost number $G$,
arbitrary statistics, and need not satisfy any reality condition.
The identities we will discuss hold for arbitrary string fields,
but at some points we will restrict ourselves to dynamical string
fields, denoted as $\Psi$, which are always grassmann even
but may have arbitrary ghost number (in the master action), and
to gauge parameters, denoted as $\Lambda$, which are Grassmann odd
and of ghost number one.

The general string product is denoted by
$$ \bigl[ B_1 , B_2, \cdots , B_n \bigr]_g . \eqn\theproduct$$
It has $n$ entries, that is $n$ states in $\H$, and is labeled by a
subscript $g$, which is related to the genus of the surfaces that
are used to define the product. The product takes the $n$ input
states in $\H$ and gives us a state in $\H$. The resulting state
satisfies the subsidiary conditions
$$(b_0-\ov b_0) \bigl[ B_1 , B_2, \cdots , B_n \bigr]_g =
(L_0-\ov L_0) \bigl[ B_1 , B_2, \cdots , B_n \bigr]_g = 0.\eqn\subprod$$
A product defines a multilinear map from the tensor product
$\H \otimes \cdots \otimes \H$ $=\H^{\otimes n}$ to $\H$:
$$\eqalign{
\bigl[ B_1 ,\cdots ,\,  B_i b + {B'}_i b'\, , \cdots , B_n \bigr]_g
= &\bigl[ B_1 ,\cdots ,  B_i, \cdots , B_n \bigr]_g\, b\,
(-)^{b(B_{i+1}+ \cdots +B_n)} \cr
+& \bigl[ B_1 ,\cdots , {B'}_i, \cdots , B_n \bigr]_g\,
b' \, (-)^{b'(B_{i+1}+ \cdots +B_n)}, \cr} \eqn\multilinearity$$
for any $i$, such that $1 \leq i \leq n$, and where $b$ and $b'$
are $c$-numbers, possibly anticommuting, which is the reason we must
include the sign factor taking into account that they must be moved
across the $B$'s that appear to their right.
The products are graded-commutative, that is
$$\bigl[ B_1 ,\cdots, B_i , B_{i+1} ,\cdots , B_n \bigr]_g
=(-)^{B_iB_{i+1}}\bigl[ B_1 ,\cdots, B_{i+1},  B_i ,\cdots , B_n \bigr]_g.
\eqn\gradedcomm$$

We will define string products for {\it all} values of $g \geq 0$ and for
{\it all} $n\geq 0$. For  $n=0$ a `product'
is not really a product (there are no input elements)
but simply gives us a special element in $\H$. They will be denoted by
$[\cdot ]_g$, and sometimes by $F_{(g)}$, that is
$$ \hbox{For}\,\, n= 0, \quad \bigl[\, \cdot \,\bigr]_g \equiv
F_{(g)} \in \H .\eqn\specialelement$$
For  $g\geq 1$ these are the states associated to surfaces of genus $g$
with one puncture, as will be defined in \S7 . For $g=0$, however,
we take it to be zero
$$ \bigl[ \, \cdot \, \bigr]_0 \equiv 0. \eqn\zeroprod$$
For $n=1$ a product takes one string field and gives us another, it
is simply a linear map. When $g=0$ this linear map will be simply
given by the BRST operator
$$ \bigl[ B \bigr]_0 = QB. \eqn\firstprod$$
All the higher products will be defined explicitly in \S7 .

The ghost number of a product of string fields is given by the
following expression
$$G(\bigl[ B_1 , B_2, \cdots , B_n \bigr]_g)= 3 +
\sum_{i=1}^n (G(B_i) -2),\eqn\gpro$$
which is $g$ independent, and holds for all the products introduced
before. Thus, for string fields of ghost number two any string product
gives us an element of ghost number three. The statistics of the
product is intrinsically odd, that is
$$\epsilon_{[B_1 , \cdots B_n]} =1+\sum_{i=1}^n \epsilon_{B_i}.\eqn\statprod$$

The fundamental identity that the products satisfy is
$$\eqalign{0= & \, Q \bigl[ B_1 ,\cdots , B_n \bigr]_g
+ \sum_{i=1}^n (-)^{(B_1 +\cdots +B_{i-1})}
\bigl[ B_1 , \cdots , QB_i , \cdots , B_n \bigr]_g \cr
&+ \sum\limits_{{g_1, g_2}
\atop{\{i_l,j_k\}, l,k}} \sigma (i_l,j_k)
\bigl[ B_{i_1},\cdots, B_{i_l},
[B_{j_1}, \cdots , B_{j_k}]_{g_2}\,\bigr]_{g_1} \cr
&+{1\over 2} \spr_{s} (-)^{\Phi_s} \bigl[ \Phi_s , \wt\Phi_s,
B_1 , B_2,\cdots ,B_n\bigr]_{g-1} ,\cr}\eqn\bidentity$$
and it holds for all $n\geq 0$. We will call this the {\it main identity}.
The second sum in the right hand side runs over
all {\it different} splittings of the set $\{ 1 , \cdots , n\}$
into a first group $\{ i_1 ,\cdots , i_l \}$ and a second group
$\{ j_1, \cdots , j_k \}$. Two splittings are the {\it same} if their
corresponding first groups contain the same integers, regardless
of their order. The following conditions should be satisfied
$$\eqalign{
{}&g_1\geq 0,\, g_2\geq 0,\quad\hbox{with} \quad g_1+g_2 = g,\cr
{}&l\geq 0, k\geq 0, \quad\hbox{with}\quad l+k = n\geq 0, \cr
{}&l\geq 1 \quad\hbox{when}\quad g_1 = 0,\cr
{}&k\geq 2 \quad\hbox{when}\quad g_2 = 0.\cr}\eqn\condsum$$
The factor $\sigma (i_l,j_k)$ appearing in \bidentity\ is defined
to be the sign picked up when one
rearranges the sequence $\{ Q,B_1, \cdots B_n\}$ into the order
$\{ B_{i_1}, \cdots B_{i_l},Q,B_{j_1},\cdots ,B_{j_k}\}$. The sum
in the last term is over states in a complete basis of the
Hilbert space $\H$. This sum is restricted to states satisfying
the $L_0 -\ov L_0$ constraint. Moreover, since
$\ket{\wt\Phi_s} = b_0^-\ket{\Phi_s^c}$, the sum can be restricted
to states $\ket{\Phi_s}$ annihilated by $b_0^-$. This happens because
the conjugate of a state $\ket{\Phi_s}$ that is not annihilated
by $b_0^-$ must necessarily be annihilated by $b_0^-$,  then
the extra $b_0^-$ in the definition of $\ket{\wt\Phi_s}$ kills the state
and the corresponding term dissapears from the sum. This identity
has its geometrical origin in the the conditions that string vertices
must satisfy so that they generate, upon the use of Feynman rules,
the complete moduli spaces of Riemann surfaces (see \S5). Its proof
will be given in \S8 .

A particularly important subcase of the above equation is the case
of $g=0$, which is relevant for the classical closed string field
theory.  This implies that $g_1=g_2=0$ and that the last term
in \bidentity\ is not present:
$$\eqalign{
0= & \, Q \bigl[ B_1 ,\cdots , B_n \bigr]_0
+\sum_{i=1}^n (-)^{(B_1 +\cdots B_{i-1})}
\bigl[ B_1 , \cdots , QB_i , \cdots , B_n \bigr]_0 \cr
&+ \sum\limits_{{\{i_l,j_k\}}
\atop{l\geq 1,k\geq 2}} \sigma (i_l,j_k)
\bigl[ B_{i_1},\cdots, B_{i_l},
[B_{j_1}, \cdots , B_{j_k}]_0\,\bigr]_0, \cr}
\eqn\cidentity$$
where all the conditions on the sum are explicitly stated.

While the above form for the identities might be most familiar,
it is convenient, for simplicity, to treat the BRST operator
$Q$ as generating the product \firstprod . This allows one to absorb
all the terms with $Q$ into the sums. One readily finds (just checking
that the signs work out) that the general identity becomes
$$0=\hskip-8pt\sum\limits_{ {g_1+ g_2=g \atop \{i_l,j_k\}; l,k \geq 0}
\atop l+k = n\geq 0 }\hskip-8pt\sigma (i_l,j_k)
\bigl[ B_{i_1},\cdots, B_{i_l},
[B_{j_1}, \cdots , B_{j_k}]_{g_2}\,\bigr]_{g_1}
+ {1\over 2}\spr_{s} (-)^{\Phi_s}
\bigl[ \Phi_s, \wt\Phi_s, B_1 , B_2, \cdots , B_n \bigr]_{g-1} ,
\eqn\bsidentity$$
where the conditions on the sum differ from \condsum\ in that
now $l$ and $k$ can take all possible values $(\geq 0)$.
The identity relevant for the classical theory becomes
$$0= \sum\limits_{ \{i_l,j_k\} ; l,k\geq 0
\atop  l + k = n \geq 0} \sigma (i_l,j_k)
\bigl[ B_{i_1},\cdots, B_{i_l},
[B_{j_1}, \cdots , B_{j_k}]_0\,\bigr]_0.
\eqn\ncidentity$$
Note that in our present setup the terms with $k=0$ vanish
automatically.

Let us give a few examples of these identities. Consider first
$g=0$, and equation \ncidentity . For the lowest possible
value of $n$, namely $n=0$, we find
$$ 0 = \l [ \cdot \, ]_0 \r_0 \quad \rightarrow \quad
Q \l \, \cdot \, \r_0 = 0 , \eqn\clfornzero$$
upon use of \firstprod\ and the fact that we defined the special
vector $\l \, \cdot \, \r_0$ to be zero. The next case is
$n=1$, and we find
$$ 0= \l [ B ]_0 \r_0 + (-)^B \l B, [\, \cdot \, ]_0 \r_0 =
\l QB \r_0 + 0 = QQB, \eqn\trivcase$$
which is the statement of the nilpotency of $Q$. The first nontrivial
identity corresponds to $n=2$ where one finds
$$0 = \l \l B_1,B_2\r_0\r_0 + (-)^{B_1} \l B_1,\l B_2 \r_0 \r_0
+ (-)^{B_2 (1+ B_1)} \l B_2 , \l B_1 \r_0 \r_0
+ (-)^{B_1 + B_2} \l B_1,B_2, \l \,\cdot\, \r_0\r_0 , \eqn\addithere$$
whereupon simplification gives
$$0 = Q \l B_1 , B_2 \r + \l QB_1 , B_2 \r + (-)^{B_1} \l B_1,
QB_2 \r . \eqn\forntwo$$
This shows that $Q$ acts as a derivation for the product $m_2$.
For $n=3$ (deleting, for brevity, the `0' subscripts from the
products) one finds
$$\eqalign{
0=&\, Q \l B_1, B_2,B_3 \r
+ \l QB_1 , B_2 , B_3 \r  +  \l B_1 , QB_2 , B_3 \r (-)^{B_1}
+  \l B_1 , B_2 , QB_3 \r (-)^{B_1+B_2} \cr
{}&+ \l B_1, \l B_2 , B_3 \r \r (-)^{B_1}
+ \l B_2, \l B_1 , B_3 \r \r (-)^{B_2(1+B_1)}
+ \l B_3, \l B_1 , B_2 \r \r (-)^{B_3(1+B_1+B_2)}.\cr}\eqn\fornthree$$
It is a good check and a straighforward exercise to verify using
the above identities that $Q(Q\l B_1,B_2,B_3 \r_0 )=0$.
In equation \fornthree , the first line corresponds to
the failure of $Q$ to be a derivation for the product $m_3$. This
failure equals the failure of the product $m_2$, in the second line,
to satisfy a Jacobi identity.

\noindent
$\underline{\hbox{A Comment on Signs}}.\,$\foot{The following comment
was prompted by a question of J. Stasheff, and obtained in collaboration
with him.} In a graded Lie algebra one
has elements $a,b, \cdots$ and a bracket $[\, , \, ]$. Every element
has a degree denoted by `deg' and the bracket satisfies
$$ [a,b] = - (-)^{\hbox{deg}\, a\cdot\hbox{deg}\, b} [b,a], \quad
\hbox{and}\quad \hbox{deg}\,([a,b])=\hbox{deg}\, a+\hbox{deg}\, b.\eqn\sglsa$$
The bracket is said to be skew (graded) commutative. In closed string field
theory, letting the degree be given by the statistics, the products are graded
commutative (see \gradedcomm ). The closed string bracket satisfies ($g=0$)
$$\l B_1 , B_2 \r = (-)^{\hbox{deg}\,B_1 \cdot \hbox{deg}\,B_2} \l B_2,B_1\r ,
\quad\hbox{and}\quad  \hbox{deg}\, (\l B_1,B_2 \r ) =1+\hbox{deg}\, B_1
+\hbox{deg}\, B_2 ,\eqn\cssglsa$$
where the second relation follows from \statprod . Thus, a priori, we
do not have the properties \sglsa\ of the bracket of a graded Lie algebra.
Nevertheless if we redefine slightly the closed string bracket and the
degree letting
$$\hbox{deg}'\, B \equiv 1+\e (B), \quad\hbox{and}\quad
\l B_1, B_2 \r ' \equiv (-)^{1+ \hbox{deg}'\,B_1}\l B_1,B_2\r ,\eqn\rfhli$$
the new bracket $\l\, , \, \r '$, with the new degree, satisfy the
conditions given in \sglsa .  The reader may verify that with the
new product, the terms in the second line in Eqn.\fornthree\ become the
standard three terms of the Jacobi identity of a graded Lie algebra.
All this indicates that if the higher products
would vanish we could obtain a standard graded Lie algebra from the
closed string product. This implies that the term  `homotopy Lie algebra'
is justified. In this work we will not use the modified products in \rfhli .
We will work with the original products which are directly motivated by
closed string theory.

Let us now consider the simplest identities for the case $g\geq 1$.
These correspond to $g=1$, $n=0$ which reads
$$0= Q \l \, \cdot \, \r_1 + {1\over 2}\spr_s
(-)^{\Phi_s} \l \Phi_s , \wt\Phi_s \r_0 ,
\eqn\forgone$$
and to $g=1$, $n=1$, which is
$$0 = Q \l B \r_1 + \l QB \r_1 + (-)^B \l B, \l\, \cdot \, \r_1
\r_0 + {1\over 2}\spr_s (-)^{\Phi_s}
\l \Phi_s, \wt\Phi_s , B\r_0 .\eqn\forgonenone$$
Let us verify the consistency of \forgone . We act with $Q$ on the left
and on the right, and we must see that acting on the right gives zero.
The first term on the right is annihilated by $Q$, we must see that the
second one is also. To this end it is sufficient to show that terms of the
following form vanish:
$$\spr_s (-)^{\Phi_s} \l Q\Phi_s , \wt\Phi_s , B_1,\cdots ,B_n \r_g
+ \spr_s  \l \Phi_s , Q\wt\Phi_s , B_1,\cdots ,B_n \r_g
=0, \eqn\qdoesnotsee$$
for any product and for arbitrary $B$'s. One may think that the
two terms cancel each other simply by symmetry properties, but this
is not quite true, in fact the second term is identical to the
first; using \mrp\
$$
-\spr_s  \l \wt\Phi_s , Q\Phi_s , B_1,\cdots ,B_n \r_g
= \spr_s (-)^{\Phi_s} \l Q\Phi_s , \wt\Phi_s , B_1,\cdots ,B_n \r_g ,
\eqn\indeedqdns$$
and \gradedcomm\  for the second step. Thus both terms are identical
$$\spr_s (-)^{\Phi_s} \l Q\Phi_s , \wt\Phi_s , B_1,\cdots ,B_n \r_g
= \spr_s  \l \Phi_s , Q\wt\Phi_s , B_1,\cdots ,B_n \r_g
=0, \eqn\qdoesnotseeit$$
At any rate it is easiest to establish this relation in the form
\qdoesnotsee\ since the relevant terms appear in the form
$$ \cdots \spr_s\l (-)^{\Phi_s} Q_1\ket{\Phi_s}_1\ket{\wt\Phi_s}_2
+ \ket{\Phi_s}_1  Q_2 \ket{\wt\Phi_s}_2 \r \cdots $$
which can clearly be written as
$$ \cdots \spr_s\l (Q_1 + Q_2 ) (-)^{\Phi_s} \ket{\Phi_s}_1\ket{\wt\Phi_s}_2
 \r \cdots = \cdots \l (Q_1 + Q_2 ) \ket{\wt {R'}_{12}}
 \r \cdots = 0 $$
since the sum of $Q$'s annihilate the sewing ket. This establishes
\qdoesnotsee\ and \qdoesnotseeit\ and therefore the consistency of
\forgone . The consistency of \forgonenone\ is checked in a
similar fashion; this time, however, one needs a particular case
of the following identity
$$\spr_s \l \cdots , \Phi_s \l \wt\Phi_s, \cdots
\r_{g_1}\r_{g_2} = 0 ,\eqn\tworref$$
where the dots denote arbitrary string fields. Another related and useful
relation is
$$\spr_{r,s} (-)^{\Phi_r}(-)^{\Phi_s} \l \Phi_s , \wt\Phi_s ,
\cdots , \Phi_r , \wt\Phi_r , \cdots \r_g = 0.\eqn\relatedanduse$$
These identities will be simple to establish in \S7.

A particularly useful case of the identities \bsidentity\ satisfied by the
products arises when all of the elements are the same string field
$\Psi$ (grassman even). We then have that splitting
$n$ string fields into two groups of $l$ and $k$ string fields
respectively can be done in $n!/l!k!$ different ways, thus
Eqn.\bsidentity\ gives
$$0= \sum\limits_{{g_1+ g_2 = g \atop l,k\geq 0} \atop l+k =n\geq 0}
{n!\over l!\, k!} \bigl[ \Psi^l , [\Psi^k ]_{g_2}\,\bigr]_{g_1}
+ {1\over 2}\spr_{s} (-)^{\Phi_s}
\bigl[ \Phi_s, \wt\Phi_s, \Psi^n \bigr]_{g-1} ,\eqn\mbsidentity$$
where the sign factor $\sigma$ dissapeared since it is always equal to one,
and where we have defined, for brevity, the object $\Psi^n$ that simply
denotes a set of $n$ entries of the field $\Psi$
$$\Psi^n \equiv \underbrace{\Psi , \Psi, \cdots \Psi}_{n\, \hbox{terms}}
\quad . \eqn\funnydef$$
Note that $\Psi^n$ is not a product.

\section{The Multilinear String Functions}

It is now convenient to use the multilinear products discussed
above, and the linear inner product introduced in \S2 to define
functions. These functions, given a set of string
fields, give us numbers. We define

$$\big\{ A, B_1, B_2 , \cdots , B_n \big\}_g \equiv \big\langle A , \l
B_1, B_2, \cdots B_n \r_g \, \big\rangle . \eqn\defmultifunc$$

\noindent
Given this definition we can actually use our basis of states
to reverse the relation and obtain the products from the multilinear
functions. We claim that the correct relation is
$$\l B_1, \cdots , B_n \r_g = \spr_s (-)^{\Phi_s} \ket{\wt \Phi_s}
\cdot \big\{ \Phi_s , B_1, \cdots , B_n \big\}_g .\eqn\oppsrel$$
We verify this by using \defmultifunc\ in the right hand side of
\oppsrel\ to obtain:
$$\eqalign{{}&=\spr_s (-)^{\Phi_s} \ket{\wt \Phi_s}
\bra{R_{12}}\Phi_s \rangle_1 \, c_0^{-(2)}\l B_1,\cdots ,B_n\r_g^{(2)},\cr
{}&= \spr_s (-)^{\Phi_s} \ket{\Phi_s} \bra{R_{12}} b_0^{-(1)}
\ket{\Phi_s^c}_1\, c_0^{-(2)}\l B_1,\cdots , B_n \r_g^{(2)},\cr
&=\spr_s\ket{\Phi_s}\cdot \bra{\Phi_s^c}\,\l B_1,
\cdots ,B_n \r_g,\cr}\eqn\deropps$$
where in the first step we used \mrp\ and then the $c_0^-$ oscillator was
moved up to the reflector and since it annihilates $\ket{\Phi_s^c}$ it
deletes the $b_0^-$ oscillator. In the final expression we recognize
the identity operator (Eqn. \completeness ), in the subspace of states
annihilated by $L_0^-$. Since the string product is in this subspace
we have obtained the desired result.

The functions we have introduced in \defmultifunc\ are multilinear
$$\eqalign{\{ B_1 ,\cdots ,\,  B_i b + {B'}_i b'\, , \cdots , B_n \}_g
= &\{ B_1 ,\cdots ,  B_i, \cdots , B_n \}_g\, b\,
(-)^{b(B_{i+1}+ \cdots +B_n)} \cr
+& \{ B_1 ,\cdots , {B'}_i, \cdots , B_n \}_g\,
b' \, (-)^{b'(B_{i+1}+ \cdots +B_n)}, \cr} \eqn\fmultilinearity$$
for any $i$, such that $1 \leq i \leq n$, and where $b$ and $b'$
are $c$-numbers, possibly anticommuting. This equation
follows from \multilinearity\ for all the arguments except the
first one. For the first one the linearity follows from the
linearity of the inner product; the sign factor arises from the
$c_0^-$ insertion in the inner product and the fact that the
multilinear products are intrinsically odd (Eqn. \statprod ).
The multilinear functions are also graded-commutative
$$\{ B_1 ,\cdots, B_i , B_{i+1} ,\cdots , B_n \}_g
=(-)^{B_iB_{i+1}}\{ B_1 ,\cdots, B_{i+1},  B_i ,\cdots , B_n \}_g,
\eqn\pgradedcomm$$
a fact that will be essentially obvious in \S7 .
Since a multilinear function gives us a number, which is of ghost number
zero, it satisfies a selection rule
$$\big\{ B_1 , B_2 , \cdots , B_n \big\}_g \not= 0 ,
\quad\hbox{implies}\quad
\sum_{i=1}^n (G(B_i) -2) = 0.\eqn\gmpro$$
The statistics of the multilinear functions is intrinsically even, that is
$$\epsilon_{ \{ B_1 , \cdots B_n \} }
=\sum_{i=1}^n \epsilon_{B_i},\eqn\statmprod$$
The lowest order multilinear functions are
$$\eqalign{
&\{ B_1 , B_2\}_g \equiv \langle B_1 , \l B_2 \r_g \rangle, \cr
&\{ B \}_g \equiv \langle B , \l \,\cdot \, \r_g \rangle , \cr
&\{ \, \cdot \, \}_g \equiv \, \hbox{Number to be defined in}\, \S7, \cr}
\eqn\deflowmult$$
where the last definition is not just a particular case of
\defmultifunc . For $g=0$ we declare
$$\{ \, \cdot \, \}_0 \equiv 0,\eqn\itiszero$$
and because of \zeroprod\ we have $\{ B \}_0 = \langle B,
\l\,\cdot\,\r_0 \rangle =0$.
The multilinear functions will enable us to write actions next.

\section{The Classical String Action and its Gauge Invariance}

In this section we give the classical action for closed string
field theory. We then give the form of the gauge transformations,
and show the invariance of the action. Our treatment just amounts
to streamlining of the proofs constructed in [\kugosuehiro ].
Finally, we compute the commutator of two gauge transformations.

The classical string action is simply given by
$$S(\Psi ) = {1\over \k^2}\, \sum_{n=2}^\infty \,
{\k^n \over n!} \{ \Psi^n \}_0.
\eqn\classaction$$
Here $\k$ is the closed string field coupling constant.
Once we recognize that $\{ \Psi , \Psi \}_0 = \langle \Psi , [\Psi ]_0
\rangle = \langle \Psi , Q \Psi \rangle$, the action can be written
in the more familiar form
$$S(\Psi ) = {1\over 2} \langle \Psi , Q\Psi \rangle
+ \sum_{n=3}^\infty {\k^{n-2} \over n!} \{ \Psi^n \}_0.\eqn\famclass$$
In fact, given that we have
$$\{ \Psi \}_0 = \langle \Psi , \l \, \cdot \, \r_0 \rangle =
\langle \Psi , 0 \rangle = 0,\eqn\wesetitup$$
and $\{ \cdot \}_0 = 0$ (from \itiszero ), we can actually take the action
to be
$$\hbox{CLASSICAL ACTION:}\quad\quad S(\Psi ) = {1\over \k^2}\,
\sum_{n=0}^\infty \, {\k^n \over n!}
\{ \Psi^n \}_0.
\eqn\notsofamca$$
This form of the action is motivated by the higher genus action, which
must receive contributions from $n=1$, and may or may not include
contributions from $n=0$.

The field equations follow from \classaction\ by simple variation
$$\delta S = \sum_{n=2}^\infty {\k^{n-2} \over n!} n
\{ \delta\Psi , \Psi^{n-1} \}_0,
=\sum_{n=2}^\infty {\k^{n-2} \over (n-1)!}
\big\langle \delta\Psi , \l \Psi^{n-1} \r_0 \big\rangle ,\eqn\varyca$$
and are given by
$$\hbox{FIELD EQUATIONS:}\quad\, 0= {\cal F}(\Psi ) \equiv
\sum_{n=1}^\infty {\k^{n-1} \over n!} \l \Psi^n \r_0
= Q\ket{\Psi} + \sum_{n=2}^\infty {\k^{n-1} \over n!} \l \Psi^n \r_0.
\eqn\eomca$$
The gauge transformations of the theory are given by
$$\hbox{GAUGE TRANSFORMATIONS:}\quad\quad\delta_\Lambda \ket{\Psi}
= \sum_{n=0}^\infty { \k^n \over n!} \l \Psi^n,
\Lambda \r_{0},\quad\quad \eqn\gaugetrca$$
which in expanded form read
$$\delta_\Lambda \ket{\Psi}=Q\Lambda +\sum_{n=1}^\infty
{ \k^n \over n!} \l \Psi^n , \Lambda \r_0.\eqn\egaugetrca$$
Let us now show the invariance of the action under the gauge
transformations. We begin with Eqn. \notsofamca\ and obtain
$$\eqalign{
\delta_\Lambda S &= \sum_{n=1}^\infty {\k^{n-2}\over (n-1)!}
\{ \Psi^{n-1},\delta\Psi \}_0,\cr
{}&= \sum_{n=0}^\infty \sum_{m=0}^\infty {\k^{n+m-1}\over n!\, m!}
\big\{ \Psi^n, \l \Psi^m, \Lambda \r_0 \big\}_0,\cr
{}&= \sum_{n=0}^\infty \sum_{m=0}^\infty {\k^{n+m-1}\over n!\, m!}
\big\{ \l \Psi^m, \Lambda \r_0 ,  \Psi^n  \big\}_0,\cr
{}&= \sum_{n=0}^\infty \sum_{m=0}^\infty {\k^{n+m-1}\over n!\, m!}
\big\langle \l \Psi^m, \Lambda \r_0 ,  \l \Psi^n \r_0  \big\rangle ,\cr}
\eqn\checkinvact$$
to find upon reordering
$$\delta_\Lambda S =
\sum_{n=0}^\infty \sum\limits_{ l\geq 0 , k\geq 0 \atop
l+k=n } {\k^{n-1}\over l!\, k!}
\big\langle \l \Psi^l, \Lambda \r_0 , \l  \Psi^k \r_0 \big\rangle.
\eqn\checkgica$$
To show that this vanishes we make use of \mbsidentity\ for the
case when $g=0$
$$0= \sum\limits_{l+k =n\geq 0  \atop l,k\geq 0} {n!\over l!\, k!}
\bigl[ \Psi^l ,
[\Psi^k ]_0\,\bigr]_0,
\eqn\trygetzero$$
and form the inner product with $\Lambda$ to get
$$0= \sum\limits_{l+k =n\geq 0  \atop l,k\geq 0} {n!\over l!\, k!}
\big\langle  \Lambda , \bigl[ \Psi^l ,
[\Psi^k ]_0\,\bigr]_0 \big\rangle ,
\eqn\trygetzero$$
and then we simply rearrange
$$\eqalign{
\big\langle \Lambda ,\bigl[ \Psi^l, [\Psi^k ]_0\, \bigr]_0 \big\rangle &=
\big\{ \Lambda, \Psi^l , \l \Psi^k \r_0 \big\}_0
=- \big\{ \l \Psi^k \r_0 , \Psi^l,\Lambda \big\}_0, \cr
{}&= -\big\langle \l \Psi^k\r_0 , \l \Psi^l , \Lambda \r_0 \big\rangle
= -\big\langle \l \Psi^l , \Lambda \r_0 , \l\Psi^k \r_0 \big\rangle ,
\cr}\eqn\arrange$$
and therefore back in \trygetzero\ we have
$$0= \sum\limits_{l+k =n \geq 0  \atop l,k\geq 0} {n!\over l!\, k!}
\big\langle \l \Psi^l , \Lambda \r_0 , \l\Psi^k \r_0 \big\rangle .
\eqn\itgtiota$$
This identity, holding for all $n\geq 0$ implies that
$\delta_{\Lambda} S$ in \checkgica\ is zero, as we
wanted to show.
\medskip
\noindent
$\underline{\hbox{Gauge Algebra}}.\,$ Let us now turn to the algebra of
gauge transformations (see also Ref.[\schubert ]).
We will see that the commutator of two gauge transformations is a gauge
transformation with field dependent parameter plus some extra terms which
vanish upon use of the equations of motion.
Consider therefore the commutator
$$ \l \delta_{\Lambda_2} ,\delta_{\Lambda_1} \r \ket{\Psi}
= \bigl( \delta_{\Lambda_2} \delta_{\Lambda_1}
-\delta_{\Lambda_1} \delta_{\Lambda_2} \bigr) \ket{\Psi},
\eqn\algofgt$$
a short computation gives us
$$ \l \delta_{\Lambda_2} ,\delta_{\Lambda_1} \r \ket{\Psi}
= \sum_{n=0}\sum\limits_{l,k\geq 0 \atop l+k=n}
{\k^{n+1} \over l!\, k!} \biggl( \l \Psi^l ,\Lambda_1 ,
\l \Lambda_2 , \Psi^k \r_0 \r_0 - (\Lambda_1 \leftrightarrow \Lambda_2 )
\,\biggr) . \eqn\itslate$$
{}From the basic identity for string products \ncidentity , consider
the relation that follows when we take the string of fields
$\Psi^n \Lambda_1\Lambda_2$ with $n\geq 0$.
We get
$$\eqalign{
0=\sum\limits_{l,k\geq 0 \atop l+k=n \geq 0}
{n! \over l!\, k!} &\biggl(
\l \Psi^l ,\Lambda_1 , \l \Lambda_2 , \Psi^k \r_0 \r_0
-\l \Psi^l ,\Lambda_2 , \l \Lambda_1 , \Psi^k \r_0 \r_0 \cr
{}&-\l \Psi^l , \l \Lambda_1 , \Lambda_2 , \Psi^k \r_0 \r_0
+\l \Psi^l ,\Lambda_1 , \Lambda_2 ,\l \Psi^k \r_0 \r_0 \biggr),\cr}
\eqn\identitytwolamb$$
where the combinatorial factor is clearly the same for all four
terms, and the signs arise from the oddness of the gauge parameters.
This equation holds for all $n\geq 0$. The first two terms appearing
in \identitytwolamb\ are precisely the ones that appear in the
right hand side of \itslate , which therefore reduces to
$$ \l \delta_{\Lambda_2} ,\delta_{\Lambda_1} \r \ket{\Psi}
= \sum_{l,k\geq 0}
{\k^{k+l+1} \over l!\, k!}
\biggl( \l \Psi^l,  \l \Lambda_1 , \Lambda_2 , \Psi^k \r_0 \r_0
-\l \Psi^l ,\Lambda_1 , \Lambda_2 ,\l \Psi^k \r_0 \r_0 \biggr) .
\eqn\itsverylate$$
It is convenient to write this right hand side as
$$\sum_{l=0}^\infty {\k^l \over l!}
\biggl( \l \Psi^l,  \sum_{k=0}^\infty {\k^{k+1}\over k!}
\l \Lambda_1 , \Lambda_2 , \Psi^k \r_0 \r_0
+\k^2\l \Psi^l ,\Lambda_2 , \Lambda_1 , {\cal F}(\Psi ) \r_0\, \biggr) ,
\eqn\solate$$
where ${\cal F}$, the field equation, was given in \eomca .
Recognizing that the first term in the above equation corresponds to
a gauge transformation (see \gaugetrca )  we finally obtain

$$\l \delta_{\Lambda_2} ,\delta_{\Lambda_1} \r \ket{\Psi} =
\delta_{\Lambda (\Psi)} \ket{\Psi} + \sum_{l=0}^\infty
{\k^{l+2} \over l!} \l \Psi^l ,\Lambda_2 ,
\Lambda_1 , \F (\Psi ) \r_0  , \eqn\lateenough$$

\noindent
where the field dependent parameter $\Lambda (\Psi )$ is given by
$$\eqalign{
\Lambda (\Psi ) &= \sum_{n=0}^\infty {\k^{n+1}\over n!}
\l \Lambda_1 , \Lambda_2 , \Psi^n \r_0 , \cr
{}& = \k \l \Lambda_1 , \Lambda_2 \r_0
+\k^2 \l \Lambda_1, \Lambda_2, \Psi \r_0 + \cdots , \cr}\eqn\theresult$$
which is indeed field dependent beyond the first term. The second term
in \lateenough\ is the term that vanishes when one uses the field
equation ${\cal F}(\Psi ) = 0$. One sees that both the field dependent
structure constants and on-shell closure of the gauge algebra
are a consequence of the theory not being cubic.

\REF\hata{H. Hata, `BRS invariance and unitarity in closed
string field theory', Nucl. Phys. {\bf B329} (1990) 698;
`Construction of the quantum action for path-integral
quantization of string field theory', Nucl. Phys. {\bf B339} (1990) 663.}

\section{The Quantum String Action and The Master Equation}

In this section we show that the identities satisfied by the
string field products and multilinear functions imply that the
full quantum action, which will be given shortly, satisfies the
Batalin-Vilkovisky master equation. This implies that the string
field action can be consistently quantized (in the path integral
formulation). Then we will give the
master transformations and the BRST transformations of the action.
The BV master equation was used earlier by Hata in the analysis of the
quantum HIKKO closed string field theory [\hata ].

Before starting, however, let us derive
an identity satisfied by the multilinear products which will be
necessary in checking the master equation. Consider
\mbsidentity\ for the case when we have a set of $n-1$ string fields
$\Psi^{n-1}$, with $n\geq 1$. The identity reads
$$0= \sum\limits_{  {g_1+ g_2=g
\atop n_1 \geq 1 ,n_2\geq 0} \atop n_1+n_2 = n\geq 1}
{(n-1)!\over (n_1-1)!\, n_2!}
\bigl[ \Psi^{n_1-1} ,
[\Psi^{n_2} ]_{g_2}\,\bigr]_{g_1}
+ {1\over 2}\spr_{s} (-)^{\Phi_s}
\bigl[ \Phi_s, \wt\Phi_s, \Psi^{n-1} \bigr]_{g-1} ,
\eqn\mbsidentityi$$
where $n_1$ and $n_2$ are treated somewhat asymmetrically. We now form the
inner product with another string field $\Psi$ to obtain
$$0= \sum\limits_{ {g_1+ g_2=g
\atop n_1 \geq 1 ,n_2\geq 0} \atop n_1+n_2 = n\geq 1}
{(n-1)!\over (n_1-1)!\, n_2!}
\big\{ \Psi^{n_1} ,
[\Psi^{n_2} ]_{g_2}\,\big\}_{g_1}
+ {1\over 2}\spr_{s} (-)^{\Phi_s}
\big\{ \Phi_s, \wt\Phi_s, \Psi^n \big\}_{g-1} .
\eqn\mbsidentityii$$
While $n_1$ and $n_2$ now appear more symmetrically in the multilinear
products the prefactor with factorials is quite asymmetric. It is
therefore useful to observe that
$$\eqalign{
\big\{ \Psi^{n_1} , [\Psi^{n_2} ]_{g_2}\,\big\}_{g_1}
{}&= \big\{ [\Psi^{n_2} ]_{g_2}, \Psi^{n_1} \,\big\}_{g_1}
= \big\langle  [\Psi^{n_2} ]_{g_2}, [\Psi^{n_1} ]_{g_1} \, \big\rangle ,\cr
{}&= \big\langle  [\Psi^{n_1} ]_{g_1}, [\Psi^{n_2} ]_{g_2} \, \big\rangle
= \big\{  [\Psi^{n_1} ]_{g_1}, \Psi^{n_2} \big\}_{g_2}, \cr
{}&= \big\{  \Psi^{n_2} ,[\Psi^{n_1} ]_{g_1}\big\}_{g_2}, \cr}\eqn\bsymmme$$
and this symmetry under the exchange of the pairs $(n_1,g_1)$ and
$(n_2,g_2)$ allows us to write the first term in \mbsidentityii\ as
$$\eqalign{{}&{1\over 2}\sum\limits_{ {g_1+ g_2=g
\atop n_1 \geq 1 ,n_2\geq 0} \atop n_1+n_2 = n\geq 1}
{(n-1)!\over (n_1-1)!\, n_2!}\biggl[ \big\{ \Psi^{n_1} ,
[\Psi^{n_2} ]_{g_2}\,\big\}_{g_1}+ \big\{ \Psi^{n_2} ,
[\Psi^{n_1} ]_{g_1}\,\big\}_{g_2} \biggr], \cr
&= {1\over 2}\biggl[ \sum\limits_{ {g_1+ g_2=g
\atop n_1 \geq 1 ,n_2\geq 0} \atop n_1+n_2 = n\geq 1}
{(n-1)!\over (n_1-1)!\, n_2!} +\sum\limits_{ {g_1+ g_2=g
\atop n_1 \geq 0 ,n_2\geq 1} \atop n_1+n_2 = n\geq 1}
{(n-1)!\over (n_2-1)!\, n_1!}\, \biggr]\, \big\{ \Psi^{n_1} ,
[\Psi^{n_2} ]_{g_2}\,\big\}_{g_1} , \cr
{}&= {1\over 2}\sum\limits_{ {g_1+ g_2=g
\atop n_1+n_2 = n\geq 1} \atop  n_1 ,n_2\geq 0}
{n!\over n_1!\, n_2!} \big\{ \Psi^{n_1} ,
[\Psi^{n_2} ]_{g_2}\,\big\}_{g_1} , \cr}\eqn\uffwaw$$
where in the last step one breaks each sum into a piece with
$n_1, n_2 \geq 1$,  plus one term; the two resulting sums simplify
neatly and the extra two terms can be absorbed into the expression
to obtain the quoted result (note that $n_1 = n_2= 0$ is not possible).
Using this back in \mbsidentityii\ we obtain the final form of
the desired identity
$$0= \sum\limits_{ {g_1+ g_2=g
\atop n_1+n_2 = n\geq 1 } \atop n_1 ,n_2\geq 0}
{n!\over n_1!\, n_2!}
\big\{ \Psi^{n_1} ,
[\Psi^{n_2} ]_{g_2}\,\big\}_{g_1} + \spr_{s} (-)^{\Phi_s}
\big\{ \Phi_s, \wt\Phi_s, \Psi^n \big\}_{g-1}.\eqn\mbsidentityiii$$

We now claim that the full quantum master action is simply given by
the obvious generalization of the classical action
$$\hbox{QUANTUM MASTER ACTION}:\quad
S(\Psi ) = {1\over \k^2}\, \sum_{g\geq 0} (\hbar \k^2)^g \sum_{n\geq 0}\,
{\k^n \over n!} \big\{ \Psi^n \big\}_g.\eqn\thefullaction$$
Note that we have included, for generality, the terms
with $g=0; n=0,1$, which are zero. We have also included, for
$g\geq 1$ the terms with $n=0$, which are simply constants, and
therefore do not enter into the Batalin-Vilkovisky master equation.
As discussed in the introduction, all such terms may play a role
in understanding background independence. The way we have defined
the coupling constants is consistent since
$${1\over \hbar} S(\Psi ) =
\, \sum_{g\geq 0} (\hbar \k^2)^{g-1}
\sum_{n\geq 0} \, {\k^n \over n!} \big\{ \Psi^n \big\}_g , \eqn\rdacc$$
where it is clear that we have a single coupling $(\hbar \k^2)$ appearing
in the theory once we let $\Psi \rightarrow k^{-1}\Psi$.
Let us now verify the master equation, which reads
$${\partial_r S\over \partial \psi^s}
{\partial_l S \over\partial \psi_s^*} + \hbar
{\partial_r\over \partial \psi^s}
{\partial_l S\over \partial \psi_s^*} = 0.\eqn\bvme$$
To this end it is convenient to define
$$S_g^n(\Psi )\equiv{1\over n!}\big\{ \Psi^n \big\}_g,\quad\rightarrow\quad
S(\Psi) =\sum_{g,n\geq 0}\hbar^g\,\k^{n+2g-2}S_g^n(\Psi ).\eqn\decompaction$$
Plugging this power expansion of the action into the BV equation we find
$$\sum\limits_{{g_1 + g_2 = g \atop n_1+n_2 = n\geq 0}\atop
n_1, n_2 \geq 0}
{\partial_r S_{g_1}^{n_1} \over \partial \psi^s}
{\partial_l S_{g_2}^{n_2} \over\partial \psi_s^*} +
{\partial_r\over \partial \psi^s}
{\partial_l S_{g-1}^{n}\over \partial \psi_s^*} = 0.\eqn\bvmexp$$
Note that for $n=0$, which implies $n_1=n_2=0$,  both terms above vanish,
since the corresponding $S$'s have no field/antifield dependence at all.
For $n=1$, the second term vanishes, and, since either $n_1$ or
$n_2$ must be zero, the first term vanishes too. For $n=2$ the selection
rules \selrules\ derived in \S3.3 imply both terms must again vanish. Thus
the relation above holds trivially for $n\leq 2$. We must show it
holds for $n\geq 3$. It is then convenient to relabel the indices
appearing above; we must show that
$$\sum\limits_{{g_1 + g_2 = g\geq 0 \atop n_1+n_2 = n\geq 1}\atop
n_1, n_2 \geq 0}
{\partial_r S_{g_1}^{n_1+1} \over \partial \psi^s}
{\partial_l S_{g_2}^{n_2+1} \over\partial \psi_s^*} +
{\partial_r\over \partial \psi^s}
{\partial_l S_{g-1}^{n+2}\over \partial \psi_s^*} = 0.\eqn\bvmexp$$

In order to be able to take the derivatives from the right or from
the left correctly we use the fact (from \S7 ) that multilinear
products are of the form
$\{ B_1 \cdots B_n \}\sim$ $\bra{\Omega} B_1\rangle\cdots \ket{B_n}$
with $\bra{\Omega}$ grassmann even. We then have
$${\partial_r S_g^n \over \partial \psi^s} = {1\over (n-1)!}
\big\{ \Psi^{n-1} , \Phi_s \big\}_g , \quad\hbox{and}\quad
{\partial_l S_g^n \over \partial \psi_s^*} = {1\over (n-1)!}
\big\{ \Psi^{n-1} , \wt\Phi_s \big\}_g (-)^{\Phi_s +1} , \eqn\howdiff$$
where the first result has no sign factor because the derivative,
coming from the right hits directly the field; in the second case
the derivative must be pushed across the state and we get the quoted
sign factor. (One has ${\partial_l \over \partial \psi_s^*} \ket{\wt\Phi_s}$
$= (-)^{\Phi_s +1} \ket{\wt\Phi_s}{\partial_l \over \partial \psi_s^*}$
since the derivative has the same statistics as the corresponding
antifield $\psi_s^*$, which
in turn has the same statistics as the state $\ket{\wt\Phi_s}$, given
that the string field is grassman even).

Let us begin the calculation with the first term of \bvmexp .  We symmetrize
explicitly in the first and second labels and using \howdiff\ we find
$${1\over 2} \sum\limits_{{g_1 + g_2 = g\geq 0 \atop n_1+n_2 = n\geq 1}\atop
n_1,n_2\geq 0} \spr_{G(\Phi_s ) \leq 2}
{1\over n_1!\,n_2!}\biggl[
\big\{ \Psi^{n_1} , \Phi_s \big\}_{g_1}
\big\{ \Psi^{n_2} , \wt\Phi_s \big\}_{g_2}
+\big\{ \Psi^{n_2} , \Phi_s \big\}_{g_2}
\big\{ \Psi^{n_1} , \wt\Phi_s \big\}_{g_1}
\biggr] (-)^{\Phi_s+1}. \eqn\eeiidd$$
The second term in the brackets can be written using \mrp\ as
$$
\spr_{G(\Phi_s ) \geq 3}
\big\{ \Psi^{n_2} , \wt\Phi_s \big\}_{g_2}
\big\{ \Psi^{n_1} , \wt\Phi_s \big\}_{g_1}(-)^{\Phi_s+1}
=\spr_{G(\Phi_s ) \geq 3}
\big\{ \Psi^{n_1} , \wt\Phi_s \big\}_{g_1}
\big\{ \Psi^{n_2} , \wt\Phi_s \big\}_{g_2}
(-)^{\Phi_s+1}, \eqn\rruuii$$
where in the last term we commuted the multilinear functions (this
produces no extra sign since the string field is even and $\Phi$
and $\wt\Phi$ are of opposite statistics). Back in \eeiidd\
we now get a complete sum representing the first term
in the BV equation \bvmexp (this term is now denoted
by (I))
$$(\hbox{I}) = -{1\over 2}
\sum\limits_{{g_1 + g_2 = g\geq 0 \atop n_1+n_2 = n\geq 1}\atop
n_1,n_2\geq 0} \spr_s {1\over n_1! n_2!}
\big\{ \Psi^{n_1} , \Phi_s \big\}_{g_1}
\big\{ \Psi^{n_2} , \wt\Phi_s \big\}_{g_2}(-)^{\Phi_s}.
\eqn\wweell$$
We now simplify the sum over states in the above expression
$$\eqalign{
\spr_s \big\{ \Psi^{n_1} , \Phi_s \big\}_{g_1}
\big\{ \Psi^{n_2} , \wt\Phi_s \big\}_{g_2}(-)^{\Phi_s}
&= \spr_s \big\{  \Phi_s ,\Psi^{n_1}  \big\}_{g_1}
\big\{  \wt\Phi_s , \Psi^{n_2} \big\}_{g_2}(-)^{\Phi_s} ,\cr
&= \spr_s \big\langle \Phi_s , \l \Psi^{n_1} \r_{g_1} \big\rangle
\big\langle \wt\Phi_s , \l \Psi^{n_2} \r_{g_2} \big\rangle (-)^{\Phi_s}.\cr}
\eqn\nnbbnn$$
In order to proceed further we use the explicit representation of the
inner product
$$\eqalign{
{}&=\spr_s \bra{R_{12}}\Phi_s\rangle_1
c_0^{-(2)}\l\Psi^{n_1}\r_{g_1}^{(2)}
\,\,\bra{R_{1'2'}}\wt\Phi_s\rangle_{1'}
c_0^{-(2')}\l\Psi^{n_2}\r_{g_2}^{(2')}(-)^{\Phi_s}, \cr
{}&=\spr_s \bra{R_{12}} c_0^{-(2)}\l\Psi^{n_1}\r_{g_1}^{(2)}
\,\bra{R_{1'2'}} c_0^{-(2')}\l\Psi^{n_2}\r_{g_2}^{(2')}
\,\ket{\Phi_s}_1 \ket{\wt\Phi_s}_{1'} (-)^{\Phi_s}, \cr
&=\bra{R_{12}} c_0^{-(2)}\l\Psi^{n_1}\r_{g_1}^{(2)}
\,\bra{R_{1'2'}} c_0^{-(2')}\l\Psi^{n_2}\r_{g_2}^{(2')}
\,b_0^{-(1')}\P_{1'}\ket{R_{11'}}, \cr}\eqn\dinnertime$$
where we moved the states to the right and recognized the
sewing ket. We take $b_0^{-(1')}$  into the bra $\bra{R_{1'2'}}$
where it becomes $b_0^{-(2')}$ and then eliminates the $c_0^-$
insertion. Moreover the projector $\P_{1'}$ is not necessary
since the state $\l \Psi^{n_2} \r_{g_2}$ is annihilated by
$L_0^-$. Thus we find
$$\eqalign{
{}&=\bra{R_{12}} c_0^{-(2)}\l\Psi^{n_1}\r_{g_1}^{(2)}
\bra{R_{1'2'}} \l\Psi^{n_2}\r_{g_2}^{(2')}
\ket{R_{11'}}, \cr
&=\bra{R_{12}} c_0^{-(2)}\l\Psi^{n_1}\r_{g_1}^{(2)}
\l\Psi^{n_2}\r_{g_2}^{(1)} , \cr
&= \big\langle \l \Psi^{n_2} \r_{g_2},\l\Psi^{n_1}\r_{g_1}\big\rangle ,\cr
&=\big\{ \Psi^{n_1} , \l \Psi^{n_2} \r_{g_2} \big\}_{g_1}, \cr}
\eqn\thelongest$$
and therefore the first term entering the BV equation (Eqn. \wweell )
is simply given by
$$(\hbox{I}) = -{1\over 2}
\sum\limits_{{g_1 + g_2 = g\geq 0 \atop n_1+n_2 = n\geq 1}\atop
n_1,n_2\geq 0} {1\over n_1!\, n_2!}
\big\{ \Psi^{n_1} , \l \Psi^{n_2} \r_{g_2} \big\}_{g_1} .\eqn\firstpart$$
The second term entering the BV equation is readily computed
$$\eqalign{
{\partial_r\over \partial \psi^s}
{\partial_l S_{g-1}^{n+2}\over \partial \psi_s^*}
&= {1\over (n+1)!} {\partial_r \over \partial \psi^s}
\big\{ \Psi^{n+1} , \wt\Phi_s \big\}_{g-1} (-)^{\Phi_s +1} ,\cr
&= {1\over n!} \spr_{G(\Phi_s) \leq 2}
\big\{ \Psi^n , \Phi_s , \wt\Phi_s \big\}_{g-1} (-)^{\Phi_s+1} ,\cr}
\eqn\secterm$$
where we got no extra sign factor in the last step
because the statistics of the derivative, which equals the
statistics of $\psi_s$ is opposite to the statistics of the state
$\ket{\wt\Phi_s}$. Now, as we have done several times by now,
the term inside the sum is written as two terms: one half of itself
plus one half of itself; one of these
terms is rewritten using \mrp , we then obtain the complete sum over
all ghost numbers. Thus finally, we have
$$ (\hbox{II}) =- {1\over 2\cdot n!} \spr_s
\big\{ \Psi^n , \Phi_s , \wt\Phi_s \big\}_{g-1} (-)^{\Phi_s}.
\eqn\sectermi$$
Now putting all together (Eqns. \firstpart\ and \sectermi )
we have that the BV equation requires

$$(\hbox{I}) + (\hbox{II}) = -
{1\over 2} \sum\limits_{{g_1 + g_2 = g\geq 0 \atop n_1+n_2 = n\geq 1}\atop
n_1,n_2\geq 0} {1\over n_1!\, n_2!}
\big\{ \Psi^{n_1} , \l \Psi^{n_2} \r_{g_2} \big\}_{g_1}
-{1\over 2 \cdot n!} \spr_s
\big\{ \Psi^n , \Phi_s , \wt\Phi_s \big\}_{g-1} (-)^{\Phi_s}=0,\eqn\palltog$$
which indeed is true on account of Eqn. \mbsidentityiii .
This concludes our verification that the master action satisfies
the quantum BV equation.
\medskip
\noindent
$\underline{\hbox{BRST transformations}}.\,$ In the BV formalism
once we have the full master action we can immediately obtain the
master transformations. These transformations become the BRST
transformations upon gauge fixing. What we want to show now is
that the master transformation for the string field takes the
form $s \ket{\Psi} = (\hbox{Equation of Motion of}\,\,\Psi)\cdot\mu$, that is,
they are proportional to the field equations arising from the master
action. This fact is familiar for the classical field theory, as
shown in [\kugosuehiro ], and was also noted in open string field
theory [\thornpr ]. We begin our derivation using Eqn.\mastertrans\
$$\eqalign{s \ket{\Psi} &= \{ \ket{\Psi} , S \} \cdot \mu ,\cr
{}&=\biggl( {\partial_r\ket{\Psi}\over \partial \psi^s}
{\partial_l S\over \partial\psi_s^*}-
{\partial_r\ket{\Psi}\over \partial \psi_s^*}
{\partial_l S\over \partial\psi^s}\biggr) \cdot \mu , \cr
{}&=\spr_{G(\Phi_s)\leq 2}\biggl( \ket{\Phi_s}
{\partial_l S\over \partial\psi_s^*}-
\ket{\wt\Phi_s}{\partial_l S\over \partial\psi^s}\biggr) \cdot \mu , \cr
{}&=\sum_{g\geq 0,n\geq 1} \hbar^g \k^{n+2g-2} \spr_{G(\Phi_s)\leq 2}
\biggl( \ket{\Phi_s}{\partial_l S_g^n\over \partial\psi_s^*}-
\ket{\wt\Phi_s}{\partial_l S_g^n\over \partial\psi^s}\biggr)\cdot\mu ,\cr}
\eqn\findmast$$
where in the last step we made use of \decompaction .
Evaluating the derivatives (as in \howdiff )
$$\eqalign{s \ket{\Psi} &= -\hskip-10pt\sum_{g\geq 0,n\geq 1}
\hskip-8pt\hbar^g\k^{n+2g-2}
\hskip-10pt\spr_{G(\Phi_s)\leq 2}{(-)^{\Phi_s}\over (n-1)!}
\biggl( \ket{\Phi_s}\big\{ \Psi^{n-1},\wt\Phi_s \big\}_g+
\ket{\wt\Phi_s}\big\{ \Psi^{n-1},\Phi_s\big\}_g\biggr)\cdot\mu ,\cr
{}&= -\sum_{g\geq 0,n\geq 1} \hbar^g \k^{n+2g-2} {1\over (n-1)!}
\spr_s(-)^{\Phi_s}\ket{\wt\Phi_s}\big\{ \Phi_s,\Psi^{n-1}\big\}_g ,\cr
{}&= -\sum_{g\geq 0,n\geq 1} \hbar^g \k^{n+2g-2} {1\over (n-1)!}
\l \Psi^{n-1} \r_g \cdot \mu, \cr}\eqn\almmast$$
where we made use of the rearrangement property \mrp , and of \oppsrel .
The minus sign can be clearly absorbed into the definition
of the anticommuting parameter $\mu$, and the sum over $n$ can be shifted
to obtain
$$\hbox{MASTER TRANSFORMATIONS:} \quad s\ket{\Psi}= \sum_{g,n\geq 0}\hbar^g
\k^{n+2g-1} {1\over n!} \l\Psi^n\r_g \cdot \mu .\eqn\alxmmast$$
This is the desired result. Indeed, for genus zero we have
$$s\ket{\Psi}= \sum_{n\geq 0} \k^{n-1} {1\over n!}
\l\Psi^n\r_0 \cdot \mu = \F (\Psi ) \cdot \mu ,\eqn\alxmmast$$
where $\F$ is the classical field equation.

In the BV formalism one obtains the BRST transformations as follows.
One requires that the antifields be specified by a relation of the form
$$\psi_s^* = {\partial \Upsilon \over \partial \psi^s}, \eqn\gferm$$
where $\Upsilon$ is called the gauge fermion. The gauge fixed
path integral is given by integrating only over fields, and substituting
on the master action $S(\psi^s, \psi_s^*)$ the antifields by use of
\gferm :
$$Z_{\Upsilon} = \int \hbox{d}\psi^s \exp \biggl[ -{1\over \hbar}
S(\psi^s , \partial\Upsilon/\partial \psi^s )\biggr] .\eqn\gfpi$$
One can verify that this gauge fixed action is independent of the
choice of gauge fermion $\Upsilon$ by use of the master equation
[\batalinvilkovisky ].
The BRST transformations of this gauge fixed theory are simply given by
the master transformations evaluated at the gauge fixing surface
$$\hbox{BRST TRANSFORMATIONS:}\quad \delta_B\,\ket{\Psi} (\psi^*=0)\equiv
(s\ket{\Psi} )\bigg|_{\psi^*=\partial\Upsilon /\partial\psi},\eqn\brstgf$$
and they are only defined for the fields (that is the reason for writing
$(\psi^*=0)$ in the left hand side). That means that the terms in the
right hand side that could be identified as transformations of antifields
are simply ignored.
The main property of this transformation is that it leaves the path
integral in \gfpi\ invariant. Neither the action nor the measure are
invariant but their variations cancel each other. The BRST transformations
are nilpotent on the mass-shell (see, for example, [\thornpr ]).

The most well-known gauge condition is the Siegel gauge, in which
one requires $b_0^+ \ket{\Psi} = 0$.\foot{One must do some work in
order to obtain this gauge choice using a gauge fermion (see [\thornpr ]
for the case of open strings). It would be interesting
to know if the Siegel gauge can always be reached, regardless of the
conformal field theory and the particular details of its cohomology.
The case of two dimensional strings, where the semirelative cohomology includes
states that are not annihilated by $b_0^+$ might be a good case study.}
For a string field annihilated by $b_0^+$ the kinetic term boils down to
$$S_{kin} = {1\over 2} \, \bra{\Psi} c_0^-c_0^+ (L_0+\ov L_0) \ket{\Psi},
\eqn\gfkt$$
and the propagator takes the form
$${b_0^- b_0^+ \over L_0+ \ov L_0} \P \ket{R_{12}}, \eqn\gfprop$$
where the operators acting on the reflector can refer to either the
first or the second Hilbert space. When the propagator is given
a sewing interpretation we use a proper time $\tau \in [-\infty , 0]$
and a twist angle $\theta \in [0,2\pi ]$. Then $b_\theta = ib_0^-$
and $b_\tau = b_0^+$ and we find that the propagator can be rewritten as
$${1\over 2\pi i}\, b_\theta b_\tau \int_{-\infty}^0 \hbox{d}\tau
\int_0^{2\pi} \hbox{d}\theta e^{(\tau + i\theta )L_0}
e^{(\tau - i\theta )\ov L_0} \cdot \ket{R_{12}}.\eqn\gfpsew$$

\REF\zwiebachoc{B. Zwiebach, `Quantum Open string theory with manifest
closed string factorization', Phys. Lett. {\bf 256B} (1991) 22; `Interpolating
string field theories', Mod. Phys. Lett. {\bf A7} (1992) 1079.}
\REF\getzlerjones{E. Getzler and J. D. S. Jones, `$A_\infty$ algebras
and the cyclic bar complex', Ill. Jour. Math. {\bf 34} (1990) 256.}

\section{Homotopy Lie Algebras and Axioms of Closed String Field Theory}

The purpose of the present section is two-fold. We will first
describe briefly a mathematical framework for homotopy Lie
algebras and then discuss the algebraic structure of the
classical closed string field theory. Closed string field
theory formulated around a conformal background corresponds to
an $L_\infty$ algebra with products $m_1,m_2,\cdots$, where
$m_n$ denotes a product whose input is $n$ string fields.
The stability of this algebraic structure under shifts of the
string field that correspond to classical solutions was
shown by Sen [\seneqm ]. Here we will consider shifts of the string
field that do not correspond to classical solutions. The gauge
invariance of the resulting theory corresponds to a homotopy
Lie algebra with products $m_0, m_1 ,m_2 \cdots$; where $m_0$
is a `product' that, without an input string field, gives us a
string field ( $m_0$ is just a special state). We give
the lowest order identities and comment on their relevance to
background independence.

Homotopy associative algebras or $A_\infty$ algebras were introduced
by Stasheff [\stasheff ]. These algebras, which are relevant to
classical open string field theory, \foot{Witten's open
string field theory is strictly associative [\witten ], but if one uses
open string stubs, which are necessary to make the closed string sector
explicit [\zwiebachoc ], the algebra of open string fields is only
homotopy associative.} were used in homotopy theory.
The main question was to understand
when a given space $X$ was homotopy equivalent to some loop space
$\Omega Y$. It was known that an $H$ space \foot{A space $X$
is an $H$-space if it has a multiplication $m: X \times X \rightarrow X$,
such that for some point $e$ one has $ex=xe=x$.} with a strictly
associative product was homotopy equivalent to a loop space, but
strict associativity, which cannot always be achieved, turned out not
to be essential. Homotopy associativity, that is, associativity up to
homotopy is. The result was that $X$ is homotopy equivalent to a
loop space $\Omega Y$ if and only if $X$ is and $A_\infty$ space
and $\pi_0(X)$ is a group. An $A_\infty$ space is a space with
a set of multiplications $m_n$, $n\geq 2$, satisfying a set of
associativity conditions. An $A_\infty$ algebra is a similar structure
on a vector space. Its definitions is reviewed in the work of
Getzler and Jones [\getzlerjones ]. The case when the multiplications
are (graded) commutative, which leads to the homotopy Lie algebras, seems to
be less familiar in the literature. We could simply take the
the homotopy Lie algebra to be defined by a set of multilinear
graded commutative products satisfying relation \ncidentity .
Alternatively, we can define the homotopy Lie algebra $L_\infty$
mimicking the definition of the $A_\infty$ algebra given in
Ref. [\getzlerjones ]. This is what we do next.

Given a graded vector space $\H$ we consider the vector space
${\bf T}(\H )$ formed by adding symmetrized products of $\H$
$${\bf T}(\H )= \sum_{n=0}^\infty S \H^{\otimes n} .\eqn\sysmhilb$$
The elements of this symmetrized space will
be linear combinations of elements denoted as
$(B_1 , B_2 , \cdots B_n) \in S \H^{\otimes n}$. Since we take
symmetrized products, the $B$'s above can be interchanged with sign factors
directly correlated to the statistics (as in the multilinear products).
The space ${\bf T}(\H)$ forms a cotensor {\it coalgebra}. That is,
it has a {\it comultiplication} $\Delta$ such that
$\Delta : {\bf T}\rightarrow {\bf T} \otimes {\bf T}$. The comultiplication
must be {\it coassociative}, that is, the following diagram must commute
$$\matrix{ {\bf T} & \buildrel \Delta \over \longrightarrow &
{\bf T}\otimes {\bf T} \cr
\mapdown{\Delta} & {} &\mapdown{\Delta \otimes {\bf 1}} \cr
{\bf T}\otimes {\bf T}&\buildrel{{\bf 1}\otimes \Delta}\over\longrightarrow &
{\bf T}\otimes {\bf T}\otimes {\bf T} \cr}
\eqn\coassociativity$$
For our case comultiplication is defined as follows
$$\Delta (B_1 ,\cdots B_n) = \sum\limits_{ \{i_l , j_k \}; l,k\geq 0
\atop l+k = n} \wh\sigma (i_l,j_k) (B_{i_1},\cdots B_{i_l})\otimes
(B_{j_1},\cdots ,B_{j_k})
,\eqn\defcomult$$
where the sum extends over inequivalent splittings (as defined below
\bidentity ) and the sign factor $\wh\sigma$ is the sign picked up
rearranging the sequence that appears in the right hand side into
the order that appears in the left hand side. One can verify that
this coderivation is coassociative. The extra ingredient to have
a homotopy Lie algebra is a coderivation $b$, with
$b: {\bf T}\rightarrow {\bf T}$, of degree $(-1)$ and such that
$b^2=0$. A coderivation satisfies a co-Leibnitz rule, that is,
the following diagram commutes
$$\matrix{ {\bf T} & \buildrel b \over \longrightarrow &
{\bf T} \cr
\mapdown{\Delta} & {} &\mapdown{\Delta} \cr
{\bf T}\otimes {\bf T}
&\buildrel {{\bf 1}\otimes b + b\otimes {\bf 1}}\over\longrightarrow
& {\bf T}\otimes {\bf T} \cr}
\eqn\coderivation$$
We want to use the multilinear products to define the coderivation.
In order to have a coderivation of degree $-1$, we will define
the degree $d$ of $B_i$ as
$$d(B_i) = 2 - G(B_i) ,\eqn\definedegree$$
then, making use of Eqn. \gpro\ we find
$$d (\bigl[ B_1 , B_2, \cdots , B_n \bigr] )= 2 -
G(\bigl[ B_1 , B_2, \cdots , B_n \bigr] )= -1 +
\sum_{i=1}^n d(B_i),\eqn\degpro$$
and now all of our products act as maps of intrinsic degree $-1$.
The BRST operator indeed increases ghost number by one, and therefore
lowers the degree by one unit. The special element $\l \, \cdot \,\r_0$
must be of degree $(-1)$, that is, of ghost number $(+3)$.
The coderivation acts linearly on the vectors in ${\bf T}$, and
is defined (on elements of ${\bf T}$) as follows:
$$b(B_1 ,\cdots B_n) = \sum\limits_{ \{i_l , j_k \} ; l,k\geq 0
\atop l+k = n} \sigma (i_l,j_k) \bigl( B_{i_1}, \cdots , B_{i_l},
\l B_{j_1},\cdots ,B_{j_k} \r_0 \bigr) ,\eqn\defcomult$$
where $\sigma$ is the same sign factor as that in \bidentity .
Note that $b$ acting on an element in $S\H^{\otimes n}$ gives
us a sum of terms in $S\H^{\otimes k}$ with $1\leq k\leq n$.
Rather than prove here that this is a coderivation and that
it satisfies $b^2=0$ let us illustrate how the conditions
$b^2=0$, reproduce the lower identities we had before (taking
$\l \, \cdot \, \r_0=0$, for simplicity).
The simplest case is $b(B_1)= (\l B_1 \r_0) = (QB_1)$, and therefore
$b^2(B_1) = b(QB_1) = (QQB_1) = 0$, by nilpotency of $Q$.
Next, we have
$$b(B_1,B_2 ) = \l B_1,B_2\r_0+(QB_1,B_2)+(-)^{B_1}(B_1,QB_2),\eqn\bontwo$$
and then if we apply $b$ again we have
$$b^2(B_1,B_2 ) = Q\l B_1,B_2\r_0+ b(QB_1,B_2)+(-)^{B_1}b(B_1,QB_2),\eqn
\bontwoi$$
using \bontwo\ to evaluate the two last terms in the right hand side
we get four terms, two of which cancel, and therefore we find
$$b^2(B_1,B_2) = Q\l B_1,B_2\r_0 +\l QB_1,B_2\r_0+(-)^{B_1}\l B_1,QB_2\r_0 =0,
\eqn\sixtwenty$$
which is the familiar condition that $Q$ is a derivation of the
product $m_2$.

\noindent
$\underline{\hbox{Shifting String Field Theory into a Non-Conformal
Background}}.\,$ Closed string field theory is currently formulated by
first choosing a conformal background, which defines a BRST operator
and to some degree the string field, since the basis of vectors that
we use to expand the string field is the basis of states of the
conformal field theory. The algebraic structure of such a classical
theory is determined by the BV equation and has been seen to be
an $L_\infty$ algebra with products $m_n$, with $n\geq 1$. If the string
field is shifted by a classical solution, one expects the resulting
string field theory to correspond to a theory formulated around
another conformal background. Indeed, upon such shift, one finds
an identical algebraic structure [\seneqm ] holding for
modified products, which is evidence that the new theory corresponds
to another conformal background. If we shift the string field
$\Psi$ by letting $\Psi \rightarrow \Psi_0 + \Psi'$ with $\Psi_0$
a string field that {\it does not} solve the classical equations
of motion, then the new string field theory, whose fluctuation
field is $\Psi'$, and whose vacuum corresponds to $\Psi'=0$,
is a string field theory formulated around some sort of background
that is not conformal. What we do next is elucidate the algebraic
structure of such theory.

It is simple to see that such theory still has gauge invariance.
If $S(\Psi)$ is invariant under
$\Psi \rightarrow \Psi+ \delta_\Lambda \Psi$ (where $\delta_\Lambda \Psi$
is defined in \gaugetrca ), then clearly $S(\Psi_0 + \Psi')$ is
invariant under
$\Psi_0+\Psi'\rightarrow \Psi_0+\Psi'+\delta_\Lambda (\Psi_0+\Psi')$,
and as a consequence it is invariant under
$\Psi'\rightarrow \Psi'+ \delta_\Lambda (\Psi_0 + \Psi')$. This
is the gauge invariance of the new action. Let us therefore begin
by finding out what happens with the action. From the expression
given in \notsofamca\ we get
$$\eqalign{
S(\Psi_0+\Psi' ) &= {1\over \k^2}\,
\sum_{n=0}^\infty \, {\k^n \over n!}
\sum_{m=0}^n { n \choose m} \big\{{\Psi'}^m \Psi_0^{n-m} \big\}_0\cr
{}&=  {1\over \k^2}\, \sum_{m=0}^\infty \sum_{n=m}^\infty
\,{\k^n\over m!(n-m)!}\big\{{\Psi'}^m \Psi_0^{n-m} \big\}_0\cr
{}&=  {1\over \k^2}\, \sum_{m=0}^\infty \sum_{n=0}^\infty
\,{\k^{n+m}\over m!n!}\big\{{\Psi'}^m \Psi_0^n \big\}_0\cr
{}&=  {1\over \k^2}\, \sum_{n=0}^\infty {\k^n\over n!}
\sum_{m=0}^\infty \,{\k^m\over m!}\big\{{\Psi'}^n \Psi_0^m \big\}_0 .
\cr}\eqn\shiftaction$$
This result suggests that we define a new set of multilinear functions
and products, denoted by primes
$$\eqalign{
\big\{ B_1, \cdots , B_n\big\}' &= \sum_{m=0}^\infty {\k^m\over m!}
\big\{ B_1, \cdots , B_n, \Psi_0^m\big\}, \cr
\l B_1, \cdots , B_n\r ' & = \sum_{m=0}^\infty {\k^m\over m!}
\l B_1, \cdots , B_n, \Psi_0^m \r ,\cr} \eqn\defprimeprod$$
which are therefore related to each other as
$$\big\{ B_0, B_1, \cdots , B_n \big\}' =
\langle B_0, \l B_1,\cdots B_n \r ' \rangle . \eqn\relprimeprod$$
Using the definition of the primed multilinear functions,
the shifted action in Eqn. \shiftaction\ simply reads
$$S(\Psi_0+\Psi' )\equiv S'(\Psi') = {1\over \k^2}\,
\sum_{n=0}^\infty \, {\k^n \over n!} \{ \Psi^n \}'.
\eqn\notsofamcai$$
The new action therefore takes the same form as the original
action (in \notsofamca ) with the multilinear functions changed
into new ones. The first two terms in the expansion are
$$\eqalign{
\{ {\Psi'}^0 \} &= \sum_{m=2}^\infty {\k^m \over m!} \{ \Psi_0^m \}
\equiv \k^2 S(\Psi_0),\cr
\{ {\Psi'}^1 \} &= \sum_{m=1}^\infty {\k^m \over m!} \{\Psi', \Psi_0^m \}
= \langle \Psi' , \sum_{m=1}^\infty {\k^m\over m!}
\l \Psi_0^m \r \rangle =
\langle \Psi' , \k \F (\Psi_0) \rangle ,\cr}\eqn\totalrecall$$
where $\F$ was defined in \eomca . Thus the shifted action, reads
$$S(\Psi_0+\Psi' )\equiv S'(\Psi')= S(\Psi_0) + \langle \Psi',
\F (\Psi_0) \rangle + \cdots , \eqn\expected$$
which, of course, is the expected form of the expansion, since
had $\Psi_0$ satisfied the field equation, the term linear in
the fluctuation field $\Psi'$ would have vanished. Let us now
consider the gauge transformations. As we discussed above
the shifted action is invariant under the transformation
$${\delta'}_\Lambda \Psi' = \delta_\Lambda (\Psi_0 + \Psi')
= \sum_{n=0}^\infty {\k^n\over n!} \l (\Psi_0+\Psi')^n , \Lambda \r ,
\eqn\eleventhirti$$
expanding out and rearranging in the same way as we did for
the action above, one finds
$${\delta'}_\Lambda \Psi'
= \sum_{n=0}^\infty {\k^n\over n!} \l {\Psi'}^n , \Lambda \r ' ,
\eqn\elehirti$$
which is just the same form of the standard gauge trasformations
but now with the primed products. The first term corresponds to
the new BRST-like operator
$$\l \Lambda \r '\equiv Q'\Lambda = \sum_{n=0}^\infty {\k^n\over n!} \l
\Lambda ,\Psi_0^n \r = Q\Lambda + \sum_{n=1}^\infty {\k^n\over n!} \l
\Lambda ,\Psi_0^n \r .\eqn\newbrst$$
As we will see this new BRST-like operator does not quite have the
standard properties.
The reader may wonder how it is
possible that the structures defining the action and gauge transformations
appear not to have changed despite the shift that did not satisfy the
field equations. Actually the structures have changed! When we began
with a conformal theory we had products $m_1, m_2 \cdots$, but for
convenience our whole analysis was done including the product $m_0$,
which was set to zero, and the associated multilinear function, that
was also set to zero. Our proof of gauge invariance, beginning in
\gaugetrca\ never assumed that the product $m_0$ (and the associated
multilinear function) was zero, we simply used the basic identity
\ncidentity\ for all values of $n$. This time $m_0$ is not zero
anymore. Nevertheless the gauge invariance of the shifted action
under the shifted gauge transformations implies that the primed
products must satisfy \ncidentity ! Thus, the structure of string
field theory formulated around a background that is not conformal
is that of an $L_\infty$ algebra with products beginning with $m_0$.
It is fun to see what happens with the identities we are so familiar
with. This time we will denote the special element defined by
$m_0$ by the name $F$,
$$\l \, \cdot \, \r ' \equiv F .\eqn\wehavemo$$
The state $F$ must be Grassmann odd and of ghost number $(+3)$.
Note that, from Eqn. \totalrecall\ we have that
$$ \{ {\Psi'}^1 \} \equiv  \langle \Psi' , \l \, \cdot \, \r ' \rangle
= \langle \Psi' , \k \F (\Psi_0) \rangle ,\eqn\amtired$$
and therefore, for our previous example $F=\k \F (\Psi_0)$.
The lowest identity in \ncidentity\ was analyzed earlier in
\clfornzero\ where we had
$$ 0 = \l [ \cdot \, ]' \r ' \quad \rightarrow \quad
Q' \l \, \cdot \, \r ' = Q'F = 0 . \eqn\clfornzero$$
That is, the special element is annihilated by the new `BRST'-like
operator. This is a nontrivial identity. The next case is
that in \trivcase\ which reads
$$ 0= \l [ B ]' \r ' + (-)^B \l B, [\, \cdot \, ]' \r ' =
\l Q'B \r' + (-)^B \l B , F \r ' = 0, \eqn\trivvcase$$
which implies that
$${Q'}^2 B + \l F , B \r ' = 0 ,\eqn\failnilp$$
and therefore the new BRST-like operator fails to be nilpotent
by a term involving the product $m_2$ and $F$.
The next identity is \addithere\ which this time gives
$$0 = Q' \l B_1 , B_2 \r ' + \l Q' B_1 , B_2 \r' + (-)^{B_1} \l B_1,
Q'B_2 \r ' + \l F,B_1,B_2 \r ' , \eqn\foxxfxo$$
that is, $Q'$ also fails to be a derivation by a term involving
$F$.

It seems that the above discussion could be useful for our
understanding of background independence. The analogy with
Einstein's gravity is interesting. In this theory, the metric
on a manifold is the dynamical variable, and the Ricci-flat
metrics are classical solutions. The analogous objects in
string theory are not clearly known, but it would seem reasonable
to believe that the dynamical variable (string field) in
some sense specifies a two-dimensional field theory, and the
classical solutions are the conformal field theories. Thus
our present knowledge is as if we knew how to write Einstein's
equations around Ricci-flat metrics, but not around
arbitrary metrics. If we were in this situation, by
shifting the metric a bit we would be simulating Einstein's
equations around a non-Ricci-flat metric. This is what
we did above for string theory. In this sense our problem
in string theory is not that we use a two-dimensional field theory to write
the string action, it is that we use a {\it conformal} field
theory. If we knew how to write the string action for a more
general class of theories it would be an advance. What we did
was exploring indirectly how such theory would look.
It would be interesting to study more directly, in two-dimesional
field theory, what is the operator $Q'$, and what is the state
$F$.

As a last point I wish to emphasize that gauge invariance of the
Einstein action expanded around an {\it arbitrary background}
works in a way exactly analogous as that discussed above for closed
string theory. Consider the Einstein action, and an arbitrary
background metric $\ov g_{\alpha\beta}$ that does not satisfy the
field equations.
Einstein's action for the metric
$g_{\alpha\beta} = \ov g_{\alpha\beta}+h_{\alpha\beta}$ would read
$$ S(g) = -{1\over \k^2} \int \sqrt{\ov g}\, \ov R\, d^4x
+{1\over \k^2} \int d^4x \sqrt{\ov g}\, \l\, \ov R^{\mu\nu}
- {1\over 2} \,\ov g^{\mu\nu}\, \ov R\, \r \, h_{\mu\nu}
+ \O (h^2), \eqn\einst$$
Now, the gauge transformations on the original metric, imply
the following transformations for the fluctuating field $h_{\mu\nu}$
$$\delta_\epsilon h_{\mu\nu} = D_\mu \epsilon_\nu+D_\nu\epsilon_\mu,
\eqn\gaugtrgrav$$
The above expansion of the action in powers of the fluctuation is
gauge invariant, this requires that the variation of the term
linear in $h$ be zero, which happens because
$$D_\nu \l \,\ov R^{\mu\nu} - {1\over 2}\,
\ov g^{\mu\nu}\, \ov R\, \r = 0, \eqn\bianchi$$
which is simply the Bianchi identity. The last three equations
are entirely analogous to our previous equations
$$S(\Psi_0+\Psi' )=  S(\Psi_0) + \langle \Psi',
\F (\Psi_0) \rangle + \cdots , $$
$${\delta'}_\Lambda \Psi' = Q' \Lambda + \cdots ,$$
$$Q' \F = 0 , \eqn\analogy$$
which were used to study the gauge invariance of the string field
theory formulated around a non-conformal background.

\chapter{String Vertices and The Geometrical Identity}

String vertices are actually defined, in most generality, by
Riemann surfaces with punctures and coordinates around the
punctures. In covariant string field theory, the vertices are
universal in the sense that they may be used for any background
around which the theory is formulated. Vertices are the Riemann
surface ingredient of the string field theory. In fact, finding
a consistent set of vertices was the main difficulty in constructing
closed string field theory.
In \S5.1 we will define precisely what vertices are, and give the basic
properties they must have in order that the field theory defined by these
vertices yield manifestly factorizable and unitary amplitudes, with
correct symmetry properties and hermiticity. The conditions that
eventually yield BRST invariance will be given in \S5.2.

\section{Definition of the String Vertices}

A string vertex, at a given genus $g$ and for a process involving
$n$ strings is a set of surfaces. This set is the subset
of the full (compactified) moduli space of Riemann surfaces
$\ov\M_{g,n}$,\foot{The compact boundaryless space $\ov\M_{g,n}$
is obtained from $\M_{g,n}$ by adding the surfaces with nodes, or
degenerate surfaces. Such surfaces, also called the compactification
divisor, are obtained by sewing ordinary surfaces in the limit of
vanishing sewing parameter.}
and as we will see in \S6, it corresponds to the surfaces that
are not produced by the sewing procedure.  The corresponding
{\it string field vertices}, which define the interactions of
string fields will be discussed in \S7. Let us now define
the {\it string vertices}.

\noindent
$\underline{\hbox{Definition}}$.  A string vertex $\V_{g,n}$ is
a set of Riemann surfaces of genus $g$ and $n$ punctures
representing a subset of $\ov\M_{g,n}$. This set does
not include surfaces arbitrarily close to degeneration: there exist
open neighborhoods of the compactification divisor of $\ov\M_{g,n}$
that $\V_{g,n}$ does not intersect. Each surface in the set
must be equipped with a specific choice of analytic local
coordinates around each of its punctures. The coordinates around
each puncture are only defined up to a constant phase. The coordinates
must be defined continuously (up to the phase ambiguity) over the
set $\V_{g,n}$.

Let us elaborate on this definition. In particular we will impose
further conditions on the vertices. The fact that local coordinates
are defined up to phases means that two sets of local coordinates
$z_i$ and $z_i'$ on the same punctured surface are said to be
equivalent if for any point $P$ around the $i$-th puncture, such
that $z_i(P)$ is defined, $z_i'(P)$ is also defined, and one has
$z_i (P) = \exp (i\theta_i) z_i'(P)$, for all $i$. The $\theta_i$'s
are puncture dependent constant phases. The set $\V_{g,n}$ may not
be connected nor simply connected so continuity of the definition
of local coordinates is understood to hold for each connected
component of the set. Continuity also holds up to the constant
phase ambiguity. It is necessary for continuity of the off-shell
amplitudes as we move in moduli space. This, in turn, is required
for BRST invariance.

It is probably impossible to get rid of the phase ambiguity in the
definition of the local coordinates in the string vertices. It is
well known that one cannot define local coordinates without phase
ambiguities all over moduli space because of a topological
obstruction [\nelson ]. Whether one can define them continuously over
the subsets $\V_{g,n}$ is not clear, but seems rather hard
since the $\V_{g,n}$'s are expected to have complicated topology.
At any rate neither quadratic differentials nor minimal area metrics
show us how to avoid the phase ambiguities.

Since the scale of the local coordinates must be fixed, we will
impose the following requirement:

\noindent
(a) For every surface $\R \in \V_{g,n}$ the local coordinates
$z_i$ around each puncture must define a disk $D_i = |z_i| \leq 1$.
The disks must not overlap, except possibly at their boundaries.

The case when the disks $D_i$ cover precisely the complete Riemann
surface, are called overlap type vertices. In general this will not
be the case, and removal of the disks leaves a surface with a number
of holes. While the classical closed string field theory can be
constructed with overlap type vertices, this is not the case for
the quantum theory, as we shall see in the next section.

The complete information about the local coordinates is actually
encoded in the disks $D_i$. \foot{I thank A. Connes for this
observation.} A disk $D_i$ whose boundary is a simple closed Jordan
curve $\g_i$ (a curve without self-intersections) allows us to define
local coordinates in a very simple way: we declare the curve $\g_i$
to be the locus $|z_i|=1$,
and the puncture to correspond to $z_i=0$. The coordinates in the
disk $D_i$ are defined by the conformal mapping to the unit disk.
The three parameter ambiguity in the conformal map relating two
disks is reduced to one via the constraint on the position of the
puncture. The left over parameter is the ambiguity due to rigid
rotations of the disk. Therefore

\noindent
$\underline{\hbox{Definition}}$. A punctured Riemann surface with
local coordinates around the punctures $p_i$, is a Riemann surface
with simple closed Jordan curves $\g_i$, called
{\it coordinate curves}, homotopic to the punctures, defining
punctured disks $D_i$, such that
$\hbox{int}(D_i) \cap \hbox{int}(D_j) = 0$ for $i\not = j$,
where int($D_i$) denotes the interior of the disk $D_i$.

Our inability to define phases corresponds to the fact that we
cannot single out a point in every coordinate curve continuously
over moduli space. Two local coordinates agree (up to phases)
if their coordinate curves are {\it identical}. While the coordinate
curves are, a priori, curves without parametrization, the map to
the unit disk gives them a canonical parametrization.

The condition that the surfaces in $\V_{g,n}$ be non-degenerate
surfaces is essential since these surfaces will give rise to the
interaction vertices of the string field theory and we insist
that all degenerate surfaces arise from propagators, in order to
have manifest factorization and unitarity. It is natural to require
for a covariant theory, that these vertices be symmetric under the
exchange of punctures, thus, even though our punctures are always
distinguishable, or labeled punctures, we demand that

\noindent
(b) The assignment of local coordinates to the punctured surfaces
is independent of the labeling of the punctures, moreover, if
a surface $\R$ with labeled
punctures is in $\V$ then copies of $\R$ with all other inequivalent
labelings of the punctures are also included in $\V$.
\medskip
\noindent
This condition has been discussed in detail in [\sonodazwiebach ], where we
also required that:
\medskip
\noindent
(c) If a surface $\R \in \V$ then $\R^* \in \V$, where $\R^*$ denotes the
mirror image of $\R$; moreover, the local coordinates in $\R$ and $\R^*$ are
related by the antiholomorphic map that relates the two surfaces.

This condition is required for manifest hermiticity of the string field
interactions defined by the string vertices [\sonodazwiebach ]. We now turn
to the conditions that will guarantee BRST invariance of the theory.

\section{Consistency Conditions on String Vertices}

The choice of vertices is not arbitrary. There is a very stringent consistency
condition represented in the following geometrical equation [\sonodazwiebach ]
\vskip .5in
$${}\eqn\twopointone$$
\vskip .5in
In this equation the left hand side denotes the set of
surfaces in $\V_{g,n}$ that lie on the boundary of the region of
moduli space this vertex fills. These are surfaces with labeled punctures. In
the right hand side the first object is a sum of {\it groups} of terms, each
such group is labeled by a choice of numbers $(g_1,n_1)$ and $(g_2,n_2)$, where
$g_1 + g_2 = g$, and $n_1+n_2=n+2$. These numbers indicate the genus and number
 of punctures of the surfaces we will deal with. Having done this we split the
$n$ labeled punctures into two groups, $(n_1-1)$ for the surfaces having $n_1$
punctures, and $(n_2-1)$ for the surface having $n_2$ punctures. The punctures
left open are to be sewn to each other. Since this splitting can be done in
in $n\choose n_1-1$ different ways this is the number of terms we obtain
for this group. For each term we now consider the sets $\V_{g_1,n_1}$ and
$\V_{g_2,n_2}$, with the labeled punctures assigned and one puncture
left over (the way to assign the labeled punctures is immaterial since
the vertices are symmetric, as demanded in property (b) \S5.1).
We then sew each surface in the set $\V_{g_1,n_1}$ to every surface
in the set $\V_{g_2,n_2}$. This sewing is done by fixing arbitrarily
the phase ambiguity in the two punctures to be sewn (marking a point
in the coordinate curves) and then sewing them according
to the relation $z_1 z_2 = t$, where the sewing parameter $t$ takes
all the values in the unit circle $|t| =1$, and where
$z_1$ and $z_2$ are the local coordinates around the punctures. It is
clear that $n_1,n_2 \geq 1$, since there must be at least one puncture
in each surface in order to sew them together. When
$g=0$, both $n_1$ and $n_2$ must be greater or equal to three.
The factor of $1/2$ preceding this first object in the right hand side
of the equation is necessary because every surface that this first object
produces is actually produced twice. This follows from the above description,
the sum double counts even when $g_1=g_2$, since then the splitting of the
labeled punctures still works as if the surfaces were of different genus.

The second term denotes the set of surfaces obtained
by taking each surface in the set $\V_{g-1,n+2}$,
choosing arbitrarily two punctures, fixing marked points on the
coordinate curves, and sewing the punctures via $z_1 z_2 = t$
with $t$ taking all values in the unit circle $|t| = 1$. This
term only exists for $g\geq 1$. A factor of $1/2$ is also necessary
in this case. This happens because once more surfaces are produced
twice. Say we label punctures `one' and `two' to be the punctures to
be sewn. Any surface in the set $\V$ can be paired to another
surface in the set which differs only by the interchange of labels on these
two punctures (this is condition (b) in \S5.1). Upon sewing, the original
surface and its corresponding pair will produce the same sets of surfaces.

Eqn.\twopointone\ applies for $n\geq 3$ when $g=0$. This is
equivalent to setting $\V_{0,n} \equiv 0$, for $n=0,1,2$.
It also applies for all $\V_{g,n}$ when $g\geq 1$ with the
exception of $\V_{1,0}$ (the one-loop cosmological constant).
Note that for this case ($g=1,n=0$), there are no candidate terms in the
right hand side. Thus the vacuum vertex $\V_{1,0}$ is not
constrained by \twopointone . All the higher vacuum vertices are.

The equality of the sets requires that the surfaces agree {\it both in their
moduli parameters and in their local coordinates around the punctures}.
Both the right hand side and the left hand side are sets of equal
dimensionality; the boundary operation subtracts one real dimension,
and sewing adds one real dimension. The set $\V$ appearing in the
left hand side (without the boundary operator) is of one complex dimension
higher than that of the sets appearing on the right (without sewing).
There are also signs in Eqn.\twopointone\ (these signs, and the symmetry
factor, which were ommitted from the graphical representation in
Refs.[\sonodazwiebach ,\zwiebachqcs ], will be included here). The boundary of
$\V_{g,n}$ in the left-hand side is some subspace of $\ov\M_{g,n}$ with an
orientation on it. The orientation is induced by that of $\ov\M_{g,n}$. The
same
subspace with an opposite orientation would be accompanied by an opposite
sign. For the terms in the right-hand side one should note that these
correspond
to Feynman diagrams built with one propagator in the limit when the propagator
collapses. This is so because the usual propagator corresponds to sewing with
parameter $t$ in the region $|t| \leq 1$. One may therefore fix the orientation
of the terms in the right-hand side of Eqn. \twopointone\ by thinking of them
as
boundaries of the regions of $\ov\M_{g,n}$ obtained with $|t|<1$.
Typically there are cancellations between terms in the right-hand side of
Eqn.\twopointone . This equation is an equation in the sense
of homology, where we can add or subtract manifolds with a given orientation.

The above equation follows if the string vertices $\V$,
generate a {\it single} cover of each moduli space $\ov\M_{g,n}$
when used together with propagators (sewing with $|t| \leq 1$)
to build the relevant Feynman graphs. The basic statement in
Eqn.\twopointone\ is that the vertex $\V_{g,n}$ and the Feynman graphs
with one propagator cover regions lying on {\it opposite\/} sides
of their common boundary.  This explains why there is a minus sign
in the right hand side of the equation.
Denoting by $R_I$ the region of moduli space covered by
all the Feynman graphs constructed with $I$ internal lines, the
above equation can be written as
$$\partial \V_{g,n} = - \partial_p R_1,\eqn\twopointtwo$$
where $\partial_p$ denotes a boundary obtained
by propagator collapse $(|t|=1$). If we have a single cover
of moduli space, the full moduli space should be
obtained by the union of all the regions covered by the
various Feynman graphs
$$\ov\M_{g,n} = \V_{g,n} \cup R_1\cup \cdots \cup R_{3g-3+n},
\eqn\twopointthree$$
where $(3g-3+n)$,  the complex dimension
of $\ov\M_{g,n}$, is also the maximum possible number of
propagators.

Equation \twopointtwo\ holds for the
lowest dimensional moduli spaces, those corresponding to
four punctured spheres and one-punctured tori. This
happens because in these two cases Eqn.\twopointthree , which
is the assumption of single cover, reduces
to $\ov\M_{g,n} = \V_{g,n} \cup R_1$, since there are only
graphs with one propagator. The absence of boundary of $\ov\M$
implies that $\partial\V = -\partial R_1$. Moreover, since the
Feynman graphs use only three point vertices, which have no
moduli, the boundary $\partial R_1$ coincides with the
propagator boundary $\partial_p R_1$.

Equation \twopointtwo\ is established in general by induction
[\sonodazwiebach ] and we elaborate on the proof next.
First note that any Feynman graph with $I$ propagators covers
some region of moduli space. This region can be parameterized
by the propagator parameters and the parameters that specify
the vertices $\V$ involved in this particular Feynman graph.
For each choice of these parameters we get a specific surface,
and different choices of parameters must correspond to different
surfaces, for otherwise we would have multiple cover of moduli
space. This implies that the boundary of the region of moduli
space that we get from this Feynman graph corresponds to the
boundary of the parameter space of the graph. Thus we can have
either propagator boundaries, denoted by $\partial_p$, when
a propagator collapses, or vertex boundaries, $\partial_v$, when
some vertex used in the Feynman graph has its parameters lying
on the boundary of the parameter space of that vertex. Thus,
for the region of moduli space produced by any Feynman graph $f_I$
with $I$ propagators we can write
$$\partial f_I = \partial_pf_I+ \partial_v f_I, \eqn\decbound$$
and therefore, for
$R_I= \sum f_I$, which is the total region of moduli space arising from
Feynman graphs with $I$ propagators, a similar equation holds
$$\partial R_I = \partial_pR_I+ \partial_v R_I, \eqn\decboundr$$
which says that the boundary of $R_I$ is also of two types, boundaries
that can be seen to arise from some graph on a propagator boundary, and
boundaries that arise from some graph on a vertex boundary.
Since $\ov\M_{g,n}$ has no boundary,
we then have
$$\eqalign{\partial \ov\M_{g,n} \equiv 0 = &\,\partial \V_{g,n} +
\partial_p R_1 \cr
{}&+(\partial_v R_1 + \partial_p R_2) + \cr
{}&\qquad\vdots \cr
{}&+(\partial_v R_{3g-4+n} + \partial_p R_{3g-3+n})\cr
{}&+\partial_v R_{3g-3+n}\cr}\eqn\twopointfour$$
We now claim that the induction hypothesis, namely that Eqn.\twopointtwo\
holds for any moduli space of dimension lower than that of $\ov\M_{g,n}$,
implies that
$$\partial_p R_I = - \partial_v R_{I-1}, \quad I\geq 2.
\eqn\twopointfive$$
We will show this next. In the left hand side, whenever a
propagator collapses, we can look at the vertices involved, these
may be just one, if the propagator joins two legs of a single vertex,
or two, if it joins two vertices. Call this the {\it collapsed subdiagram}.
It is a subdiagram since the graph must have at least another
propagator ($I\geq 2$). The collapsed
subdiagram corresponds to some $\ov\M_{g',n'}$, where $n'$ is the
number of legs attached to the vertex (or vertices) without counting the legs
involved with the collapsed propagator, and $g'$ is the genus of the vertex
(or sum of genera of the vertices). The dimensionality of
$\ov\M_{g',n'}$ must be lower than that of $\ov\M_{g,n}$, since
the graph has additional parameters that those involved in the
subdiagram (at least the parameters of the extra propagator).

Consider then any given surface in $\partial_pR_I$. The
collapsed subdiagram is of the form $\partial_p f_1$, that is, a
propagator boundary of some Feynman graph with one propagator. Consider
now keeping fixed all the parameters of the graph that {\it do not} correspond
to the subdiagram. This includes the parameters of the propagators
joining into the subdiagram, or the parameters of propagators
that may be joining two legs of the subdiagram; such propagators are
not, in general, at their boundary values. Since using Feynman rules
one must include all possible diagrams, it is clear that keeping
these parameters fixed, we must find in $\partial_p R_I$ the diagram
in question, with the collapsed subdiagram not only taking the value
$\partial_p f_1$ but rather running over the whole set $R_1$
associated with the moduli space $\ov\M_{g',n'}$. Therefore we can
use the induction hypothesis, Eqn.\twopointtwo\ to replace the
collapsed subdiagram $\partial_pR_1$ by a vertex $-\partial_v\V_{g',n'}$
at a vertex boundary. It should be noted that Eqn.\twopointtwo , or its
graphical representation in Eq.\twopointone , can be used even if the $n$
external states are actually not external (as is our case now), but are
connected via sewing to other vertices. This is so because we can think
of Eqn.\twopointone\ as an equation relating surfaces with boundaries, given
the control we have over the coordinate systems at the punctures.
In this process we have obtained a diagram with one less propagator and
with a vertex at a vertex boundary.  This is just the statement of
Eqn.\twopointfive .

One now makes use of Eqn.\twopointfive\ to cancel all pairs
of terms in parenthesis in Eqn.\twopointfour . Moreover
the last term vanishes since the corresponding Feynman graph
is built with three point vertices only and those have no
vertex boundary. It therefore follows that the first
two terms in the right hand side of Eqn. \twopointfour\ must add up to
zero. This shows Eqn.\twopointtwo\ holds to next order in the
dimensionality of moduli space, and therefore in general.

Let us note the significance of the geometrical equation \twopointone .
This is a consistency condition for sets of Riemann surfaces with local
coordinates, and has nothing to do with a specific choice of conformal field
theory, or string theory. As we will see later the existence
of $\V_{g,n}$'s satisfying this condition is the main
geometrical input to the full off-shell field theory of closed strings.
This equation encodes the Riemann surface
aspect of string theory, to all orders in perturbation theory.
Given that the existence of a set of string vertices satisfying
the equation has such important consequences, it is essential
to make sure such consistent string vertices exist. This will be the
subject of section \S6, where we will use minimal area metrics
to find what appears to be the simplest
possible solution of the consistency conditions. As it turns out
a vertex is necessary for every moduli space. Our strategy will
be to show that we have a single cover of moduli space when
we choose suitable $\V_{g,n}$'s. Therefore, as we discussed above,
the string vertices will satisfy the consistency conditions.

Equation \twopointone\ may describe some interesting algebraic
structure on the space of Riemann surfaces, with the subsets
$\V_{g,n}$ representing generators and the product operation
being sewing. Our use of \twopointone , however, will be
based on the fact that it will help us define, in \S7, a
useful algebraic structure on {\it string fields}, that is, on the
state space of conformal field theories.

\chapter{Minimal Area String Diagrams}

In the present section we will discuss the basic ideas
concerning minimal area metrics. The main purpose is to determine
a canonical set of string vertices $\V_{g,n}$ that satisfy the consistency
condition \twopointone . Our analysis will be different from that
of [\zwiebachma ] in that we will not emphasize the relation to quadratic
differentials. Using the results of Ranganathan [\ranganathan ], Wolf and
the author [\wolfzwiebach ] and some further results given here we will be
able to give a characterization of the string vertices $\V_{g,n}$ which is
intuitively compelling, and far more concrete than any previous description.
Moreover, we will show how to obtain the unique Feynman graph corresponding
to any surface by inspection of its minimal area metric.

\REF\thurston{W. Thurston, private communication (1991).}
\FIG\convexfig{Consider the family of metrics $\rho_t$ with
$t\in [0,1]$. The area function $\A (\rho_t )$ is a strictly convex function
of $t$. This can be used to show the uniqueness of the minimal area metric.}

\section{Properties of the Metric of Minimal Area}

A (conformal) metric $\rho$ on a Riemann surface $\R$ defines
a length element $dl= \rho |dz|$ and an area element
$\rho^2 dx \wedge dy$. Here $\rho$ must be positive semi-definite,
and since $dl^2 = \rho^2(dx^2+dy^2)$, $\rho^2$ is recognized to be
simply the Weyl factor of the metric. Whenever we talk about metrics
on Riemann surfaces we refer to such conformal metrics. For any
surface $\R$ we are interested in the metric of least possible area under
the condition that all nontrivial closed curves on $\R$ be longer than or
equal to $2\pi$ [\zwiebachma ]. This minimal area problem, in contrast
with the related problems studied earlier in the mathematical literature
(see [\strebel ]) does not require specifying some homotopy classes of
curves on the Riemann surface on which we impose length conditions.
The length conditions are imposed on all curves. This is why
this is a problem defined on moduli space, and why the resulting metrics
have nice factorization properties (= nice behavior on the compactifying
divisor). A metric is called {\it admissible} if it satisfies the length
condition
$$ \int_\g \rho |dz| \geq 2\pi , \eqn\condadm$$
for all nontrivial closed curves $\g \in \R$. A useful property of the
space of admissible metrics is that it is a convex space; if
$\rho_0$ and $\rho_1$ are admissible metrics then
$$\rho_t = (1-t) \rho_0 + t \rho_1, \eqn\convexspace$$
is an admissible metric for all $t \in [0,1]$, as it follows directly from
the above two equations.
The area functional $\A$ has a very nice property: it is
strictly convex. This means that for the above one-parameter
family of metrics, with $\rho_0 \not= \rho_1$, one has
$$\A (\rho_t ) < (1-t) \A (\rho_0 ) + t \A (\rho_1 ),\eqn\strictcon$$
for $t \in (0,1)$. The inequality is strict since we do not include
the endpoints $t=0,1$ corresponding to $\rho_0$ and $\rho_1$. The
above relation is quite familiar, and is derived using the Schwartz
inequality. It simply means that the area, as a function of $t$,
is convex (see Fig. \convexfig ). An immediate consequence of this fact
is the uniqueness of the minimal area metric (if it exists).
This is argued as follows. Assume there are two different metrics
$\rho_0$ and $\rho_1$ on the surface $\R$, both admissible and
both having the (same) least possible area $A$. It follows from convexity
of the space of admissible metrics that $\rho_t$ (for any
$t\in (0,1)$ ) as defined
above is admissible, and due to the convexity of the area functional
$\A (\rho_t) < (1-t) A + tA = A$, in contradiction with the assumption
that $A$ was the least possible value for the area.
The uniqueness of the minimal area metric is fundamental for our
purposes. Another simple consequence of convexity of the area
is that given two admissible metrics $\rho_0$ and $\rho_1$
such that $\A (\rho_0) > \A (\rho_1)$, we then have
$\A (\rho_t) < A (\rho_0)$ for $t\in (0,1)$ (see Fig. \convexfig ).

A metric solving the minimal area problem is expected to
give rise to closed geodesics of length $2\pi$ that foliate
the surface completely. The reason for this was given in
[\zwiebachqcs ] and elaborated in [\wolfzwiebach ]: through every
point where
the metric is continuous there must exist nontrivial closed
curves of length arbitrarily close to $2\pi$, otherwise the
metric could be lowered at this point without destroying
admissability. While further discussion is necessary to prove
the existence of a geodesic of length precisely $2\pi$,
it seems clear intuitively that this curve must exist.
One can show that such curves must foliate the surface since
they cannot intersect whenever they are of the same homotopy
type (see below). We shall call the length $2\pi$ geodesics by
the name {\it saturating geodesics}.
They can be thought to describe the closed strings on the Riemann surface.
We can actually go a long way towards understanding the
properties of the minimal area metric by the use of the
following Lemma, suggested to me by W. Thurston [\thurston ].

\noindent
$\underline{\hbox{Lemma 6.1}}.\,$  Two saturating geodesics
that intersect at a point where the metric is smooth cannot
intersect elsewhere.

\FIG\thursfig{Two saturating geodesics $\C_1$ and $\C_2$ intersect
at a point $p_1$ where the metric is smooth. One can show their
intersecting elsewhere (as in the figure) leads to a contradiction.}

\noindent
$\underline{\hbox{Proof}}.\,$ Consider two saturating geodesics
$\C_1$ and $\C_2$. By definition they are nontrivial simple closed
curves of length $2\pi$. They may or may not be homotopic to
each other. Pick the intersection point $p_1$  where the metric
is smooth, and assume there are more intersection points. Find another
intersection point $p_2$ in $\C_1$ such that there is a segment
$\overline{p_1p_2}$ in $\C_1$ containing no other intersection points.
Denote this segment by $a'$ and the remaining of $\C_1$ by $a$.
The points $p_1,p_2$ also determine two segments on $\C_2$, they
will be denoted as $b'$ and $b$ (see Fig.\thursfig ).

The symbols $a,a',b,b'$ will denote both the segments and their
lengths; $a \sim b$ will mean that the two segments are homotopic,
$a \{ >, < , = \} b$ will mean relations between their lengths.
We have
$$a+ a' = 2\pi, \quad b+ b' = 2\pi, \eqn\basic$$
We claim now that neither $a$ nor $a'$ can be homotopic to
$b$ or $b'$. Suppose $a \sim b'$, then we have that
two new nontrivial closed curves: $a' + b' \sim a' + a$
and $b+ a \sim b + b'$. If $b'<a$ then the curve $a' + b'$
would be smaller than $2\pi$. If, on the contrary $a<b'$
then the curve $b+a$ would be smaller than $2\pi$. Therefore
$a = b'$, and the curves $a+ b'$ and $b+a$ would be saturating
geodesics. Their lengths, however, can be shortened at the
point $p_1$, where these geodesics have corners. This shows
that $a \not\sim b'$. The other cases are entirely analogous.

Given two open curves with identical endpoints which are not
homotopic to each other, they define a nontrivial closed curve.
Thus we must have four nontrivial closed curves
giving the following conditions:
$$\eqalign{
a \not\sim b \quad \rightarrow \quad a+b \geq 2\pi, \cr
a \not\sim b' \quad \rightarrow \quad a+b' \geq 2\pi, \cr
a' \not\sim b \quad \rightarrow \quad a'+b \geq 2\pi, \cr
a' \not\sim b' \quad \rightarrow \quad a'+b' \geq 2\pi. \cr}
\eqn\newclosed$$
Adding pairs from the above equations and using \basic\ we then find that
$\{ a, a' , b , b' \} \geq \pi$, which implies that
actually $\{ a , a' , b , b' \} = \pi$, and that the four
curves above are saturating geodesics. Since they all have
corners their lengths can all be reduced, and this is a
contradiction to admissibility. This concludes our proof
of the Lemma. \hfill $\square$

Let us now consider Riemann spheres with punctures. Let us understand
why the minimal area metrics here always arise from Jenkins-Strebel
quadratic differentials. Since any saturating geodesic on a genus
zero surface with punctures must cut the surface into two separate
pieces, if two saturating geodesics cross they must do it
at least in two points. The above lemma indicates that this
cannot be a generic situation, thus we are led to expect that
the surface will be foliated by bands of geodesics that do not
intersect. This is precisely what happens with Jenkins-Strebel
quadratic differentials, the horizontal trajectories of the
quadratic differential are the saturating geodesics. The surface
is then foliated by bands of geodesics, the so-called ring-domains
of the quadratic differential. The horizontal trajectories
can intersect in the critical graph of the quadratic differential,
but this graph is of zero measure on the surface.

It also becomes clear from the lemma that in higher genus Riemann
surfaces one can very easily have saturating bands of geodesics
that cross. This already happens at genus one, where curves
along the ``a-cycle'' and ``b-cycle'' are both nontrivial and
intersect only once. Thus higher genus minimal area metrics
do not always arise from quadratic differentials. One concrete
example showing such nontrivial crossing of foliations was given
in [\wolfzwiebach ].

\section{Results and Open Questions in Minimal Area Metrics}

Let us try to make clear what we still do not know about
minimal area metrics. Still missing is a proof of existence of such
metrics. The evidence in favor of existence is very strong; the minimal
area metrics are known for all Riemann spheres and for large subsets
of the higher genus moduli spaces (the restricted Feynman graphs
of Ref. [\zwiebachma ]). We now have even examples of minimal area metrics
that do not arise from quadratic differentials [\wolfzwiebach ]. Still an
abstract proof of existence is not available yet. Another important
property is that of completeness of the minimal area metric. This
has to do with the presence of punctures; since curves surrounding
punctures must also be longer than or equal to $2\pi$, we expect
that neighborhoods of the punctures must look like tubes going off
to infinity. Thus we expect that any curve reaching the puncture
would be of infinite length. This is the statement of completeness
of the metric. It is simple to convince oneself that an incomplete
metric is innefficient as far as minimizing area, but a proof of
completeness is not available yet. Finally, we need smoothness of
the minimal area metric in some neighborhood of each puncture.
Again, it is simple to convince oneself that discontinuous metrics
are inefficient, but a proof is required. In summary, we have
the following assumption

\noindent
$\underline{\hbox{Assumption}}.\,$ We assume the existence of
a complete minimal area metric smooth in some neighborhood of
each puncture.

At the present time we can prove, with the above assumption, that
the minimal area metrics have the requisite conditions to define
the string diagrams of covariant closed string field theory.
The requisite conditions were essentially those of flatness and
amputation [\zwiebachqcs ].

Flatness requires that there should exist a neighborhood of the
punctures that it isometric to a flat semiinfinite cylinder of
circumference $2\pi$. This is established by showing first that
completeness and smoothness implies that there must exist a
neighborhood of each puncture which is foliated by a single
band of geodesics, all homotopic to the puncture (Lemma 2.1 [\wolfzwiebach ]).
Then one uses the result of Ranganathan [\ranganathan ], that in a region
foliated by a single homotopy type of geodesics the metric is flat.
A similar result was established in [\wolfzwiebach ] where it was also shown
that the neighborhood is isometric to a flat semiinfinite cylinder
of circumference $2\pi$.

The basic tool for the proof of flatness in [\wolfzwiebach ] was
a lemma showing how to construct local deformations of admissible
metrics. One begins with a metric $\rho_0$ smooth over a domain foliated
by (posibly several bands of) saturating geodesics
and perturbs it into $\rho_0 + \e h$ where $h$ is a smooth metric with
compact support in the domain and such that no saturating geodesic
becomes smaller than $2\pi$ in the new metric. The new metric is
not yet admissible, since the old saturating geodesics, which are
still fine, may not be the shortest curves in the new metric. One can show
that the addition of an $\e^2$ term to the perturbed metric restores
admissibility. This means there is a constant $K$ such that the metric
$\rho_\e = \rho_0 +\e h + \e^2 K\chi$, where $\chi = \rho_0$ over a
compact domain, is admissible. The useful consequence is that the
derivative with respect to $\e$ of the area of the metric $\rho_\e$
is simply controlled by the sign of $\int \rho_0 h$. In order to show a metric
is not of minimal area one must find a suitable $h$ which makes the
above integral negative, which implies that for sufficiently small $\e$ the
new admissible metric $\rho_\e$ has less area.

\FIG\cancorfig{(a) A minimal area metric must be isometric to a
semiinfinite cylinder of circumference $2\pi$ for some neighborhood
of each puncture. This cylinder must end somewhere: the boundary
of the maximal region with saturating geodesics homotopic to the puncture
is the curve $\C_0$. A curve $\C_l$ is the saturating geodesic in
the cylinder a distance $l$ away from $\C_0$. The coordinate curves
will be taken to be the $\C_\pi$ curves. (b) The property of amputation
states that if we amputate the semiinfinite cylinder along $\C_l$
for $l >0$, the metric of minimal area on the cut surface is simply
the restriction of the original metric on the complete surface.}

The flatness property implies that the minimal area metric defines
a canonical domain around each puncture. The metric must cease to
look like a flat semiinfinite cylinder at some point. We define the
curve $\C_0$ to be the boundary of the maximal region foliated only by
geodesics homotopic to the puncture. We let $\C_l$ be the saturating geodesic
in the cylinder a distance $l$ away from $\C_0$. The maximal region is a
canonical domain which
will be used to define the local coordinates around the punctures (see
Fig.\cancorfig\ (a)).

The amputation property (Fig.\cancorfig\ (b)) states that given a minimal area
metric on a surface $\R$ with punctures, amputation of a semiinfinite
cylinder along the curve $\C_l$ for $l>0$ yields a surface with
boundary whose minimal area metric (under the assumption that it
be smooth in a neighborhood of the boundary) is simply the one
induced by the original metric in the surface $\R$ (see [\wolfzwiebach ],
Thm. 5.1). This property is essential to show that the plumbing
of surfaces with minimal area metrics gives a surface with a
minimal area metric [\zwiebachqcs ,\wolfzwiebach]. The basic idea is
simple, given two surfaces to be sewn together (or a single surface
with two legs to be sewn), one first amputates the semiinfinite cylinders
corresponding to the relevant legs.  Given the amputation property the
resulting surfaces with boundaries have minimal area metrics. If we glue
together the open boundaries and by doing this we do not create closed
curves that are smaller than $2\pi$ (this is the reason for stubs, as
we will see) then the resulting surface inherits a minimal area metric.
This is so because any candidate metric with less area would have to
have less area in at least one of the surfaces that were glued, but this
is impossible given that the amputated surfaces have minimal area metrics.
The reader interested in a rigorous argument should consult
Ref. [\wolfzwiebach ].

\section{Heights of Foliations and Cutting}

We need a couple of additional results in order to give a simple
characterization of the string vertices $\V_{g,n}$.
It is convenient to define the height of a foliation.
Consider a band of saturating geodesics that make up a foliation. Such
a band corresponds to an annulus, and this annulus has two boundaries.
Consider the set $\Gamma$ of all open curves $\g_o$ totally contained
in the annulus and whose endpoints lie on the two boundary components.
We define the height $h$ of the foliation to be
$h = \inf\limits_{\Gamma} l_\rho (\g_o )$. The height is essentially
the shortest distance, along the annulus, between the two boundary
components. We can now show that
\medskip
\noindent
$\underline{\hbox{Lemma 6.3}}.\,$ A foliation $F$ with a smooth metric
and with height $h$ greater than $2\pi$ is isometric to a flat cylinder
of circumference $2\pi$ and height $h$.
\medskip
\noindent
$\underline{\hbox{Proof}}.\,$ Denote the saturating geodesics that
foliate the annulus $F$ by $\g_f$. We first claim that there cannot be
another saturating geodesic $\g_i$ different from the $\g_f$ geodesics
going through the interior of $F$.
Suppose $\g_i$ is in the same homotopy type as the $\g_f$ geodesics, but does
not coincide with any $\g_f$ geodesic. Then pick a point $p \in \g_i$
such that $p\in F$, and consider the saturating geodesic $\g_f(p)$ going
through $p$. Since the metric is smooth at $p$, and $\g_i$ and $\g_f(p)$,
being homotopic, must intersect elsewhere \foot{The two geodesics cannot
be paralell at $p$ since they would have to coincide elsewhere, given that
the metric is smooth at $p$.} this is in contradiction with
Lemma 6.1. If $\g_i$ is in some other homotopy class it cannot be
completely contained in the interior of $F$ (the curves going around the
annulus more than once are longer than or equal to $4\pi$). It is also
clear that $\g_i$ cannot intersect any boundary of $F$ more than once,
since in that case, there would be $\g_f$ curves intersecting
$\g_i$ at more than one point (simply pick a $\g_f$ curve sufficiently
close to the boundary in question). Therefore $\g_i$ must intersect each
boundary once and only once. As a consequence a subcurve of $\g_i$ is totally
contained in $F$ and goes from one boundary component to the other. This
subcurve, by definition must be longer than or equal to the height $h$ of $F$.
Since we assumed $h > 2\pi$, the subcurve is too long and $\g_i$ cannot be a
saturating geodesic. Since there are no other saturating geodesics within $F$
the arguments of Refs. [\ranganathan ,\wolfzwiebach ] apply and imply that the
metric is flat, and isometric to that of a finite cylinder, which must
therefore be of height $h$.\hfill $\square$

We now need to establish a slight variation of the amputation theorem.
This is a ``cutting theorem'', and applies to internal annuli.
While most of the proof is analogous to that of amputation in
[\wolfzwiebach ] there are some differences in the analysis of curves, so we
give the proof below. Our later arguments will only use the
statement of the theorem, therefore the proof, which is not
so straightforward, may be skipped by the reader only interested
in the general arguments.

\noindent
$\underline{\hbox{Theorem 6.3}}.\,$ (Cutting Theorem)  Consider a Riemann
surface $\R$ with a minimal area metric $\rho_0$ and a foliation
$F$ of height $2\pi + 2\delta$ with $\delta >0$. Let $\C$ be
the middle geodesic in $F$ and let $\R_c$ denote the
cut surface (or surfaces) obtained by cutting $\R$ along $\C$. Assume the
cut surface(s) $\R_c$ has a minimal area metric $\rho$ continuous
in a neighborhood of the boundaries created by cutting.
Then $\rho$ is the restriction of $\rho_0$ to $\R_c$.

\FIG\cutfig{This figure is used to prove the cutting theorem. Shown
in (a) is a surface $\R$ which has a foliation $F$ of height
$2\pi + 2\delta$. This surface is cut along the middle curve $\C$
to obtain the cut surface $\R_c$ shown in (b). In (c) we show a
typical curve crossing $\C$.}

\noindent
$\underline{\hbox{Proof}}.\,$ The general setup for the
proof is given in Fig.\cutfig , where we show the surface $\R$
with the foliation $F$. This foliation defines an annulus,
which must be isometric to a flat cylinder of circumference
$2\pi$ and height $h= 2\pi + 2\delta$ (Lemma 6.3). We will use the
canonical flat coordinates on this cylinder where the original minimal
area metric $\rho_0 = 1$. When the surface is cut along the curve
$\C$, it may result in two surfaces
or in one surface. The proof below does not distinguish these two
possibilities and we will simple denote the cut surface by $\R_c$.

Since every nontrivial closed curve in $\R_c$ is actually a
nontrivial closed curve in $\R$,\foot{If it were not it would
have to bound a disk $D$ in $\R$. Since $D$ cannot be fully
contained in $\R_c$ (otherwise the curve would be trivial in $\R_c$)
$D$ must either intersect the cutting curve $\C$ or contain it
completely. Both possibilities are impossible, the first because
the original curve did not intersect $\C$, and the second because
$C$ is a nontrivial closed curve and therefore cannot be inside a disk.}
it follows that the restriction
to $\R_c$ of $\rho_0$ is an admissible metric on $\R_c$. We just
have to show it is of minimal area.
Assume there is a metric $\rho_1$ admissible in
$\R_c$, continuous near the boundaries produced by
the cutting procedure, and with less area than $\rho_0$.
Form now the metric
$$\rho_\e = (1-\e ) \rho_0 + \e \rho_1, \eqn\defrhoe$$
which is admissible due to linearity, and it has lower area
than $\rho_0$ by convexity. One has
$$A(\rho_\e ) = A (\rho_0) - k \e + k' \e^2, \eqn\vararea$$
where $k,k' >0$ are constants (independent of $\e$).

The idea, as in [\wolfzwiebach ] is to see how close is $\rho_\e$ to
be an admissible metric on the original surface $\R$.
If $\rho_\e$ were admissible on $\R$ we would be done
since we would have a contradiction with the fact that $\rho_0$
is supposed to be the minimal area metric on $\R$.

Let $\C^l$ and $\C^r$ denote the boundaries of the foliation
$F$, as indicated in the figure. Consider now the metric
$\rho_\e$ on $\R$. The closed curves that do not touch $\C$ must be
long enough since they are still nontrivial closed curves on $\R_c$.
Consider now curves that touch $\C$. They may be completely contained in
$F$ or may not be completely contained  in $F$. Consider
the second case first; such curves either intersect $\C^l$,
$\C^r$ or both. Suppose $\g$  intersects $\C^l$ only, then $\g$
must travel twice across the foliation $F_l$
and therefore using \defrhoe\ we find
$$l_{\rho_\e} (\g ) \geq (1-\e) l_{\rho_0} (\g ) \geq (1-\e) (2\pi
+2\delta),\eqn\whynonum$$
and choosing $\e < \pi /\delta$ such curves are clearly long enough.
If $\g$ intersects both $\C^l$ and $\C^r$ its $\rho_0$-length
must also exceed $(2\pi + \delta)$ and therefore the same
argument applies.

We now come to the curves that do not work automatically. These
are the curves that cross $\C$ and are completely contained in
the foliation $F$. Consider such a curve $\g$, we will show now
how to give a lower bound for its length. Consider now a subfoliation $\N$
of $F$ of height $2\e^{1/4}$ symmetrically centered around $\C$ (as shown
in part (b) of Fig.\cutfig ). We define the constant $K$ via
$$1 + K = \sup\limits_{\N} \rho_1 ,\eqn\defkk$$
where $\rho_1$ is evaluated in the canonical coordinates defined
by $\rho_0$. The constant $K$ cannot be negative, because otherwise
it would mean that $\rho_1 <1$ throughout $\N$ and the $\rho_1$
could not be admissible. We will therefore consider only
$K \geq 0$ and bounded (since we assumed the minimal area metric
is continuous in some neighborhood of the boundary).

Let us now show that for curves $\g$ completely contained
in the foliation $F$ one must have $l_{\rho_1}(\g ) \geq 2\pi (1-K)$.
(Recall that on the surface $\R$ the metric $\rho_1$ is possibly
discontinuous across $\C$.) Consider a curve $\g$, its intersections with
$\C$ divide the curve into segments. Let $\ov a_i$ and $\ov b_i$
denote the pieces of $\g$ on $F_l$ and on $F_r$ respectively.
Moreover, let $a_i$ and $b_i$ denote the segments on $\C$ homotopic
to $\ov a_i$ and $\ov b_i$ respectively. We now have two nontrivial
closed curves in addition to $\g$, the first is obtained
joining the $\ov a_i$ segments to the $b_i$ segments, and
the second joining the $\ov b_i$ segments with the $a_i$ segments.
Both curves are nontrivial curves on $\R_c$ (since they are homotopic
to the boundary $\C$), and therefore
admissibility of $\rho_1$ gives the following conditions
$$\eqalign{\sum_i \left[ l_{\rho_{1l}}(\ov a_i) +
l_{\rho_{1l}}( b_i) \right] &\geq 2\pi , \cr
\sum_i \left[ l_{\rho_{1r}}(\ov b_i) +
l_{\rho_{1r}}( a_i) \right] &\geq 2\pi , \cr }\eqn\beginder$$
where $\rho_{1l}$ and $\rho_{1r}$ denote the metric $\rho_1$
restricted to $F_l$ and $F_r$ respectively. We therefore
obtain
$$\sum_i \left[ l_{\rho_1}(\ov a_i) +
l_{\rho_1}(\ov b_i) \right] \geq 4\pi -
\sum_i \left[ l_{\rho_{1r}}( a_i) +
l_{\rho_{1l}}( b_i) \right] \eqn\nextder$$
and therefore
$$l_{\rho_1} (\g ) \geq 4\pi - {\hbox{max}}
\,\sum_i \left[ l_{\rho_{1r}}( a_i) +
l_{\rho_{1l}}( b_i) \right].\eqn\nnextder$$
{}From the definition of $K$ we have that
$$\eqalign{l_{\rho_{1r}} (a_i) &\leq (1+K) l_{\rho_0}(a_i),\cr
l_{\rho_{1l}} (b_i) &\leq (1+K) l_{\rho_0}(b_i),\cr} \eqn\nnnextder
$$
and therefore
$$ \sum_i \bigl( l_{\rho_{1r}}(a_i) + l_{\rho_{1l}}(b_i)\bigr)
\leq (1+K) \sum_i \bigl( l_{\rho_0}(a_i) + l_{\rho_0}(b_i)\bigr)
= (1+K) 2\pi ,\eqn\nnnnder$$
Back in \nnextder\ we obtain
$l_{\rho_1} (\g ) \geq 4\pi - (1+K) 2\pi = 2\pi (1-K)$
as we wanted to show. This is a powerful result since it tells how
small a curve can get by using cleverly the discontinuity of the
metric $\rho_1$ across $\C$. Note that in order to have good control,
the metric $\rho_\e$ weights in $\rho_1$ weakly.

We can now reduce the space of curves that are problematic.
Consider the subfoliation $\N '$ of $F$, which is of height
$2\e^{1/3}$. Note that $\N' \subset \N$. We now claim that
for $\e$ sufficiently small all curves crossing $\C$ and
extending beyond $\N'$ are good. The reason is that such
curves extend at least a distance $\e^{1/3}$ in the foliation
and therefore their $\rho_0$-lengths satisfy
$l_{\rho_0}(\g) \geq 2\pi \sqrt{1+ \e^{2/3}/\pi^2}$ (Lemma 5.1 of
[\wolfzwiebach ]).  It is
then possible to show (see Eqns. (5.20-5.23) of [\wolfzwiebach ]) that
for sufficiently small $\e$ one has $l_{\rho_\e}(\g ) > 2 \pi$.
Therefore the only problematic curves are the ones completely
contained in $\N'$. Following [\wolfzwiebach ] one introduces a new metric
$\rho_\e ' \geq \rho_\e$ everywhere such that the remaining
problematic curves now work; $\rho_\e '$ is admissible.
The area of $\rho_\e '$
differs from that of $\rho_\e$ by terms of order $\e^\chi$
with $\chi >1$ and therefore \vararea\ implies that for
sufficiently small $\e$ the area of $\rho_\e '$ is lower
than that of $\rho_0$ (the interested reader should consult
[\wolfzwiebach ] for details). This is the contradiction that completes
our proof of theorem 6.3. \hfill $\square$

\REF\brusteinalwis{R. Brustein and S. P. De Alwis, Nucl. Phys. B352 (1991)
451.}
\REF\brusteinroland{R. Brustein and K. Roland, `Space-Time versus
World-Sheet Renormalization Group Equation in String Theory',
Nucl. Phys. {\bf B372} (1992) 201.}
\REF\giddingsmartinec{S. Giddings and E. Martinec, `Conformal
geometry and string field theory', Nucl. Phys. {\bf B278} (1986) 91.}

\section{Towards a Concrete Description of All String Vertices}

We have a very explicit description of the minimal area metrics
for genus zero, and of the
string vertices for genus zero; these are simply given by the
restricted polyhedra [\saadizwiebach ,\kks ]. The polyhedra represent
the critical graphs of a Jenkins-Strebel (JS) quadratic differential.
For any surface, with at least one puncture, one can find a
unique JS quadratic differential with second order poles at the
punctures, all of equal (real and negative) residue,
and with a number of ring domains equal to the number of punctures
of the surface.\foot{The reader who is not familiar with
quadratic differentials may consult Refs.
[\strebel ,\saadizwiebach ,\zwiebachma ,\giddingsmartinec ].}
This simply means that any surface with $n$ punctures can be
constructed by gluing together the open boundaries of
$n$-semiinfinite cylinders of circumference $2\pi$ in a unique way.
The pattern of gluing is encoded on the `polyhedron' of the surface,
which is the critical graph of the quadratic differential. Thus any
surface has an associated polyhedron. The metric induced on the surface
by the quadratic differential ($\rho = |\phi (z)|^{1/2}$, where
$\phi(z) dz^2$ is the quadratic differential) is flat except at
isolated points with negative curvature singularities. The main result of
Refs.[\saadizwiebach ,\kks ] was that the string vertices $\V_{0,n}$
are given by the surfaces whose polyhedron is a {\it restricted} polyhedron,
that is, a polyhedron such that all of its nontrivial closed paths
are larger than or equal to $2\pi$. Such vertices were shown to give
a single cover of the moduli spaces $\ov\M_{0,n}$ relevant to the
classical theory [\kks ,\zwiebachma ].

No such concrete description is available yet for higher genus, basically
because we do not know completely what type of curvature singularities the
higher genus metrics have. A fairly concrete description is possible, however,
in terms of the structure of the foliations by geodesics that we expect to
exist. The basic result indicating how complicated the pattern of
foliations can be was given in Lemma 6.1. An interesting
counting problem would be to determine which is
the maximum number of foliations that can go through a point
in a surface of genus $g$ and with $n$ punctures. This would
require counting the maximum number of simple closed Jordan
curves of different homotopy all of which have a single common
point $p$ and intersect nowhere else. We expect the different
patterns of foliations as we move in $\ov\M_{g,n}$ to correspond
to different ``cells'' in a decomposition of the moduli space.
We will not try to elaborate on this point at present.
Instead we claim that a description of the string
vertices can be given as follows:

\noindent
$\underline{\hbox{The String Vertex}\,\, \V_{g,n}}.\,$ Consider
a Riemann surface $\R$ of genus $g\geq 0$ and $n\geq 0$ punctures
($n\geq 3$ for $g=0$) equipped with the metric of minimal
area. Then, $\R \in \V_{g,n}$ if and only if the heights of
all internal foliations are less than or equal to $2\pi$. If
$\R$ is in $\V_{g,n}$ the coordinate curves are defined to be
the $\C_\pi$ curves around each puncture.

In the above, internal foliation refers to a foliation
that does not correspond to a punctured disk, namely, it is not
one of the semiinfinite foliations around the punctures.
An internal foliation defines an internal annulus. We have placed the
coordinate curves leaving stubs of length $\pi$ for each
puncture. This is necessary in order to make sure that the
plumbing procedure produces candidate metrics; if we had
no stubs the plumbing of two punctures on a single surface
with a propagator of small length could introduce curves
shorter than $2\pi$ ruining the construction. While the stubs
are not necessary for the classical theory, the quantum theory
requires them and we must therefore use them here.

This actually means that the $\V_{0,n}$'s determined by this
rule do not agree with the restricted polyhedra! This is
actually good, for it was clear all along that the restricted
polyhedra were problematic at the loop level. We will elaborate
on this point in the next subsection. At any rate it is nice
that a simple generalization of the restricted polyhedra gives
a correct description of all string vertices.

Let us verify briefly that this definition for the string vertices
satisfies conditions (a),(b), and (c) required in \S5.1. Condition
(a) is clearly satisfied since the disks determined by the coordinate
curves cannot ever intersect. The string vertices are not of the
overlap type. Condition (b), that the assignment of local coordinates
be independent of the labelling of the punctures is also manifest;
the minimal area metric does not depend on the labels we assign to
the punctures, and therefore the coordinate curves do not either.
Condition (c), that the local coordinates in mirror pairs be related
by the antiholomorphic map between the surfaces, is slightly less
trivial. We claim that given a mirror pair $\R$ and $\R^*$, the
minimal area metrics on these surfaces must be precisely related
by the antiholomorphic map. First note that given an admissible metric
on $\R$ the map gives us an admissible metric on $\R^*$, and viceversa.
This shows that the minimal area metrics on $\R$ and $\R^*$ must have
the same area. Uniqueness of the minimal area metric then implies that
the metrics are related by the map. Since the map takes geodesics
to geodesics, it must, in particular take the coordinate curves on
one surface to the coordinate curves on the other, since these are
determined by the last saturating geodesics homotopic to the punctures.
Thus the coordinate disks, and as a consequence, the local coordinates are
mapped into each other by the antiholomorphic map relating the mirror pair.

The case of the one-loop vacuum graph is special. Suppose, as usual,
that we build tori using length $T$ cylinders of circumference $2\pi$,
and gluing together the boundaries with a twist angle $\theta$.
If we do this for all values of $T$ and $\theta$ ($-\pi \leq \theta <\pi$)
such that all closed curves on the resulting torus are longer than or
equal to $2\pi$, this will produce exactly one copy of the fundamental
domain [\zwiebachma ]. This happens because $\tau = (iT+\theta )/2\pi$,
and the length $l_c$ of the closed geodesics along the tube is
$l_c = 2\pi \sqrt{T^2 + \theta^2}$. Then, $l_c\geq 2\pi$ implies that
$|\tau | \geq 1$ (the condition on $\theta$ gives the standard condition
on the real part of $\tau$). In this construction we have a single foliation,
and its height is the length $T$ of the cylinder we started with. The
definition of the string vertices, applied to $\V_{1,0}$ would imply that
this vertex is not zero; it includes all the tori built with cylinders of
length $T\leq 2\pi$ (and having some nonzero twist angle).
This result, from the viewpoint of Feynman rules, would only seem to make
sense if we use a cutoff propagator. Using such cutoff propagator is
exactly the same as using stubs, except for this vacuum vertex (more on
this in \S6.5). In the cutoff propagator picture the string vertices
are the same as in our definition above, with the difference that the
coordinate curves would be the $\C_0$ curves. While we will use the
stubbed vertices viewpoint in this paper, we should keep in mind that
the alternative viewpoint could turn out to be relevant.

\noindent
$\underline{\hbox{Reconstructing Feynman Graphs}}.\,$
Before checking that the above is the correct string vertex let us
show how to reconstruct the Feynman graph associated to a surface from
the knowledge of its minimal area metric. Given a surface $\R$
with a minimal area metric first consider the semiinfinite foliations.
They correspond to the external legs of the diagram. Then consider the
foliations of height bigger than $2\pi$. If there are none, then the
surface is an elementary interaction $\V_{g,n}$, where
$n\geq 0$ is the number of infinite height foliations, and $g$ is the
genus of the surface. The foliations of height greater than $2\pi$,
which must be flat cylinders, give rise to the propagators. On each such
cylinder mark two closed geodesics, each a distance $\pi$ from each boundary
of the cylinder. Mark one curve on each semiinfinite foliation,
a distance $\pi$ from its boundary. Cut the string diagram open along
all these curves. The surface breaks up into a number of semi-infinite
cylinders, the external legs, a number of finite cylinders, the propagators,
and a number of surfaces with boundaries, that correspond
to the elementary interactions. Each elementary interaction corresponds
to an element of the set $\V_{g,n}$ where $g$ is the genus of the surface
with boundary and $n$ is the number of boundaries. This defines in a
unique way the Feynman graph associated to the surface. In a similar
way, a Feynman graph determines a unique minimal area metric; given
the data associated to the propagators and the vertices $\V$ used,
we can build the surface. It follows that two different Feynman graphs
must built surfaces with different patterns of foliations, and therefore
different metrics.

\noindent
$\underline{\hbox{Consistency of the Choice of}\, \V_{g,n}}.\,$
Let us now return to the problem of verifying the consistency
of our ansatz for the string vertices.
Two things must be checked. First, no surface in $\V$ must be produced by
sewing. This is clear since due to the stubs of length
$\pi$ present in every vertex, sewing creates foliations
of height greater than or equal to $2\pi$. When the height is
greater than $2\pi$ the surface is clearly not in $\V$ (by definition),
when the height is equal to $2\pi$ we have the case when the
boundary of the set $\V$ coincides with the collapsed propagator
boundary of the sewing domain. The uniqueness of the metric
of minimal area plus the compatibility of sewing with minimal
area implies that the metrics agree in this boundary, and so
do the coordinate curves, as a consequence.
The second point that must be checked is that all surfaces which
are not in $\V$ are produced by sewing, and produced only once.
The first part is clear because
for surfaces not in $\V$ there is at least one foliation whose height is
greater than $2\pi$. By the cutting theorem we can cut all the foliations
open and obtain a surface(s) with minimal area metric(s).
Restoring the semiinfinite cylinders around all boundaries one obtains
the Riemann surface(s) whose plumbing gives us the original surface.
Finally, no surface could have been produced more than once, since
different Feynman diagrams correspond to different metrics, which
by uniqueness of the minimal area metric, cannot correspond to the
same Riemann surface.

This concludes our proof that $\V$ is indeed the string vertex.
Since we have established that the $\V$'s generate a single
cover of all relevant moduli spaces, it follows, by the arguments
in \S5 that they satisfy the recursion relations \twopointone .

It is important to note, for our applications in \S9 that our arguments
hold also for string vertices defined by the condition that all internal
foliations be of height less than or equal to $2l$, where $l$ is any length
satisfying $l\geq \pi$. In this case the coordinate curves defined by
$\C_l$ with $l \geq \pi$. This means that the stubs are now of lenght
$l$. We therefore have generated a one parameter family of consistent
string vertices. Thinking of the stub length as a ultraviolet cutoff parameter
the basic identity \twopointone\ can be shown to imply a renormalization group
equation for string field theory [\brusteinalwis ,\brusteinroland].

\REF\sonodazw{H. Sonoda and B. Zwiebach, `Covariant closed string
theory cannot be cubic', Nucl. Phys. {\bf B336} (1990) 185.}
\REF\zembazwiebach{G. Zemba and B. Zwiebach, `Tadpole graph in
covariant closed string field theory', J. Math. Phys. {\bf 30} (1989) 2388.}

\section{Rearranging the Classical Theory}

Consider the classical theory, that is restrict to the set of vertices
$\V_{g=0,n}$, with $n\geq 3$. The consistency condition for such vertices does
not involve the second term in the right hand side of the recursion relation
\twopointone . Given such vertices one can construct a classical field theory
that has a `BRST' invariance and a gauge invariance. One can show that
under the condition of symmetry (condition (b) in \S5.1) there is
no solution with only a three-vertex and no higher vertices [\sonodazw ].
The simplest known solution is given by the restricted polyhedra described
in the previous subsection.

Let us see what is the problem in trying to extend
this to loops. If the classical vertices were
sufficient we should be able to see that the
remaining consistency conditions are satisfied.
Actually, if we could show that all the vertices
$\V_{1,n}$ at genus one vanished, then all the
higher ones would vanish too. Let us look at the
simplest vertex, namely $\V_{1,1}$. The consistency
condition requires that
$\partial \V_{1,1} = -R_1(loop) = 0$
where $R_1(loop)$ is the set of surfaces obtained
sewing two of the legs of the three string vertex.
This set of surfaces does not vanish when we use the naive
three string vertex. In fact the set of surfaces we get corresponds
to noded surfaces! This happens because the naive closed string
vertex is an overlap, and
is just another manifestation of the fact that the
three-vertex together with the standard propagator
does not give the correct one loop tadpole [\giddingsmartinec ,\zembazwiebach
].
The same problem would take place for all the vertices
$\V_{1,n}$ since in the right hand side of the consistency equation we
would find a classical vertex with two legs identified, and since all
classical vertices are overlaps this is bound to introduce
degenerate surfaces. Let us therefore deform the
original vertices in order to obtain sensible ones by using
two simple steps discussed in [\sonodazwiebach ] and [\zwiebachqcs ].
This procedure, applied to the original classical
vertices, which give problems at loops, gives us the string
vertices $\V$ discussed in the previous section.

Imagine we begin with a set of Feynman rules such
that the propagator involves tubes with length $l$ such
that $l\geq l_0$. Moreover assume
the vertices have legs all of which are of length $a$,
for example, overlap vertices but with cylindrical stubs of
length $a$ included. Note that when two vertices are
joined by a propagator, the intermediate tube has
total length, including stubs, $l \geq 2a+l_0$. Suppose
we want to change the propagator so that it includes lengths
$l\geq l_+ \geq l_0$. If we now join vertices, the intermediate
tubes will have lengths $l\geq 2a + l_+$, and therefore we
are going to miss some surfaces. One can compensate by
changing the stub length $a$, one amputates them, by an
amount which is half the difference between $l_+$ and $l_0$.
Since any intermediate line includes twice the stub length,
this will restore the missing surfaces. In the same way,
if $l_0$ is changed into $l_-$, which is smaller than $l_0$,
the intermediate lines become too short, and this is compensated
by increasing the length of every stub by half the difference
between $l_0$ and $l_-$. This latter operation is always possible
since we can always add stubs to the vertices. The former
operation cannot always be done since we cannot remove lengths
greater than $a$ to the stubs.

In fact, the first step we must do is of the problematic type.
We begin with propagators that go from zero to infinity and
with contact type vertices (zero length stubs). We want to
change the propagator to include only $l\geq l_0 > 0$. In
the above procedure this would involve amputating stubs, which
is not possible. The solution was given in [\sonodazwiebach ]. One leaves
the three-point vertex as it is. Then in calculating four
point amplitudes we would miss the Feynman graphs involving
two three-point vertices and an intermediate tube of length
less than $l_0$. We then define these Feynman graphs to be
an extra contribution to the original four-point vertex. This
procedure can be continued. The general answer given in [\sonodazwiebach ]
is that at any order the new vertices are given by the old
ones plus all the Feynman graphs built with the lower point
{\it original} vertices and all intermediate tubes of length
less than $l_0$. This is the first step.

The second step consists of restoring the original propagator.
As discussed above this is never a problem and it is done
simply by adding stubs of length $l_0/2$ to every one of the
vertices (including the three-point vertex) obtained by step
one. In this way we have obtained a new solution of \twopointone .
If we begin with classical overlap vertices, we obtain classical
vertices that are not overlaps.
In building up the full solution to higher
genus by using minimal area, it is convenient to take
$l_0=2\pi$. Thus the rearranged classical vertices will
have cylindrical stubs of length $\pi$, and we have recovered
the result of the previous subsection. The vertex $\V_{0,4}$,
for example does not only include the restricted tetrahedra
(with stubs) but also string diagrams built with two
(unstubbed) three string vertices and a propagator of
length less than $2\pi$ joining them.

\REF\schiffer{M. Schiffer and D. C. Spencer, ``Functionals of
Finite Riemann surfaces'', Princeton University Press, Princeton,
New Jersey, 1954.}
\REF\gardiner{F. P. Gardiner, ``Schiffer's Interior Variation and
quasiconformal mappings''. Duke. Math. Jour. {\bf 42} (1975) 371.}
\REF\nag{S. Nag, `Schiffer variation of complex structure and
coordinates for Teichmuller spaces', Proc. Indian Acad. Sci.
(Math. Sci.) {\bf 94} (1985) 111.}

\chapter{Construction of the String Field Interactions}

In the present section we use the geometrical structures
of \S5 and \S6 to construct the string field products, and
the string field vertices, or multilinear functions.
The main input is the set of string vertices $\V_{g,n}$.
Since such vertices correspond to subsets of the moduli spaces
$\ov\M_{g,n}$ we will have to concern ourselves
with integration. As we will see, we want to integrate over
a section of a certain bundle $\wh\P$ over moduli space (the integrand
is not defined on moduli space because we work off-shell). Our first
step will be to understand the various spaces involved. Then we show how
a tangent representing a deformation of a surface with local
coordinates, can be constructed concretely with a Schiffer
variation. Such variations are done with vector fields on the
surface, and these vector fields are relevant to define the
antighost insertions. We define a family of forms of different
degree on $\wh\P$, all labeled by a set of state vectors of the conformal
field theory. These forms are proven to be well defined if the
state vectors satisfy the subsidiary conditions $b_0^- = L_0^- = 0$.
We then show that the BRST operator acts as an exterior derivative
on the above forms. We define the string field multilinear
functions and multilinear producs, and establish some of their
properties. Much of the technology necessary for us has been presented earlier
in Refs.[\alvarez,\nelson ], where closely related issues were studied.

\section{The Schiffer Variation}

Our basic data are Riemann surfaces with punctures and with
a canonical set of local coordinates around the punctures.
As usual the surfaces are considered as points of $\ov\M_{g,n}$,
and the space of surfaces with local coordinates at the punctures
is denoted as $\P_{g,n}$. This may be called the space of decorated
surfaces. Thus a surface with local coordinates, or a decorated surface,
represents a point in $\P_{g,n}$. The space $\P_{g,n}$ is infinite
dimensional, since the space of local coordinates at a point is
infinite dimensional.
We are interested in forming a coset space
$\wh \P_{g,n} = \P_{g,n}/\sim$, where $\sim$ indicates that
two identical punctured surfaces, with local coordinate systems that
differ by a constant phase around each puncture should be identified.
The space $\wh\P_{g,n}$ is still infinite dimensional. Recall, from the
discussion in \S5 that local coordinates are defined by ``coordinate
curves'', Jordan closed curves homotopic to the punctures, with
a marked point on them. The curves correspond to the locus $|z| =1$,
and the marked point is $z=1$, with $z$ the local coordinate.
The space $\P_{g,n}$ corresponds to the space of surfaces with
these marked coordinate curves. The space $\wh\P_{g,n}$ is simply the
space of surfaces with
unmarked coordinate curves.

Given a point in $\P_{g,n}$ there
is an obvious projection $\pi '$ to $\wh\P_{g,n}$, namely the point
is mapped to the coset where it belongs, in other words we forget
about the marked point on the coordinate curves. There is another projection
$\pi$ from $\wh\P_{g,n}$ to $\ov\M_{g,n}$ consisting of
forgetting about the local coordinates. Finally there is a
projection from $\ov\M_{g,n}$ to $\ov\M_{g,0}$ which simply means
forgetting about the punctures. In \S5 we actually
gave a map $\sigma$ from $\ov\M_{g,n}$ to
$\wh\P_{g,n}$ since we found a way to define canonical local coordinates
(up to phases) on any punctured Riemann surface using minimal area metrics.
The local coordinates are defined in terms of the maximal domain foliated
by geodesics homotopic to the punctures.
We summarize the above information in the following diagram

$$
\matrix{\P_{g,n}\cr
\mapdown{\pi '}\cr
\wh\P_{g,n} \cr
\sigma\,\mapup{} \,\,\mapdown{\pi} \,\, \cr
\ov\M_{g,n} \cr
\mapdown{}\cr
\ov\M_{g,0}\cr} \eqn\projectmap$$

The map $\sigma$, defines a canonical section in $\wh\P_{g,n}$.
\foot{Technically speaking $\P_{g,n}$ is a principal fiber bundle with
base space $\ov\M_{g,n}$, projection $(\pi \circ \pi ' )$, and with
fiber $F$ the possible local coordinates, which is just the topological
group $G$ of maps of the form $z \rightarrow  \sum_{n=1} a_n z^n$
with ($a_1 \not= 0$).} Clearly $(\pi \circ \sigma)$ is the identity
map on the space $\ov\M_{g,n}$.

We must understand the various tangent spaces appearing here.
Consider a point $P \in \P_{g,n}$ corresponding to a decorated surface.
Let $\wh P = \pi ' (P) \in \wh\P_{g,n}$ denote the corresponding equivalence
class, $\wt P = \pi (\wh P) \in \ov\M_{g,n}$ denote the underlying punctured
surface and $\vec P = f (\wt P ) \in \ov\M_{g,0}$ denote the underlying
surface.
Let us introduce the following
notation:
$$\eqalign{
V_1, V_2, \cdots  &\in T_{P}(\P_{g,n}), \cr
\wh V_1 , \wh V_2 , \cdots &\in T_{\wh P}(\wh\P_{g,n}), \cr
\wt V_1 , \wt V_2 , \cdots &\in T_{\wt P}(\ov\M_{g,n}), \cr
\vec V_1 , \vec V_2 , \cdots &\in T_{\vec P} (\ov\M_{g,0}). \cr}
\eqn\deftangents$$

The tangent vectors in $T_{\vec P}(\ov\M_{g,0})$ correspond to changes
in moduli of the unpunctured surface $\vec P$. The tangent vectors
in $T_{\wt P}(\ov\M_{g,n})$ include the deformations that change the
positions of the punctures and the underlying moduli of the
punctured surface $\wt P$. The tangent vectors
in $T_{\wh P}(\wh\P_{g,n})$ include the possibility of changing the unmarked
coordinate curves (changing the local coordinates by more than a phase)
in addition to changing the positions of the punctures and the
underlying moduli of $\wh P$. Finally the tangent vectors in
$T_{P}(\P_{g,n})$ correspond to all possible deformations of the
marked coordinate curves, changes in the positions of the punctures
and changes in the underlying moduli of $P$.

The projections between the various spaces in \projectmap\ induce
obvious projections between the respective tangent spaces. The
projections can map many vectors to the same one, due to the
existence of ``vertical'' vectors, which are those that project to
zero. They are the following; in $T(\P_{g,n})$, the vectors that only
change the marked points on the coordinate curves; in $T(\wh\P_{g,n})$
the vectors that only change the unmarked coordinate curves;
and in $T(\ov\M_{g,n})$ the
vectors that only change the positions of the punctures. Such vectors are
the ones that change precisely the data that is forgotten by the
projection maps.

Our concrete objective is to
understand how given a decorated surface $P \in \P_{g,n}$ we
can implement concretely on the surface the variations that
correspond to the tangent space $T_{\P}(\P_{g,n})$. The appropriate
tool is the Schiffer variation [\schiffer ,\gardiner ,\nag ], which
uses vector fields on the surface to generate the changes in data.
The main ideas have been summarized in the physics literature in
[\alvarez ,\nelson ].
Let $P$ correspond to the punctured surface $\R$, with punctures
$p_1 , p_2 , \cdots p_n$ and coordinate curves $\g_1 , \g_2 \cdots
\g_n$. We are now given an $n$-tuple of vector fields
$${\bf v} = (v^{(1)} , v^{(2)} , \cdots v^{(n)} ),\eqn\ntuple$$
where the vector $v^{(i)}$ is a holomorphic vector defined
(a priori only) in an annular neighborhood of the
coordinate curve $\g_i$ surrounding the puncture $p_i$ and defining
the disk $D_i$ (some of the vectors $v^{(i)}$ may be zero). One defines,
for each of the punctures the maps
$$z_i \rightarrow z_i' = z_i + \e v^{(i)}(z_i)\eqn\modvariat$$
which for sufficiently small $\e$ map the coordinate curves $\g_i$
into simple closed Jordan curves $\g_i^\e$.
The new surface $\R'$ is constructed by gluing
the interior of $\g_i^\e$ (the domains which include the respective
punctures) to the exterior of $\g_i$, for each
puncture:
$$\R'=\big\{\sum_i\hbox{Int}(\g_i^\e )\big\}\cup\big\{\R-\sum_iD_i\big\}.
\eqn\newsurface$$
This gluing is done by identifying points as follows. Let $p$ be a point
on the boundary of $\R - \sum D_i$, and $p'$ be a point on the boundary
of the new disk defined by $\g_i^\e$. We identify the two points
$$p \in \g_i \, \longleftrightarrow \, p'\in \g_i^\e \quad
\hbox{when} \quad z_i' (p) = z_i (p').\eqn\schiffident$$
More explicitly, we remove the disk defined by the
coordinate curve, namely the $|z_i| = 1$ disk $D_i$ and map its boundary
into a curve in the $z_i'$ plane defining a new disk $D_i'$. It is crucial
to note that while we are only mapping the boundary, this defines a
new disk. Any two punctured disks are conformally equivalent, but
here we have two disks whose boundaries are mapped into each other
in a prescribed analytic way.  Under this condition it is not clear
that this map extends to the complete disks. We then glue the
disks $D_i'$ back into $\R -\sum D_i$. This gives us the new surface.
Intuitively it is very clear what
the variation is doing, it leaves $\R-\sum D_i$ unchanged but is
sticking new disks back (this is the reason for the term
``Schiffer's interior variation'').

We now need to understand what happens to
the punctures and the local coordinates. The new local coordinates
are simply the $z_i'$ coordinates. This is a well defined coordinate
in the new disk $D_i'$.
The new puncture $p_i'$ is located at the origin
of the new coordinate system, namely at $z_i' = 0$. Since the
vectors $v^{(i)}$ do not necessarily extend holomorphically inside
the disk, there is no other sensible alternative for the new coordinates
or new puncture. It should be emphasized
that the new canonical coordinates are simply the $z_i'$ coordinates
and therefore the boundary of the disk $D_i'$ does not coincide
with the new coordinate curves. Those are now the $|z_i'|=1$ curves.
An alternative way to get the new coordinate curves directly would be to
do the Schiffer variation with curves $\wh\g_i$ chosen such that under
the variation $|z_i'(\wh\g_i )| = 1$. Then the new coordinate curves
are simply the $\wh\g_i$ curves, which correspond to the boundaries of
the amputated surface $\R- \sum D_i$.

The curve we use to do the Schiffer variation can be changed.
We can begin with another curve homotopic to the original curve
and the resulting surface is identical as long as one curve
can be deformed into the other one without encountering singularities
of the vector field. Suppose we have a curve $\g$ defining a punctured
disk $D$ with puncture $p$ and a vector
field $v$ that has a singularity inside $D$ at some point $q\not=p$
and possibly a singularity at $p$.  Then by contour deformation
the original variation using $\g$ is equivalent to doing two variations,
one with a contour enclosing $q$ and not $p$, and another with a contour
enclosing $p$ but not $q$. This means that the most general situation
we need to consider is really that of a Schiffer variation with a vector
field that inside the countour may become singular only at the puncture.

\section{Using the Schiffer Variation to Generate a Tangent}

Let us begin by discussing the conditions under which data do not
change. Suppose all the mappings  $z_i \rightarrow z_i'$ extend
holomorphically
to the interior of the disks, namely all $v^{(i)}$ are holomorphic in
$D_i$. Then the disks $D_i$ and $D_i'$ are completely equivalent
and when reinserted back, if we forget the
punctures, the surface has not changed. Thus such vectors do not
change the moduli of the underlying unpunctured surface.
Suppose, in addition that a vector $v^{(k)}$ vanishes
at $z_k =0$, then the position of the puncture $p_k$ is not changed.
This is clear because in this case the disks $D_k$ and $D_k'$ are
completely equivalent as punctured disks since $z_k =0$ is mapped
to $z_k' =0$ and the puncture has not moved.

Suppose a vector $v^{(i)}$ extends holomorphically
into the disk $D_i$ and does not vanish at the origin $z_i=0$.
Then the conformal map relating the $D_i$ and $D_i'$ disks maps the
original puncture into $z_i' = \e v^{(i)} (z_i =0) \not= 0$.
Since the new puncture is now at $z_i'=0$ this indicates the
possibility that the puncture {\it may} have moved. We will
see presently how to decide if it moved or not.

\FIG\trivfig{This figure is used for the proof that a Schiffer vector
that extends holomorphically away from the puncture does not change
any data. In (a) we show the surface $\R$ broken up into a disk $D$ and
the rest. In (b) we show the surface $\R'$ obtained by variation.
In (c) we show once more the surface $\R$ broken up in a different way.}

If all the vectors $v^{(i)}$ comprising ${\bf v}$ arise from a
single globally defined holomorphic vector field on the whole of
$\R - \sum p_i$ (singularities are permitted at the punctures)
then the variation, performed around all the punctures changes
nothing at all! It does not change the
underlying moduli, nor the position of the punctures, nor the
coordinate curves, nor their marked point. Let us derive this.
Consider Fig.\trivfig\ where we show a surface with one puncture,
for simplicity. The puncture is surrounded by the curve $\g$
and we will do a variation using a vector field $v$ that extends
holomorphically to the exterior of $\g$, but not necessarily to
the interior of $\g$, where it may have singularities. We will
show that this variation does not change any data.
In part (a) we show the original surface decomposed into
the disk $D$, and  $\R-D$, to the right (both shaded).
In part (b) we have the new surface $\R'$ obtained by the variation,
decomposed into the disk $D'$ and $\R-D$. The respective coordinates
are denoted as
$z'$ and $\wt z{}'$ and the gluing is done by identifying points
$s' \in \partial D'$ with $\wt s{}'\in \partial (\R-D)$ via
$z'(s')= \wt z{}' (\wt s{}') + \e v(\wt z{}' (\wt s{}'))$.  Let us now show
how to construct a conformal map between the punctured surfaces $\R$
and $\R'$. First we rearrange the construction of $\R$ given in
(a) into what is shown in (c); we have changed the shape of the
initial disk $D$ into that of $D'$ by adding and subtracting regions
and compensating for this on $\R-D$ (this is just the same as cutting
the original surface $\R$ along the image of the original curve $\g$
under the variation.). The coordinate systems for the disk and
the remainder are called $z$ and $\wt z$, and the gluing of points
$s$ and $\wt s$ in their boundaries is via $z(s)= \wt z(\wt s)$.
Now the mapping between $\R'$ and $\R$ is very simple, we use
Fig.\trivfig\ (b) and (c). For the points on the disks we simply take
the identity map
$$ q' \rightarrow q \quad \hbox{when}\quad z(q) = z'(q') \eqn\mapsidentic$$
Note that this map maps the new puncture into the old puncture and
furthermore the coordinates of points mapped into each other are
identical! If we show this map can be extended to the complete
surfaces we will have shown that the puncture has not moved and that
the coordinates around it have not changed.
For the remainder of the surfaces we take
$$ \wt q' \rightarrow \wt q \quad \hbox{when}\quad
\wt z(\wt q) =\wt z'(\wt q')+ \e v(\wt z'(\wt q')).\eqn\mapsid$$
This map exists because the vector field was assumed to extend to the
remainder of the surface. It is simple to check that the complete map
{}from $\R'$ to $\R$ defined by the last two equations respects the
identifications along the cutting lines, and is therefore a well
defined map of the surfaces. This concludes the argument.
It is also clear how to extend the argument for the case when
there are several punctures and the vectors around the punctures
arise from a single holomorphic vector field on $\R -\sum p_i$.

Actually, the complete result is that an $n$-tuple of vectors
defined around the punctures does not change any data {\it if and
only if} it arises from a globally defined holomorphic vector field on
the surface minus the punctures. We showed above that such vectors
do not change the data, now we argue that any vector which is not
of this type must actually change the data.

Consider a vector field defined around a puncture, assume it extends
holomorphically up to the puncture, where it vanishes, but does not
extend holomorphically to the rest of the surface. This vector does
not change the underlying moduli, nor the position of the puncture,
but it does change the local coordinate around the puncture. The map
between the original surface and the new surface is simply the map
between the two disks, induced by the vector field, plus the identity
map on the leftover of the surfaces. It is clear that that the coordinates
have changed, unless there are
continuous conformal self-mappings of the whole
surface into itself. Such mappings only exist for $g=0, n=1,2,3$ and
$g=1, n=0$, and are generated by globally defined vector fields
(conformal Killing vectors). However, since these maps would restore back
the coordinates they should agree with the vector that generated the variation
around the puncture, in contradiction with the fact that
the variation vector does not extend. A similar argument would show
that a vector that extends holomorphically into the puncture without
vanishing there, and does not extend to the rest of the
surface must change the position of the puncture, unless the surface is
one of the above exceptional cases, in which case at least the local
coordinates are changed.

Consider changing the underlying moduli of the surface.
It is possible to show that the $3g-3$ complex moduli of the
underlying surface
can be changed by variations around a single puncture [\schiffer ].
That is, there are vector fields around a puncture which do not extend
holomorphically to the rest of the surface nor to the puncture,
that can be used to represent the $3g-3$ (complex) dimensional vector space
$T(\ov\M_{g,0})$. The Weierstrass gap theorem shows that if we specify the
leading singularity
$z^{-n}$, with $n>0$, of a vector field around a point $p$ (with $z$ a local
coordinate) there are $3g-3$ values for $n$ such that the vector cannot
be extended holomorphically to the rest of the surface.
\foot{For genus one, a vector field whose leading singularity
is $z^{-1}$ cannot be extended. If it could, then
working on the presentation of the torus as a lattice in the $z$-plane
the vector field would be of the form $v(z) (dz)^{-1}$ where actually
$v(z)$ must be a {\it function} on the torus. But it is well-known that
there are no functions on a torus having a single first order pole
(the famous $\P$ function of Weierstrass has a double pole).
For genus zero, all vectors with leading singularity $z^{-n}$
with $n>0$ can be extended.}
These are the vectors that will produce the underlying
change of moduli of the surface. Every vector $v = z^{-n}$, with
$n$ one of the special values, must generate a nontrivial tangent
in $T(\ov\M_{g,0})$. The tangents associated with all the
special values of $n$ must give a basis for $T(\ov\M_{g,0})$. If this
were not the case, we would not be able to generate all possible
tangents by Schiffer variations. This shows that any vector that
does not extend (in and out) must generate a deformation of the
underlying moduli. We summarize all the above information in
the following table. The first two columns tell us whether the
Schiffer vectors extend holomorphically into the punctures (extends in)
and/or to the rest of the surface (extends out); $v(p)$ denotes the
value of the Schiffer vector at the puncture. To the right we
show the type of data they change.

\noindent
$\underline{\hbox{Table}}.\,$ Schiffer variations.

$$\hbox{\vbox{\offinterlineskip
\def\strut{\hbox{\vrule height 15pt depth 10pt width 0pt}}
\hrule
\halign{
\strut\vrule#\tabskip 0.1in&
\hfil$#$\hfil &
\vrule#&
\hfil$#$\hfil &
\vrule#&
\hfil$#$\hfil &
\vrule#&
\hfil$#$\hfil &
\vrule#&
\hfil$#$\hfil &
\vrule#\tabskip 0.0in\cr
& \hbox{Extends in} && \hbox{Extends out}
&& \matrix{\hbox{Changes}\cr\hbox{moduli}\cr} && \matrix{\hbox{Moves}\cr
\hbox{punctures}\cr} && \matrix{\hbox{Changes}\cr\hbox{coordinates}\cr}
& \cr\noalign{\hrule}
& \hbox{Yes/No} && \hbox{Yes} && \hbox{No} && \hbox{No} &&\hbox{No}
& \cr\noalign{\hrule}
& \hbox{Yes}, v(p) = 0 && \hbox{No} && \hbox{No} && \hbox{No}&&\hbox{Yes}&
\cr\noalign{\hrule}
& \hbox{Yes}, v(p)\not= 0 && \hbox{No} && \hbox{No} && \hbox{Yes}^*
&& -- & \cr\noalign{\hrule}
& \hbox{No} && \hbox{No} && \hbox{Yes}^{**} && -- && -- &
\cr\noalign{\hrule}}}}$$

\noindent
The asterisk on the third row indicates that the positions of the punctures
change unless $n=1,2,3$ with $g=0$, or $n=g=1$, in which case the local
coordinates must change. The double asterisk indicates that the underlying
moduli change unless the surface is of genus
zero and there are no such moduli. If this is the case the vector
must change the position of the punctures, unless $n=1,2,3$, in which case
the local coordinates must change.

We can explain now how to find the variations and the associated
vectors on the surface corresponding to a given tangent vector
in $\P_{g,n}$. We use a given puncture to generate the underlying
moduli deformations. We then move the positions of the punctures
by a Schiffer variation around each puncture with a vector field
that extends holomorphically inside the disk but does not extend
outside. If that vector does not vanish at the puncture it will change the
position of the puncture. We finally change the coordinate
system around the punctures. This is also done with a variation around
each puncture using a vector that extends holomorphically
inside the disk, vanishes at the puncture and does not extend outside.
Two successive infinitesimal Schiffer variations with vectors
$\e v$ and $\e v'$, are equivalent,
to order $\e$, to a single variation with vector $\e (v+v')$. Since
we are interested in reproducing a specified tangent in $\P_{g,n}$, all
variations are infinitesimal, and therefore the
$n$ variations moving the punctures and the $n$ variations modifying
the coordinate systems can be put together into $n$ variations, each
around a different puncture. Once we have learned how to do the
variations there is no sense of order in which the underlying moduli
variations and the puncture variations must be done.

To summarize, consider a specified tangent vector in $\P_{g,n}$.
It can be generated by the sum of
$(3g-3)$ variations of the type
$$  (v^{(1)}_k, 0,0, \cdots, 0),\quad k=1,2, \cdots ,3g-3
\eqn\modvarx$$
all based on the same puncture (and each changing a complex modulus)
, and $n$ variations of the type
$$  (0,0,\cdots , v^{(k)}, \cdots , 0), \quad k=1,2,\cdots ,n
\eqn\punctvarx$$
each around a different puncture (changing the position of the
puncture and the local coordinates). Thus {\it a given tangent in}
$\P_{g,n}$ {\it maps into the vector} ${\bf v}$ {\it which is simply the
sum of the} $3g-3+n$ {\it vectors listed above}.
Any other possible vector ${\bf v}'$ representing the same
tangent in $\P_{g,n}$ can only differ from ${\bf v}$ by
a vector ${\bf t}$ arising from a globally defined holomophic
vector on the surface minus the punctures.

\REF\sonoda{H. Sonoda, `Sewing Conformal Field Theories: I,II', Nucl. Phys.
{\bf B311} (1988) 401,417.}

\section{Construction and Properties of Forms on $\wh\P_{g,n}$}

Our aim in this section is to construct a set of differential forms
on the space $\wh\P_{g,n}$. The degree of these forms will go from
zero to infinity, since the space $\wh\P_{g,n}$ is infinite dimensional.
The most relevant forms, however, are those of degree less or equal to
$6g-6+2n$, the dimensionality of the moduli space of surfaces of the
corresponding genus and number of punctures. These forms can be
naturally integrated over the minimal area section, or subspaces of this
section.  The form of degree $6g-6+2n$ is the basic ingredient in the
definition of string field vertices and string field products,
as we will see in the next subsection. The forms will be labeled by
$n$ states. These will be denoted as
$$\ket{B_1}\otimes\ket{B_2}\otimes \cdots \otimes\ket{B_n}
\equiv \ket{\vec B} \in {\H}^{\otimes n}. \eqn\deflabel$$
Here $\ket{B_k}$ denotes the state to be inserted on the
$k$-th puncture.
\medskip
\noindent
$\underline{\hbox{Basic Objects in the Operator Formalism}}.\,$
We must recall some basic properties and constructions of the
operator formalism which are necessary for our purposes. Consider
our Schiffer variations, in a surface with one puncture, for
simplicity. We write
$$(1 + \e \delta_v) z \equiv z + \e v(z), \eqn\thealgvar$$
where this is the way the variation acts on a local coordinate.
It follows by a simple computation that these variations satisfy
the algebra
$$\l \delta_{v_1} , \delta_{v_2} \r = \delta_{[ v_2,v_1 ]} ,\eqn\sixnine$$
as verified by acting on a local coordinate. For the case of several
punctures, the vectors become the boldface vectors ${\bf v}_i$, and
\thealgvar\ is naturally generalized by acting on an $n$-tuple of
local coordinates. We then have that
$$\l \delta_{{\bf v}_1} , \delta_{{\bf v}_2} \r
= \delta_{[ {{\bf v}_2},{{\bf v}_1} ]} .\eqn\sixnine$$
Associated to a point $P$ in $\P_{g,n}$, which represents a surface $\Sigma$
of genus $g$ with $n$ punctures (here $n\geq 1$) with local coordinates,
the operator formalism gives us a state  $\ket{\Sigma}\in \H^{\otimes n}$
(defined up to normalization). The corresponding conjugate state is obtained
by use of the reflector
$${}_{1\cdots n}\bra{\Sigma} = \bra{R_{11'}}\cdots \bra{R_{nn'}}\Sigma
\rangle_{1'\cdots n'}.\eqn\refsurf$$
We must understand how the state $\ket{\Sigma}$ varies as we change
the data of the surface, in particular, if we make a Schiffer variation
with a vector ${\bf v}= (v^{(1)} ,\cdots v^{(n)})$. To this end we must
use the stress tensor $(T(z),\ov T (\ov z ))$ of the theory.  Let us define
$${\bf T}({\bf v})
= \sum_{i=1}^n \bigg(
\oint T^{(i)}(z_i)v^{(i)}(z_i) {dz_i\over 2\pi i} + \oint \ov T^{(i)}(\ov z_i )
\ov v^{(i)}(\ov z_i ) {d\ov z_i\over 2\pi i} \bigg) , \eqn\deftbold$$
where the operator valued fields $T^{(i)}(z_i)$
and $\ov T^{(i)}(\ov z_i)$ refer to the $i$-th Hilbert space (we use
$\oint dz/2\pi i z = \oint d\ov z /2\pi i \ov z = 1$).
The vectors $\ov v^{(i)}$ are simply the complex conjugates
of $v^{(i)}$ and the integrals are taken using contours surrounding the
punctures and lying in the domain of definition
of the local coordinates and the $n$-tuple of vectors.
We then have that
$$\l {\bf T}({\bf v}_1),{\bf T}({\bf v}_2)\r = -{\bf T}
\bigl( \l {\bf v}_1 , {\bf v}_2 \r \bigr), \eqn\emalg$$
which follows from the operator product expansion
$$T(z) T(w) \sim  {2\over (z-w)^2} T(w) + {1\over (z-w)} \partial T(w)
+ \cdots ,\eqn\opese$$
the analogous relation for the antiholomorphic modes, and the fact
that the various Hilbert spaces in \deftbold\ are independent.
It is convenient at this point to introduce the antighost analog of
\deftbold . We define ${\bf b}({\bf v})$ as
$${\bf b}({\bf v}) = \sum_{i=1}^n \bigg(
\oint b^{(i)}(z_i)\, v^{(i)}(z_i) \, {dz_i\over 2\pi i}
+ \oint \ov b^{(i)}(\ov z_i )
\ov v^{(i)} (\ov z_i ) {d\ov z_i \over 2\pi i} \bigg) . \eqn\defbbold$$
Making use of the operator product expansion
$$T(z) b(w) \sim  {2\over (z-w)^2} b(w) + {1\over (z-w)} \partial b(w)
+ \cdots ,\eqn\opeantigh$$
we find that
$$\l {\bf T}({\bf v}),{\bf b}({\bf v}_k)\r = {\bf b}
\bigl( \l {\bf v}_k , {\bf v} \r \bigr). \eqn\ebhmalg$$
Similarly, making use of Eqn.\frbb\ we readily obtain
$$\{ \sum_{i=1}^n Q^{(i)}, {\bf b}({\bf v}) \}
= {\bf T}({\bf v}).\eqn\qwithb$$

Returning now to the stress tensor, consider its BPZ conjugation property.
Since the stress tensor behaves on a Riemann surface like a
quadratic differential (a dimension two primary field) its behavior
on the reflector must be identical as that of the antighost field
(\qonref ), we therefore have
$$\bra{R_{12}} (L_n^{(1)} - L_{-n}^{(2)}) = 0.\eqn\vironref$$
Using the mode expansion for the stress tensor and that for an arbitrary vector
$$T(z) = \sum_n {L_n\over z^{n+2}} ;\quad v(z)= \sum_n {v_n\over z^{n-1}},
\eqn\modextv$$
we find that
$$T(v) = \int T(z) v(z) {dz\over 2\pi i} = \sum_n L_nv_{-n}.\eqn\tvprop$$
Using \vironref\ and \tvprop\ we find that
$$\bra{R_{12}}\l T^{(1)}(v) + T^{(2)} (v^T) \r = 0, \eqn\connectse$$
where the vector $v^T(z)$ must be given by
$$v^T(z) = -\sum_n {v_{-n}\over z^{n-1}} = -z^2 v \bigl({1\over z}\bigr).
\eqn\vectcontj$$
It is a straightforward calculation, using the above definition of
$v^T$ in terms of $v$ to find that
$$\l v_1^T(z) , v_2^T(z) \r = \l v_1(z),v_2(z) \r^T.\eqn\commutevs$$
We can now address the issue of deforming the state $\ket{\Sigma}$.
Following [\nelson ,\alvarez ] one must have
$$\delta_{\bf v} \ket{\Sigma}={\bf T}({\bf v}^T)\ket{\Sigma}.\eqn\defrmstate$$
One can make a simple consistency check
$$\eqalign{
\l \delta_{{\bf v}_1} , \delta_{{\bf v}_2} \r \ket{\Sigma}
&= \l {\bf T}({\bf v}_1^T) , {\bf T}({\bf v}_2^T) \r \ket{\Sigma}\cr
&= {\bf T} \bigl( \l {\bf v}_2^T , {\bf v}_1^T \r \bigr) \ket{\Sigma}\cr
&= {\bf T} \bigl( \l {\bf v}_2 , {\bf v}_1 \r^T \bigr) \ket{\Sigma}\cr
&= \delta_{ [{\bf v}_2 , {\bf v}_1 ] } \ket{\Sigma},\cr}\eqn\seconsist$$
in agreement with the algebra of the $\delta_{\bf v}$ operators in
\sixnine . In fact, the other candidate:
$\delta_v\ket{\Sigma} = T(v)\ket{\Sigma}$ would have also passed this
consistency check. We can see that \defrmstate\ must be the correct form
by a consistency argument. Consider a surface with one puncture,
with local coordinate $z$ giving a state $\ket{\Sigma}$. The one point
function of a physical state $\ket{\Psi}$ is essentially given by
$\bra{\Psi} \cdots \ket{\Sigma}$.  Suppose we change $z$ by the
addition of a vector field $v(z) = \e z^{n+1}$ ($n\geq 1$). In this case
$T(v^T) = -\e L_{-n}$. If we change the coordinates in this way, the
coupling to a physical state $\bra{\Psi}$ must not change. This
implies that $\bra{\Psi} \cdots L_{-n}\ket{\Sigma}$ must be zero,
and this indeed can happen because for physical states we have
$L_{n\geq 1} \ket{\Psi} = 0$.

It follows from \refsurf ,\defrmstate , and \connectse\ that
$$\eqalign{
\delta_{{\bf v}} \, {}_{1\cdots n}\bra{\Sigma}
&\equiv \bra{R_{11'}}\cdots \bra{R_{nn'}} \delta_{{\bf v}}
\ket{\Sigma}_{1'\cdots n'}\cr
{}&= \bra{R_{11'}}\cdots \bra{R_{nn'}} {\bf T}'({\bf v}^T)
\ket{\Sigma}_{1'\cdots n'}\cr
{}&= -\bra{R_{11'}}\cdots \bra{R_{nn'}} {\bf T}({\bf v})
\ket{\Sigma}_{1'\cdots n'}\cr
{}&= -{}_{1\cdots n}\bra{\Sigma}{\bf T}({\bf v}) ,\cr}\eqn\wawot$$
and therefore we have that
$$\delta_{{\bf v}}\bra{\Sigma} = - \bra{\Sigma} {\bf T}({\bf v}).
\eqn\varbradef$$
This equation will be useful to us shortly.
Let us conclude this brief discussion of the operator formalism
with two useful properties. For ${\bf t}$ a Borel vector (a vector
whose data along the punctures extends holomorphically to the rest of
the surface) we have that
$$\bra{\Sigma}{\bf T}({\bf t})= 0, \quad
\bra{\Sigma}{\bf b}({\bf t})= 0,\eqn\conservedch$$
as explained in [\alvarez ]. Moreover the state representing
the surface is annihilated by the BRST operator
$$\bra{\Sigma} (Q_1 + \cdots Q_n ) = 0.\eqn\brstannih$$
{}From the operator formalism we know that the state $\ket{\Sigma_P}$ is a
state with (first quantized) ghost number $6g -6 + 6n$
$$ \Sigma \in \P_{g,n} \quad \rightarrow\quad
G\ket{\Sigma} = (6g-6+6n)\ket{\Sigma}, \eqn\ghostof$$
and therefore, in our conventions, the bra $\bra{\Sigma}$ has the
same ghost number. Since this is an even ghost number, the bra
$\bra{\Sigma}$ {\it is Grassmann even}.

It is important for us to have the normalization
of the state $\bra{\Sigma}$ unambiguously determined. We can fix
by hand the normalization of the surface corresponding to the three
punctured sphere (the three-string vertex) by defining the string
coupling constant $\k$.  The normalization of the sewing
states was fixed in \S2. Then using sewing and the three punctured
sphere we can define the normalization of states for higher number
of punctures and higher genus since it is clear that we can
produce some surfaces for each $\M_{g,n}$ (not all because the
vertex is fixed), and then using the transport equation \defrmstate\
we can extend the definition of normalization to the remaining
surfaces in the moduli space (see [\alvarez ] and [\sonoda ] for
further analysis of the consistency of sewing).
\medskip
\noindent
$\underline{\hbox{Defining the Differential Forms}\,\,\Omega_{\vec B}}.\,$
We will define now a differential form ${\Omega_{}^{}}_{\vec B}^{(0)g,n}$
labelled by $n$ states $B_i$, of real degree $(6g-6+2n)$ on the tangent
space $T_{\wh P}(\wh \P_{g,n})$
based at the point $\wh P$ (recall \projectmap ) by
the following expression
$${\Omega_{}^{}}_{\vec B}^{(0)g,n} (\wh V_1,\cdots \wh V_{6g-6+2n}) \equiv
N_{g,n}\bra{\Sigma_P}\, {\bf b}({\bf v}_1)
\cdots {\bf b}({\bf v}_{6g-6+2n}) \, \ket{\vec B},\eqn\dfinally$$
where the necessity for the normalization constant $N_{g,n}$
$$N_{g,n} = \bigl( 2 \pi i \bigr)^{-d_{g,n}} =
\bigl( 2 \pi i \bigr)^{-(3g-3+n)}, \eqn\normcon$$
first pointed out by Kugo and Suehiro [\kugosuehiro ] in the context
of the classical action, will become apparent in discussing sewing and
the main identity.\foot{The apparent disagreement with the
constant given in [\kugosuehiro ] is due to our different
convention for $b_0^-$ and different definition for the string field.}
Moreover, if this constant were not present, the string field vertices
would not be real (see \S7.4). Similarly we can define other forms of
lower degree, all of them labeled by $n$ states:
$${\Omega_{}^{}}_{\vec B \,}^{(-r)g,n} (\wh V_1,\cdots \wh V_{6g-6+2n-r}) =
N_{g,n}\bra{\Sigma_P} {\bf b}({\bf v}_1) \cdots
{\bf b}({\bf v}_{6g-6+2n-r}) \, \ket{\vec B}.\eqn\lowerform$$
This is a form of degree $(6g-6+2n-r)$, it can be therefore
used to integrate over submanifolds of codimension $r$ in the
section $\sigma (\D_{g,n})$ (thus the label $(-r)$).

Given that the state $\bra{\Sigma_P}$ has ghost number $6g -6 + 6n$,
the $6g-6+2n$ antighost insertions in ${\Omega_{}^{}}_{\vec B}^{(0)g,n}$
reduce the ghost number to $4n$. In order to get a nonzero answer we need a
total ghost number of $6n$, since in between two vacua
$\bra{{\bf 1}} \cdots \ket{{\bf 1}}$
for each Hilbert space there must be a total ghost number of six. Thus
for the form ${\Omega_{}^{}}_{\vec B}^{(0)g,n}$ to be nonvanishing
the sum of the ghost numbers of the $B$'s must be $2n$, that is
$\sum_i (G(B_i)-2) = 0$. The form $\Omega_{\vec B}^{(-r)}$
can be nonvanishing only if $\sum_{i=1}^n G_i = 2n-r$.

The bra $\bra{\Sigma_P}$, lives in
${\cal H}^{\otimes n}$, where each Hilbert space corresponds to each of
the labeled punctures. We can label the Hilbert spaces more
explicitly as $\bra{\Sigma_P (1,2,\cdots n)}$.
We may also add to the states the label of the Hilbert space where
they are inserted, say $\ket{B}_{(k)}$, indicates that the
state $B$ will be associated to the $k$-th puncture. Operationally,
the tensor product in \deflabel\ means that in calculating an overlap
we use
$$ \ket{\vec B} = \ket{B_1}_{(1)} \ket{B_2}_{(2)} \cdots \ket{B_n}_{(n)},
\eqn\wwmean$$
and therefore equation \dfinally , for example, is evaluated as
$${\Omega_{}^{}}_{\vec B}^{(0)g,n} (\wh V_1,\cdots \wh V_{6g-6+2n}) =
N_{g,n}\bra{\Sigma_P (1,\cdots n)}\, {\bf b}({\bf v}_1)
\cdots {\bf b}({\bf v}_{6g-6+2n}) \, \ket{B_1}_{(1)}
\cdots \ket{B_n}_{(n)}.\eqn\dfinallyi$$

Let us explain the various ingredients entering
into the definition of the $\Omega_{\vec B}$ forms,
noting as we go along the conditions that must hold so that
it they are well defined forms on $T_{\wh P}(\wh \P_{g,n})$.

The forms $\Omega$ are labelled by the $n$ states $\ket{B_i}$
that are introduced at the punctures.
Given a set of $(6g-6+2n)$ real vectors
$\wh V_k \in T_{\wh P}(\wh \P_{g,n})$, the form $\Omega$
must give us a number.
One must first choose a point $P \in \P_{g,n}$ such that it
projects down to $\wh P$ under the map $\pi '$ in \projectmap .
(Must check that: (i) the choice of $P$ does not affect the result.)
Given a vector $\wh V_k$ one must now find a vector
$V_k \in T_{P}(\P_{g,n})$ that
projects via $\pi'$ into $\wh V_k$. (Must show that: (ii) that the
choice of $V_k$ does not affect the result.) Then we must choose
an $n$-tuple ${\bf v}_k = (v^{(1)}_k , \cdots , v^{(n)}_k)$
that represents the tangent $V_k$ as a Schiffer variation. (Must
check that: (iii) the choice of ${\bf v}_k$ does not affect the
result.)

Let us show that (i), (ii) and (iii) hold. Let us begin with
(iii); that is showing that the ambiguity in the choice of
the $n$-tuple ${\bf v}_k$ that represents the tangent $V_k$ does
not matter. Indeed the only ambiguity, as we discussed in \S7.2,
is the addition of an $n$-tuple `${\bf t}$'
arising from a globally defined vector field holomorphic except
at the punctures. This ambiguity requires that the replacement of
${\bf b}({\bf v}_k)$ by
$({\bf b}({\bf v}_k) + {\bf b}({\bf t}))$ should not alter the
value of the form $\Omega$.  Indeed, since
$\bra{\Sigma_P}{\bf b}({\bf t}) = 0$ (\conservedch ), the extra term can be
brought all the way to the bra $\bra{\Sigma}$ and gives no contribution.

Let us now turn to (ii). Given a vector $\wh V_k$ the vector $V_k$ is
determined up to a vertical vector $V$, which as explained in the
previous subsection, can only change the marked point in the coordinate curve.
This means that the vector only changes coordinate systems up to a phase.
Since we can change the phases of the local coordinates around each puncture
independently, the general situation corresponds to changing the phase around
one of the punctures. The associated vector
${\bf v} = (0,  \cdots , v^{(i)}, \cdots , 0 )$ for the Schiffer variation
must implement the variation
$$z_i' =  \exp (i\e ) z_i \sim \,  1 + i\e z_i, \quad \rightarrow
\quad v^{(i)} = iz_i . \eqn\firstcheck$$

Thus changing $V_k$ to $(V_k+\e V)$ changes
${\bf v}_k$ into $({\bf v}_k + \e {\bf v})$ and as a consequence will change
the relevant antighost insertion in $\Omega_{\vec B}$ as follows
$${\bf b}({\bf v}_k)\,  \rightarrow
{\bf b}({\bf v}_k) + \e \,{\bf b}({\bf v}) .\eqn\pairch$$
Equations \defbbold\ and \firstcheck\ give us
${\bf b}({\bf v}) = i(b_0^{(i)}- \ov b_0^{(i)})$,
and therefore the extra term in \pairch\ will not affect the
evaluation of $\Omega_{\vec B}$ if
$$  (b_0^{(i)} - \ov b_0^{(i)} )\ket{B_i} = b_0^{-(i)}\ket{B_i}
= 0 ,\eqn\weknewit$$
which must hold for all $i$. This is the standard condition that
states should be annihilated by $b_0^{-}$, which we must require
(this completes the verification of (ii)).

Let us finally verify that (i) gives no problems. As we change
$P$ into $P'$ (both projecting to $\wh P$) there are
two types of changes involved in \dfinally\ : the bra $\bra{\Sigma_P}$
is changed into $\bra{\Sigma_{P'}}$, and, corresponding to each of the
vectors $\wh V_k$ we must now find new vectors $V_k' \in T_{P'}(\P_{g,n})$
(the old vectors $V_k$ are in  $T_P(\P_{g,n})$) and the new $n$-tuples
${{\bf v}'}_k$ corresponding to the new tangents.
For any two points $P$ and $P'$ related by an infinitesimal
tangent $\e V \in T_P(\P_{g,n} )$ the variation of the
$\bra{\Sigma_P}$ is given by
$$\bra{\Sigma_{P'}} = \bra{\Sigma_{P}} - \e\, \bra{\Sigma_{P}}
{\bf T}({\bf v}), \eqn\vvvx$$
where ${\bf v}$ (\firstcheck ) is the Schiffer $n$-tuple representing the
tangent $V$. In order to understand the second type of change, consider the
original Schiffer variation ${\bf v}_k$. What we need now is a new variation
${{\bf v}'}_k$ that implements the same change in the curves used for the
variation, despite the fact that the coordinates have changed. The answer is
$$ {{\bf v}'}_k = {\bf v}_k + \e\, [ {\bf v}_k , {\bf v} ].\eqn\lievar$$
This is derived beginning with the fact that if point $p$ is mapped
into $p'$ (near a puncture $p_i$)  we want this to happen both with
the initial variation and with the new variation:
$$\eqalign{z_i (p) + \e v^{(i)}_k (z_i (p)) &= z_i (p'), \cr
{z'}_i (p) + \e {v'}^{(i)}_k ({z'}_i (p)) &= {z'}_i (p'), \cr}\eqn\showlie$$
where the new and old coordinates are related by
${z'}_i(p) = z_i (p) + \e v^{(i)} (z_i (p))$. A short calculation
shows that ${v'}^{(i)}_k = v^{(i)}_k + \e [v^{(i)}_k , v^{(i)}]$,
which is the content of \lievar\ around the puncture $p_i$.
As a consequence each antighost insertion changes into
$${\bf b} ({\bf v}_k) \, \rightarrow  \,
{\bf b} ({\bf v}_k) + \e {\bf b} ([ {\bf v}_k , {\bf v} ]).\eqn\newbinsert$$
Equations \vvvx\ and \newbinsert\ tell us how to vary
\dfinally . The strategy is to move the ${\bf T}$ terms all the way
towards the states. In commuting ${\bf T}$ and ${\bf b}$'s we use \ebhmalg .
The extra variations \newbinsert\ and the extra terms arising from
the commutators cancel. At the end we simply have ${\bf T}$ acting
on the states
$ {\bf T}({\bf v})\ket{\vec B}$. But given the explicit form of
${\bf v}$ in \firstcheck , (supported only on the $i$-th puncture)
we have that
$${\bf T}({\bf v}) = i (L_0^{(i)} - \ov L_0^{(i)})
,\eqn\whouw$$
which simply means that the form $\Omega_{\vec B}$ is well defined
if the states in $\ket{\vec B}$ satisfy the usual $L_0 - \ov L_0 = 0$
condition. This concludes our proof that the form $\Omega_{\vec B}$
is well defined on $T_{\wh P}(\wh\P_{g,n})$.

\REF\spivak{M. Spivak, `A comprehensive introduction to differential
geometry', Publish or Perish Press, Berkeley, 1979.}

\noindent
$\underline{\hbox{BRST Action on the Forms}\, \Omega_{\vec B} }.\,$
We have introduced above a set of forms $\Omega_{\vec B}^{(-r)}$
of degree $\hbox{deg} [\Omega^{(-r)}]= \hbox{Dim}\, [\M_{g,n}]-r$, where
$r \leq \hbox{Dim}\,\M_{g,n}$. Since $\wh\P$ is infinite dimensional,
there is no top form, and $r$ can take all possible negative values.
Our minimal area section $\sigma$ in $\wh\P_{g,n}$ is of dimension
equal to that of $\M_{g,n}$, and therefore $r=0$ gives us top forms
for integration over the section. For $r\geq 0$ the forms can be integrated
over some relevant codimension $r$ subspace of the section. The form
$\Omega_{\vec B}^{(-r)}$ is nonvanishing only if $\sum_{i=1}^n G(B_i)=2n-r$.
The forms with $r=-1$ are expected to be relevant to string field
redefinitions.

The result that we want to establish is that the BRST
operator acts in a simple way on the above forms. We will show that

$$\Omega_{\sum Q_i \ket{\vec B}} ^{(*)} =
(-)^*\hbox{d} \,\Omega_{\ket{\vec B}}^{(*-1)}, \eqn\qaction$$

\noindent
where `d' is the exterior derivative in $\wh \P_{g,n}$, $Q$ is
the closed string BRST operator, and `$*$' is just a number. This
relation holds for all relevant values of $g$ and $n$, and we have
therefore ommitted those labels on the forms.

In order to prove this \qaction\ we need to establish some preliminary
results. Let us first recall that the exterior derivative of any
$(k-1)$-form $\Omega$ on some manifold can be defined with the help
of $k$ vector {\it fields} on the manifold \foot{See, for example,
Ref. [\spivak ] Vol.1, Ch.7, Thm. 13.}
$$\eqalign{\hbox{d}\Omega (\wh V_1,\cdots , \wh V_k) &= \sum_{i=1}^k(-)^{i+1}
{\wh V_i}\,\Omega (\wh V_1,\cdots ,\wh V_i^\circ ,\cdots ,\wh V_k) \cr
{}& + \sum\limits_{1\leq i < j \leq k}(-)^{i+j}\, ([\wh V_i,\wh V_j],\wh V_1,
\cdots ,\wh V_i^\circ ,\cdots ,\wh V_j^\circ ,\cdots ,\wh V_k).\cr}
\eqn\extderdef$$
In the first term in the right hand side, the differential operator
${\wh V_i}$ is acting on the function appearing to the right ($\Omega$
acting on the vector fields is a function on the manifold).  The superscript
`$\circ$' on a vector field indicates that the vector field is ommitted.
Therefore, while the form $\hbox{d}\Omega$ has $k$ entries, the form
$\Omega$ appearing in the right hand side shows $(k-1)$ entries, as it should.

In applying \extderdef\ we will be working on the space $\wh\P_{g,n}$.
Pick a reference point $\Sigma \in \P_{g,n}$ corresponding to a surface
with coordinate curves. We consider a set of vector fields
$\wh V_1, \cdots \wh V_k$ defined on the tangent spaces corresponding
to points in a neighborhood of $\Sigma$. We introduce for convenience
a set of coordinates $u^I$ around the point $\Sigma$. Then a tangent
vector at $\Sigma$ will be denoted as
$$\wh V_i\big|_{\Sigma} \equiv V_i^I {\partial\over \partial u^I}
\bigg|_{\Sigma},\eqn\dvect$$
where the sum over $I$ is implicit, $\partial/\partial u^I |_{\Sigma}$
is the usual set of basis vectors, and $V_i^I$ are the components of the
vector. When we want to emphasize that this is a vector field we write
$$\wh V_i\big|_{\Sigma '} \equiv V_i^I(u^I(\Sigma '))
{\partial\over \partial u^I}\bigg|_{\Sigma '} ,\eqn\dvectfi$$
where $\Sigma '$ is any point in the neighborhood of $\Sigma$.
Associated with these vectors there are Schiffer variations. We define
$$ {\bf v}_{{}_I}|_{{}_\Sigma} \equiv {\bf v} \biggl( {\partial
\over \partial u^I} \bigg|_\Sigma \biggr) ,\eqn\schidefb$$
as the Schiffer vector that implements the deformations associated
to the indicated basis vector based at point $\Sigma$. By linearity, we have
that
$${\bf v} \bigl( \wh V_i|_\Sigma ) = V_i^I{\bf v}_{{}_I}|_{{}_\Sigma}
\equiv {\bf v}_i|_{{}_\Sigma} , \eqn\schifefvf$$
where the last relation has introduced a further piece of notation.

There is a nontrivial relation between the Schiffer vectors and
the tangent vectors due to the fact that the algebra of Schiffer
variations must give a representation of the algebra of tangent
vectors. We must have that
$$\l \delta_{{\bf v}(\wh V_i)},\delta_{{\bf v}(\wh V_j)}\r
=\delta_{{\bf v}([\wh V_j ,\wh V_i ])}, \eqn\algcond$$
with the particular ordering of tangent vectors on the right hand side
arising because the algebra of transformations shows a reversal of
ordering with respect to the algebra of the underlying vectors
(as in \sixnine ).
Let us evaluate the left hand side of Eqn.\algcond\ acting on a local
coordinate $z$. We begin by calculating
$$\eqalign{
(1+\e\delta_{{\bf v}(\wh V_i)}) (1+ \e\delta_{{\bf v}(\wh V_j)}) \, z &=
(1+\e\delta_{{\bf v}(\wh V_i)})\,
\bigl( z+\e\,{\bf v}_j|_{{}_\Sigma}(z)\bigr)\cr
&= z+\e\, {\bf v}_{j}|_{{}_\Sigma}(z) +\e\, {\bf v}_{i}|_{{}_{\Sigma_j}}
\bigl( z+ \e\, {\bf v}_{j}|_{{}_\Sigma}(z)\bigr),\cr}\eqn\gntident$$
where $\Sigma_j \in \P_{g,n}$ denotes the point obtained from $\Sigma$
after the variation corresponding to the tangent $\e \wh V_j$. Note that
in the second variation we have used the Schiffer vector corresponding
to the new basepoint. By further expansion we get
$$(1+\e\delta_{{\bf v}(\wh V_i)}) (1+ \e\delta_{{\bf v}(\wh V_j)}) \, z
= z+\e\, {\bf v}_{j}|_{{}_\Sigma}(z)
+\e\, {\bf v}_{i}|_{{}_{\Sigma_j}}(z)
+\e^2\, {\bf v}_j \partial {\bf v}_i + \O(\e^3), \eqn\lhssimp$$
where in the last term $\partial \equiv \partial /\partial z$, and the
Schiffer vectors in that term, to the approximation we are working, can be
taken to be based at $\Sigma$. Exchanging $i$ and $j$ in the above,
subtracting the resulting equation from \lhssimp , and substituting back
in \algcond , we find
$$ ({\bf v}_j |_{{}_\Sigma} - {\bf v}_j |_{{}_{\Sigma_i}} )
+  ({\bf v}_i |_{{}_{\Sigma_j}} - {\bf v}_i |_{{}_\Sigma} )
= \e \, \l {\bf v}_i , {\bf v}_j \r + \e \,
{\bf v}\bigl( [\wh V_j ,\wh V_i ] \bigr)+ \O(\e^2), \eqn\tentwenone$$
where in first bracket in the right hand side the Schiffer vectors
are operators on the surface. This relation will be used shortly.

We can now begin in earnest our proof of Eqn.\qaction .
Consider the case when the form in the left hand side is a $k$-form, so
in particular, $*= k-\hbox{Dim}\,\M_{g,n}$. Moreover, since \qaction\ is
homogeneous and does not involve changes in $g$ or $n$, the normalization
factor $N_{g,n}$ is irrelevant and we will
not show it below.  Our strategy will be as follows. We will begin with
the form
$$\Omega (\wh V_1, \cdots , \wh V_{k-1} ) = \bra{\Sigma}
{\bf b} \bigl({\bf v}(\wh V_1)\bigr)
\cdots {\bf b} \bigl({\bf v}(\wh V_{k-1})\bigr) \ket{\vec B}, \eqn\bformbe$$
where, for clarity, we have written the Schiffer vectors as function
of the tangent vectors. Then we will use Eqn.\extderdef\ to evaluate
$\hbox{d} \Omega$. Finally we will evaluate, by direct computation
the left hand side of \qaction .

Consider therefore the first term in the right hand side of
\extderdef\ which corresponds to
$$\sum_{i=1}^k (-)^{i+1} \wh V_i \,\bigl(\, \bra{\Sigma}
{\bf b} \bigl({\bf v}(\wh V_1)\bigr)
\cdots {\bf b}^\circ \bigl({\bf v}(\wh V_i)\bigr)\cdots
{\bf b} \bigl({\bf v}(\wh V_k)\bigr)\ket{\vec B}\,\bigr).\eqn\tdertone$$
In order to take this directional derivative it should be emphasized
that not only the state changes, but also the antighost insertions
change, since they depend on vector fields. The derivative can be
evaluated as the limit
$$\eqalign{
\sum_{i=1}^k (-)^{i+1} \lim\limits_{\e\rightarrow 0} \,{1\over \e}\,
&\biggl[ \, \bra{\Sigma_i} {\bf b} \bigl({\bf v}(\wh V_1)\bigr)
\cdots {\bf b}^\circ \bigl({\bf v}(\wh V_i)\bigr)
\cdots {\bf b} \bigl({\bf v}(\wh V_k)\bigr) \ket{\vec B}
\bigg|_{\Sigma_i}\cr
{}&-\bra{\Sigma} {\bf b} \bigl({\bf v}(\wh V_1)\bigr)
\cdots {\bf b}^\circ \bigl({\bf v}(\wh V_i)\bigr)
\cdots {\bf b} \bigl({\bf v}(\wh V_k)\bigr) \ket{\vec B}
\bigg|_\Sigma \,\,\biggr] ,\cr}\eqn\ederform$$
where $\Sigma_i$, as defined earlier, denotes the point in $\P_{g,n}$
obtained by varying $\Sigma$ with the tangent $\e\wh V_i$. We know
that
$$\bra{\Sigma_i}=\bra{\Sigma}\bigl( 1-\e {\bf T}({\bf v}(\wh V_i))\, \bigr)
+ \O(\e^2), \eqn\varbrat$$
and, in analogy, we define for the antighosts
$${\bf b}({\bf v}(\wh V_j))\big|_{{}_{\Sigma_i}} \equiv
{\bf b}({\bf v}(\wh V_j))\big|_{{}_\Sigma} +\e\,\Delta_i{\bf b}_j+\O(\e^2).
\eqn\varanti$$
With a short computation we find that \ederform ; which is simply the
first term in the right hand side of \extderdef , can be written as
$$\eqalign{
{}&\quad\sum_{i=1}^k(-)^i\bra{\Sigma}{\bf T}({\bf v}(\wh V_i))
\,{\bf b} \bigl({\bf v}(\wh V_1)\bigr)
\cdots {\bf b}^\circ \bigl({\bf v}(\wh V_i)\bigr)
\cdots {\bf b} \bigl({\bf v}(\wh V_k)\bigr) \ket{\vec B}\cr
{}&+\hskip-8pt\sum\limits_{1\leq i < j \leq k}\hskip-8pt (-)^{i+j}
\bra{\Sigma} (\Delta_j{\bf b}_i-\Delta_i{\bf b}_j)
\,{\bf b}\bigl({\bf v}(\wh V_1)\bigr)
\cdot\cdot {\bf b}^\circ \bigl({\bf v}(\wh V_i)\bigr)
\cdot\cdot {\bf b}^\circ \bigl({\bf v}(\wh V_j)\bigr)
\cdot\cdot {\bf b} \bigl({\bf v}(\wh V_k)\bigr)\ket{\vec B}.\cr}\eqn\sfirts$$
On the other hand the second term in \extderdef\ gives
$${}\hskip-8pt\sum\limits_{1\leq i < j \leq k}\hskip-8pt (-)^{i+j}
\bra{\Sigma} {\bf b}\bigl( {\bf v}(\l\wh V_i,\wh V_j\r )\bigr)
\,{\bf b}\bigl({\bf v}(\wh V_1)\bigr)
\cdots {\bf b}^\circ \bigl({\bf v}(\wh V_i)\bigr)
\cdots {\bf b}^\circ \bigl({\bf v}(\wh V_j)\bigr)
\cdots {\bf b} \bigl({\bf v}(\wh V_k)\bigr)\ket{\vec B}\eqn\sfixrts$$
The last term in \sfirts\ and the term in \sfixrts\ combine naturally
and suggest that we simplify the quantity
$\Delta_j{\bf b}_i-\Delta_i{\bf b}_j +{\bf b}\bigl( {\bf v}
(\l\wh V_i,\wh V_j\r )$.
Making use of Eqn.\varanti\ and \tentwenone\ we find
$$\eqalign{\Delta_j{\bf b}_i-\Delta_i{\bf b}_j +{\bf b}\bigl( {\bf v}
(\l\wh V_i,\wh V_j\r ) \bigr)
&= {1\over \e}{\bf b}\biggl( {\bf v}_i|_{{}_{\Sigma_j}}-{\bf v}_i|_{{}_\Sigma}
-{\bf v}_j|_{{}_{\Sigma_i}} +{\bf v}_j|_{{}_\Sigma} + \e\,{\bf v}
(\l\wh V_i,\wh V_j\r ) \biggr)\cr
{}& = {\bf b} \bigl( \l {\bf v}_i , {\bf v}_j \r \bigr) .\cr}\eqn\athed$$
Using this result to add up Eqns.\sfirts\ and \sfixrts , we obtain the
complete expression for $\hbox{d}\Omega$
$$\eqalign{
\hbox{d}\Omega (\wh V_1 ,\cdot\cdot , \wh V_k )&=
\sum_{i=1}^k(-)^i\bra{\Sigma}{\bf T}({\bf v}_i)
\,{\bf b} ({\bf v}_1)
\cdots {\bf b}^\circ ( {\bf v}_i)
\cdots {\bf b} ( {\bf v}_k) \ket{\vec B}\cr
{}&+\hskip-8pt\sum\limits_{1\leq i < j \leq k}\hskip-8pt (-1)^{i+j}
\bra{\Sigma}{\bf b} \bigl( \l {\bf v}_i , {\bf v}_j \r \bigr)
{\bf b}( {\bf v}_1) \cdots {\bf b}^\circ ( {\bf v}_i)
\cdots {\bf b}^\circ ( {\bf v}_j)
\cdots {\bf b}({\bf v}_k) \ket{\vec B},\cr}\eqn\fpartd$$
where we have gone back to the notation ${\bf v}_k = {\bf v}(\wh V_k)$,
and all the vectors are based at $\Sigma$.
Having evaluated the form $\hbox{d}\Omega$ appearing in the
right hand side of \qaction , we now must evaluate the left hand
side.  We begin with
$$\bra{\Sigma_P} {\bf b}({\bf v}_1) \cdots {\bf b}({\bf v}_k)
\,(Q^{(1)} + \cdots Q^{(n)} )\ket{\vec B},\eqn\ssddff$$
and move the sum of BRST operators all the way to the left using \qwithb ,
until the sum of BRST operators hits the bra $\bra{\Sigma}$ and annihilates it.
In doing so we obtain
$$\sum_{i=1}^k (-)^{k-i} \bra{\Sigma_P} {\bf b}({\bf v}_1)
\cdots {\bf b}^\circ ({\bf v}_i) {\bf T}({\bf v}_i) \cdots
{\bf b}({\bf v}_k)\ket{\vec B} , \eqn\irsttype$$
where the sign factor arises because $\sum Q$ was commuted through $(k-i)$
antighosts. We must now move the ${\bf T}$'s all the way to the left. In doing
so we obtain two type of terms; terms where ${\bf T}$'s have reached
the bra $\bra{\Sigma}$
$$(-)^k\sum_{i=1}^k(-)^i\bra{\Sigma_P}{\bf T}({\bf v}_i){\bf b}({\bf
v}_1)\cdots
{\bf b}^\circ ({\bf v}_i)\cdots{\bf b}({\bf v}_k)\ket{\vec B},\eqn\firsttype$$
and terms arising from the commutators of ${\bf T}$ with ${\bf b}$
(see \ebhmalg )
$$(-)^k\sum\limits_{1\leq i < j \leq k}  (-)^{i+j}
\bra{\Sigma_P} {\bf b}([ {\bf v}_i , {\bf v}_j ]) \, {\bf b}({\bf v}_1)
\cdots {\bf b}^\circ ({\bf v}_i) \cdots {\bf b}^\circ ({\bf v}_j) \cdots
{\bf b}({\bf v}_k) \, \ket{\vec B},\eqn\secondtype$$
where in obtaining this from \irsttype\ we first relabeled $i$ by $j$,
and the antighost factor ${\bf b}([ {\bf v}_i , {\bf v}_j ])$ was moved
across $(i-1)$ antighosts to get to the leftmost position.

The above two groups of terms are readily seen to correspond to those
shown in \fpartd\ up to the sign factor $(-)^k$. \foot{Note that it
would have been incorrect to naively identify the expressions
in \firsttype\ and \secondtype\ with the first and second terms in
equation \extderdef .} Given that $k= *+\hbox{Dim}\,\M_{g,n}$,
and all moduli spaces have even real dimension, the sign factor equals
$(-)^*$. This proves the result in \qaction .
Note that the sign factor is irrelevant when the form in the left hand side
is the top form $\Omega^{(0)}$ on the section $\sigma$.

\section{Defining the String Field Multilinear Functions and Products}

We have done all the preparatory work to be able to define precisely the
string field multilinear functions, and the string field products.
We will first discuss the multilinear functions, give their definition and
basis properties. These multilinear functions, when used for the
dynamical string field $\Psi$, give the string field vertices entering
the string field actions. We show that these string field vertices are real.
Then we define the string field products and prove some of their basic
properties. We leave for \S8 the proof of the main identity.

\noindent
$\underline{\hbox{The String Multilinear Functions}}.\,$ The form
${\Omega_{}^{}}_{\vec B}^{(0)g,n}$ is of degree $(6g-6+2n)$ and therefore
can be integrated on the section $\sigma (\M_{g,n}) \in \wh \P_{g,n}$.
Let $\V_{g,n} = \sigma (\D_{g,n})$ denote the subspace of the section
corresponding to the surfaces $\D_{g,n}\subset \M_{g,n}$ comprising string
vertex. The string {\it field} vertex or multilinear function associated
to this subset of surfaces is defined to be
$$\big\{ B_1 , B_2 , \cdots , B_n \big\}_g
=\int_{\V_{g,n}}{\Omega_{}^{}}_{B_1\cdots B_n}^{(0)g,n}.\eqn\stringfvertex$$
It is simply the integral over the subset of surfaces defining
the string vertex, using the minimal area section. Thus given
$n$ string fields, the multilinear product gives us a number.
This number, of course, is an ordinary number, complex, in general
times a set of spacetime fields (recall the usual expansion of the
string field in \setupsf ).

As we discussed below Eqn.\normcon\ the form entering $\Omega_{\vec B}^{(0)}$
entering the definition above vanishes unless $\sum_i (G(B_i)-2) = 0$, and
therefore, the multilinear function vanishes unless this is the case
(as stated in \gmpro ). Since the bra $\bra{\Sigma}$ entering
$\Omega_{\vec B}^{(0)}$ is Grassmann even and the number of antighost
insertions is even, the statistics of the multilinear function is simply
the statistics of the product of the $B$'s, in other words, the
multilinear functions are intrinsically even (Eqn. \statmprod ).
It is also obvious from the above definition that the functions
are indeed multilinear, in the form stated in Eqn. \fmultilinearity .

It is convenient to be more explicit as far as the
states to be inserted. Using \dfinally\ we are led to define
$$\bra{\Omega^{(0)g,n}}\equiv N_{g,n}\bra{\Sigma_P}{\bf b}({\bf v}_1)
\cdots {\bf b}({\bf v}_{6g-6+2n}), \eqn\intronnot$$
which allows us to write
$$ \Omega_{\vec B}^{(0)g,n} =\bra{\Omega^{g,n}}B_1\rangle
\cdots \ket{B_n}.\eqn\seoutnot$$
The bra $\bra{\Omega^{g,n}}$ is simply introduced for convenience of
notation. It is a form on
$T_{\wh P}(\wh\P_{g,n})$, but valued in the tensor product of
$n$ copies of the Hilbert space $\H$. It is not uniquely defined
as such, since as we saw in the previous section, the various choices
one has to make affect $\bra{\Omega^{g,n}}$, but do not affect
the value of $\Omega_{\vec B}^{(0)g,n}$ due to the subsidiary conditions
on the states $\ket{B_i}$. Whenever we will use $\bra{\Omega^{g,n}}$
it will appear contracted with suitable states. Using this notation
we write
$$\big\{ B_1 , B_2 , \cdots , B_n \big\}_g
= \int_{\V_{g,n}} \bra{\Omega^{g,n}} B_1\rangle\ket{B_2}\cdots\ket{B_n}.
\eqn\mesvertex$$
Let us consider now the graded-commutativity property of the multilinear
functions. By construction, the string vertices $\V_{g,n}$ used to define
the multilinear functions include a set of Riemann surfaces with labeled
punctures, with the condition (stated in \S5.1 as condition (b)) that the
local coordinates are assigned in a labelling independent way, and if a surface
$\R$ is in the set, the copies of $\R$ with all other inequivalent labelings of
the punctures are also included in the set. This implies that the result of
the multilinear product does not depend on which state is assigned to which
puncture. Using the notation explained below Eqn. \dfinally\ this implies that
$$\int_{\V_{g,n}} \bra{\Omega^{g,n}(1,2,\cdots ,n)}
\cdots \ket{B_i}_{(i)} \cdots \ket{B_j}_{(j)}\cdots
=\int_{\V_{g,n}} \bra{\Omega^{g,n}(1,2,\cdots ,n)}
\cdots \ket{B_i}_{(j)} \cdots \ket{B_j}_{(i)}\cdots ,\eqn\symmpropmp$$
and effectively means that the object
$$\int_{\V_{g,n}} \bra{\Omega^{g,n}(1,2,\cdots ,n)} , \eqn\oacting$$
acting on states, is symmetric under any exchange of labels of the
Hilbert spaces.\foot{The symmetry may not be manifest, but it
must hold given the properties of the states representing the surfaces.
For example, the antighost insertions cannot be generically done
treating the punctures in a symmetric way, however, it is clear that
using Borel vectors the insertions can be moved around to show the
symmetry of the bra $\bra{\Omega^{g,n}}$.} This means that
the multilinear product simply picks up a sign corresponding to the
statistics of the states that one commutes, that is
$$\big\{ B_1 ,\cdots, B_i , B_{i+1} ,\cdots , B_n \big\}_g
=(-)^{B_iB_{i+1}}\{ B_1 ,\cdots, B_{i+1},  B_i ,\cdots , B_n \}_g,
\eqn\pgradedcomm$$
Note that the multilinear function we have introduced need always
at least one argument, that is $n\geq 1$, otherwise there is no
state $\bra{\Sigma_P}$ to work with. In fact, since we did not
introduce the string vertices $\V_{g=0,n}$ for $n=0,1,2$, so far we only
have multilinear products defined for
$n\geq 3$ when the genus is zero.  When we studied the geometrical
recursion relations we observed that the subsets $\V_{g,0}$ for
$g\geq 2$ were all defined by the recursion relations. The case of
$g=1$ is peculiar in that the subset $\V_{g=1,0}$ is not defined by
the recursion relations. We will simply assume it is determined
somehow (see \S6.4).
Let us therefore extend the definition of the multilinear products
to $g \geq 1$ and $n=0$. The idea is that the multilinear product,
denoted in this case by $\{ \, \cdot \, \}_g$ must give us a number.
Physically, this number should just be the result for the corresponding
vacuum graph, but since the operator formalism does not give us a
state for surfaces without punctures, we must use the same idea that
was used in [\alvarez ] to calculate partition functions. Consider then
the subset $\V_{g,0}$, and note that since we have no punctures and therefore
no local coordinates, this should be thought as a subset of $\ov\M_{g,0}$.
We then find a corresponding subset
$\wt\D_{g,1} \in \ov\M_{g,1}$ such that when we forget about the extra
puncture we obtain the set $\V_{g,0}$.  Then we define
the form $\Omega^{g,0}$ on $T_{\vec P}(\ov\M_{g,0})$ by the expression
$$\Omega^{g,0} (\vec V_1,\cdots \vec V_{6g-6})
= \bra{\Sigma_P} {\bf b}({\bf v}_1)\cdots {\bf b}({\bf v}_{6g-6})\ket{{\bf 1}},
\eqn\vacvert$$
where the surface $P\in \sigma (\wt\D_{g,1})$ projects into the
surface $\vec P \in \ov \M_{g,0}$ as we forget about the puncture. Note
that ghost number works out correctly,
since, compared to the top form on $\wh\P_{g,1}$ we have two less antighost
insertions, but at the same time instead of inserting a state of ghost
number two, we are inserting the vacuum, of ghost number zero (for $g=1$
one must use two antighost insertions). It was shown in [\alvarez ] that
this is a well defined form on $\ov\M_{g,0}$. Now we simply set
$$\big\{\,\cdot\,\big\}_g=\int_{\V_{g,0}\subset\ov\M_{g,0}}\Omega^{g,0}.
\eqn\multnzero$$
For genus zero we will set
$$\big\{ \, \cdot \, \big\}_0 =0.\eqn\asmenav$$
The multilinear products for $g=0$ and $n=1,2$ are taken to be
$$\big\{ \, B \big\}_0 \equiv 0, \quad\hbox{and}\quad
\big\{ B_1 , B_2 \big\}_0 \equiv \langle B_1 , Q B_2 \rangle .\eqn\deflowdc$$
\medskip
\noindent
$\underline{\hbox{Hermiticity of the String Field Vertices}}.\,$
Recall now from \S4 that the string field vertices were the key elements
in constructing the string action. We had that
$$S_g^n (\Psi ) \equiv {1\over n!} \big\{ \Psi^n \big\}_g.\eqn\sfvreal$$
We want to show that given a dynamical string field $\ket{\Psi}$ satisfying
the reality condition $\bra{\Psi_{hc}} =-\bra{\Psi}$ imposed in \S3.1,
the string field vertex $\{ \Psi^n \}_g$ is real.
The key geometrical input that guarantees the hermiticity of the
string field vertex is condition (c) in \S5.1, which requires that
given a surface $\Sigma \in \V$ then the mirror $\Sigma^*$ is also in $\V$,
and that the local coordinates in $\Sigma$ and $\Sigma^*$ be related by the
antiholomorphic map that relates the two surfaces. The minimal area
area problem that we use to define the local coordinates indeed gives us
string vertices satisfying those conditions (\S6.4). In terms of the
states representing the surfaces the above conditions imply that
$$\bra{\Sigma_{hc}} = \bra{\Sigma^*},\eqn\conjopstate$$
which can be established by using the conserved charges method of the
operator formalism [\alvarez ]. In order to show that the string
vertex is real we must consider the form we are integrating around
the surface $\Sigma$
$$(\Sigma ) = N_{g,n}\bra{\Sigma}{\bf b}({\bf v}_1)\cdots {\bf b}({\bf v}_r)
\ket{\vec\Psi},\eqn\firstsurf$$
and the form around the surface $\Sigma^*$
$$(\Sigma^* ) = N_{g,n}\bra{\Sigma^*}{\bf b}({\bf v}_1^*)\cdots {\bf b}
({\bf v}_r^*)\ket{\vec\Psi},\eqn\mirrsurf$$
where the Schiffer vector ${\bf v}_i^*$ is defined by the condition
that the deformations it produces on the coordinates of $\Sigma^*$
be the mirror images of the deformations the Schiffer vector
${\bf v}_i$ produces on $\Sigma$. Under this condition, we must
show that the number in \mirrsurf\ is just the complex conjugate of
that in \firstsurf . This will imply the hermiticity of the string field
vertex since we can then break up the total integral into two pieces
one the hermitian conjugate of the other.

The condition that the coordinates be related by the antiholomorphic map
means that if the map takes a point $p\in \Sigma$ in the domain of the local
coordinate $z$ into the point $p^*\in \Sigma^*$, then the local coordinate
$z^*$ satisfies
$$z^*(p^*) = \ov {z(p)},\eqn\cccoord$$
where the bar will denote complex conjugation. Then given a Schiffer vector
$v(z)$ in $\Sigma$, deforming $z$ as $z(p) \rightarrow z(p) + v(z(p))$,
complex conjugating, we have
$$\ov {z(p)} \rightarrow \ov {z(p) + v(z(p))},$$
and using \cccoord\ it implies that
$$z^* (p^*) \rightarrow z^* (p^*) + \ov v (z^*(p^*)),\eqn\cschuff$$
and we deduce that $v^*(z) = \ov v (z)$. More explicitly
$$v(z) = \sum {v_n\over z^{n-1}} \quad \rightarrow\quad
\ov v (z) = v^*(z)= \sum {\ov v_n\over z^{n-1}}.\eqn\moreexpuuff$$
This defines the Schiffer vectors ${\bf v}^*$ entering in \mirrsurf .
Let us now begin by taking the hermitian conjugate of \firstsurf\ to find
$$\eqalign{\ov {(\Sigma )} &= \ov N_{g,n}\bigl( \bra{\Sigma}{\bf b}({\bf v}_1)
\cdots {\bf b}({\bf v}_r)\ket{\vec\Psi}\bigr)^\dagger ,\cr
{}&= (-)^n \ov N_{g,n}\bra{\vec\Psi_{hc}}[{\bf b}({\bf v}_r)]^\dagger
\cdots [{\bf b}({\bf v}_1)]^\dagger\ket{\Sigma_{hc}},\cr
{}&= (-)^{d_{g,n}}{N_{g,n}}\bra{\vec\Psi}[{\bf b}({\bf v}_r)]^\dagger
\cdots [{\bf b}({\bf v}_1)]^\dagger\ket{\Sigma_{hc}},\cr
{}&= N_{g,n} \bra{\vec\Psi}[{\bf b}({\bf v}_1)]^\dagger
\cdots [{\bf b}({\bf v}_r)]^\dagger\ket{\Sigma_{hc}},\cr}\eqn\sohardtopr$$
where in the first step we got a factor of $(-)^n$ since there are
$n$ Hilbert spaces and each one gives a minus sign due to \conjherm . In the
second step we used the reality condition on the string field, which
gives us another $(-)^n$, and the definition of the
normalization factor in \normcon . Finally, in the last step we rearranged the
antighost insertions; since there are $r=2d_{g,n}$ of them, this
cancels the previous sign factor. The last equation can be written in the form
$$\ov{(\Sigma )}= N_{g,n}\bra{\vec\Psi}{\bf M}\rangle ,\eqn\stall$$
where the Grassmann even ket $\ket{{\bf M}}$ is given by
$$\ket{{\bf M}} = [{\bf b}({\bf v}_1)]^\dagger
\cdots [{\bf b}({\bf v}_r)]^\dagger\ket{\Sigma_{hc}}.\eqn\sotopr$$
It follows from \symmbpz\ that
$$\ov{(\Sigma )}=N_{g,n}\bra{{\bf M}}{\vec\Psi}\rangle ,\eqn\stalli$$
and therefore we must simply evaluate
$$\bra{{\bf M}}=\bra{R_{11'}}\cdots\bra{R_{nn'}}[{\bf b}({\bf v}_1)]^\dagger
\cdots [{\bf b}({\bf v}_r)]^\dagger\ket{\Sigma_{hc}}.\eqn\musteval$$
Using the definition of $b(v)$ in \defbbold\ and the expansion of the vector
$v$ in \moreexpuuff\ we have
$$b(v) = \sum (b_n v_{-n} + \ov b_n \ov v_{-n}), \quad\hbox{and}\quad
b(v)^\dagger = \sum (b_{-n} \ov v_{-n} + \ov b_{-n} v_{-n}),\eqn\daggbx$$
and therefore using \qonref\ we have
$$\eqalign{\bra{R_{ii'}}[b_{(i')}(v)]^\dagger &=\bra{R_{ii'}}\sum
(b_{-n(i')}\ov v_{-n}+\ov b_{-n(i')}v_{-n}),\cr
{}&=\bra{R_{ii'}}\sum (b_{n(i)} \ov v_{-n} + \ov b_{n(i)} v_{-n}),\cr
{}&=\bra{R_{ii'}} [b(v^*)]_{(i)}.\cr}\eqn\funnyes$$
Using this result, together with the Grassmann even property of
$\ket{\Sigma_{hc}}$ one finds that Eqn. \musteval\ gives
$$\bra{{\bf M}}= \bra{\Sigma_{hc}} {\bf b}({\bf v}_1^*)
\cdots {\bf b}({\bf v}_r^*).\eqn\mustevali$$
Finally, using Eqn.\conjopstate\ and substituting back into \stalli\ we get
$$\ov{(\Sigma )}= N_{g,n} \bra{\Sigma^*} {\bf b}({\bf v}_1^*)
\cdots {\bf b}({\bf v}_r^*)\ket{\vec\Psi}= (\Sigma^* ),\eqn\stallii$$
which is the result we wanted to establish. This concludes our proof
that the vertices of the string field theory are hermitian.
Having shown in \S3.2 the hermiticity of the kinetic term, this establishes
the hermiticity of the full string field theory.

\noindent
$\underline{\hbox{String Field Products}}.\,$
Having defined the multilinear string field functions let us now define the
closely related string products. These products take a set of string fields,
and give us another string field. In the same way as the multilinear functions
they are labeled by the genus $g$. We define
$$\l B_1, \cdots , B_{n-1} \r_g \equiv \spr_s (-)^{\Phi_s}\cdot
\int_{\V_{g,n}} \bra{\Omega^{g,n}}\Phi_s\rangle_0 \ket{\wt\Phi_s}_e
\ket{B_1} \cdots  \ket{B_{n-1}} ,\eqn\defstringprod$$
where the sum extends over a basis of states $\Phi_s$ complete in the subspace
of states annihilated by $(L_0-\ov L_0)$. The label $e$, standing for
external, on the state $\ket{\wt\Phi_s}_e$ is the label for the Hilbert space
of the resulting product. This state is {\it not} contracted with
$\bra{\Omega^{g,n}}$. The Hilbert spaces of the bra $\bra{\Omega^{g,n}}$ are
labeled by $0,1,\cdots,n-1$. The state $\ket{\Phi_s}_0$ is contracted with the
$0$-th Hilbert space, and each state $\ket{B_i}$ is contracted with the $i$-th
Hilbert space. One easily rewrites the product in the following forms
$$\eqalign{\l B_1,\cdots ,B_{n-1}\r_g &=\spr_s(-)^{\Phi_s}\ket{\wt\Phi_s}_e
\cdot\int_{\V_{g,n}}\bra{\Omega^{g,n}}\Phi_s\rangle_0\ket{B_1}\cdots
\ket{B_{n-1}},\cr
&= \spr_s (-)^{\Phi_s}\ket{\wt\Phi_s}_e\cdot
\big\{ \Phi_s , B_1, \cdots , B_{n-1} \big\}_g , \cr
&=\spr_s\int_{\V_{g,n}} \bra{\Omega^{g,n}}b_0^{-(e)} \ket{\Phi_s}_0
\ket{\Phi_s^c}_e\ket{B_1} \cdots  \ket{B_{n-1}} ,\cr
&=  \int_{\V_{g,n}} \bra{\Omega^{g,n}} \wt {R'}_{0e}\rangle
\ket{B_1} \cdots  \ket{B_{n-1}} .\cr}\eqn\altformprod$$
These are obtained from \defstringprod\ as follows. In the first
relation we simply used the facts that $\bra{\Omega^{g,n}}$ is Grassmann
even and that $\ket{\Phi_s}$ and $\ket{\wt\Phi_s}$ commute with each
other. In the second form we made use of the definition of the
multilinear functions.  In the third form we made explicit the
$b_0^-$ factor in the definition of $\ket{\wt\Phi_s}$. In the last form
we simply recognized the form of the reflector.
Multilinearity of the product is manifest from the definition. Moreover
it is also clear that the product is intrinsically Grassmann odd, because
the multilinear functions are Grassmann even and the sewing ket
$\ket{\wt {R'}}$ entering the definition of the product is Grassmann odd.
Finally the multilinear products are clearly graded commutative just
as the multilinear functions.

The multilinear functions and the multilinear products are simply
related to each other via the linear inner product. In fact, using
the last expression in \altformprod\ we have
$$\eqalign{
\big\langle A , \big\{ B_1 , \cdots , B_{n-1}\big\}_g \big\rangle
&={}_e\bra{A} c_0^{-(e)} \cdot
\int_{\V_{g,n}} \bra{\Omega^{g,n}} b_0^{-(e)}\ket{{R'}_{0e}}
\ket{B_1} \cdots  \ket{B_{n-1}}, \cr
&= \int_{\V_{g,n}} \bra{\Omega^{g,n}}
\bigl( {}_e\bra{A} c_0^{-(e)}b_0^{-(e)}\ket{{R'}_{0e}}\bigr)
\ket{B_1} \cdots  \ket{B_{n-1}}, \cr
&= \int_{\V_{g,n}} \bra{\Omega^{g,n}}
\bra{A} R_{0e}\rangle \ket{B_1} \cdots  \ket{B_{n-1}}, \cr
&= \int_{\V_{g,n}} \bra{\Omega^{g,n}}A\rangle \ket{B_1}\cdots\ket{B_{n-1}},\cr
&=\big\{ A, B_1,\cdots ,B_n \big\}_g .\cr}\eqn\relateprod$$

Let us conclude this section by establishing two properties
(given in \tworref\ and \relatedanduse\ )that
arise because of symmetry properties of the states. The first one is
$$\spr_s \l \cdots , \Phi_s \l \wt\Phi_s, \cdots
\r_{g_1}\r_{g_2} = 0 ,\eqn\tworrefi$$
where the dots denote arbitrary string fields. This expression amounts
to
$$ \spr_s\int_{\V_{g_1,n}} \bra{\Omega^{g_1,k} (0,\cdots k,l')}
\wt {R'}_{0e}\rangle \cdots \ket{\Phi_s}_{(k)} \cdot
\int_{\V_{g_2,m}} \bra{\Omega^{g_2,m} (l, k' \cdots )}\wt {R'}_{ll'}\rangle
\ket{\wt\Phi_s}_{(k')} \cdots \eqn\prorrworrefi$$
pushing the state $\ket{\Phi_s}_{(k)}$ in front of the state
$\ket{\wt\Phi_s}_{(k')}$ we get a sign factor $(-)^{\Phi_s}$ (since
the operator in between is odd) and we recognize the sewing ket
$$ \spr_s\int_{\V_{g_1,n}} \bra{\Omega^{g_1,k} (0,\cdots k,l')}
\wt {R'}_{0e}\rangle \cdots
\int_{\V_{g_2,m}} \bra{\Omega^{g_2,m} (l, k' \cdots )}\wt {R'}_{ll'}\rangle
\ket{\wt {R'}_{kk'}}. \cdots \eqn\orrefi$$
This term vanishes because the product of kets
$\ket{\wt {R'}_{ll'}}\ket{\wt {R'}_{kk'}}$ is antisymmetric under the Hilbert
spaces label exchange $k\leftrightarrow l' ; k'\leftrightarrow l$, but the
rest of the expression in \orrefi\ is symmetric under this exchange.
For the case of the identity
$$\spr_{r,s} (-)^{\Phi_r}(-)^{\Phi_s} \l \Phi_s , \wt\Phi_s ,
\cdots , \Phi_r , \wt\Phi_r , \cdots \r_g = 0.\eqn\rexlanduse$$
the left hand side amounts to
$$\int_{\V_{g,n}} \bra{\Omega^{g,n}(\cdots , l,l',\cdots , k,k'\cdots )}
\cdots \ket{\wt {R'}_{ll'}}\ket{\wt {R'}_{kk'}}\cdots ,\eqn\newrel$$
and this clearly vanishes for the same reason as in the case above.

\chapter{Establishing the Main Identity}

In sections \S5 and \S6 we learned how to find subsets $\V_{g,n}$,
called string vertices,
satisfying a nontrivial set of recursion relations and conditions.
The reason they satisfy such relations is that sewing of the
string vertices with the standard Feynman combinatorial rules gives
a single cover of all relevant moduli spaces. In \S7 we learned
how to define string products as integrals of forms over the $\V_{g,n}$
subsets. In the present section we will use the recursion
relations satisfied by the string vertices in order to prove the
identity satisfied by the string products. This identity, called the
main identity, and given in \bidentity , guarantees the consistency of the
string field theory, as we have established in \S4. It relates the failure of
the BRST operator to act as a derivation on the string products to the failure
of the string products to satisfy Jacobi type identities.

We begin our work with
the expression for the violation of the derivation property
$$(\hbox{I})\,\, \equiv Q \l B_1 , \cdots , B_n \r_g
+ \sum_{i=1}^n (-)^{(B_1 + \cdots B_{i-1})} \l B_1, \cdots
QB_i, \cdots , B_n \r_g , \eqn\fderv$$
and now using the expression for the string field product given
in the last right hand side of Eqn. \altformprod\ we write (I) as
$$\eqalign{
(\hbox{I}) &= \int_{\V_{g,n+1}}\hskip-2pt \bra{\Omega^{g,n+1}}
\biggl( Q_e |\wt {R'}_{0e}\rangle \ket{B_1}\cdots\ket{B_n}
+ \sum_{i=1}^n (-)^{(B_1+\cdots B_{i-1})}|\wt {R'}_{0e}\rangle\ket{B_1}\cdots
Q\ket{B_i} \cdots \ket{B_n} \biggr) ,\cr
&= -\int_{\V_{g,n+1}} \bra{\Omega^{g,n+1}}
\bigl( Q_0 + \sum_{i=1}^n Q_i ) \ket{\wt{R'}_{0e}} \ket{B_1}\cdots\ket{B_n},\cr
&= \spr_s (-)^{\Phi_s+1} \int_{\V_{g,n+1}} \bra{\Omega^{g,n+1}}
\bigl(\sum_{i=0}^n Q_i \bigr)
\ket{\Phi_s}_0 \ket{\wt\Phi_s}_e \ket{B_1}\cdots\ket{B_n},\cr}
\eqn\itssat$$
where use was made of the fact that $\bra{\Omega}$ is Grassmann even to
commute $Q_e$ through it, and in the second step we used the BRST
property of the sewing ket (Eqn.\qprim ) and brought all the
BRST operators to the same position. In order to be able to use
the BRST properties of the forms derived earlier it is convenient to
break up the sewing ket and take out of the integrand the state with
external label. Due to the
presence of the BRST operators we get an extra sign factor of
$(-)^{\Phi_s+1}$ and therefore we obtain
$$\eqalign{
(\hbox{I})&= \spr_s \ket{\wt\Phi_s}_e \cdot
\int_{\V_{g,n+1}} \bra{\Omega^{g,n+1}}
\bigl( \sum Q_i ) \ket{\Phi_s}_0 \ket{B_1}\cdots\ket{B_n},\cr
&=\spr_s \ket{\wt\Phi_s}_e \cdot
\int_{\V_{g,n+1}} {\Omega_{}}_{(\sum Q)\Phi_s B_1\cdots B_n}^{(0)g,n+1} ,\cr}
\eqn\readthenyt$$
where we used our standard notation for forms in $\wh\P_{g,n+1}$. We can
now use the BRST property \qaction\ and Stokes theorem to obtain:
$$(\hbox{I}) =\spr_s \ket{\wt\Phi_s}_e \cdot
\int_{\V_{g,n+1}} \hbox{d}\Omega_{\Phi_s B_1\cdots B_n}^{(-1)g,n+1}
\, = \, \spr_s \ket{\wt\Phi_s}_e \cdot
\int_{\partial\V_{g,n+1}}\Omega_{\Phi_s B_1\cdots B_n}^{(-1)g,n+1} .
\eqn\gettingthere$$
But now recall that in \S5 it was seen that (Eqn. \twopointtwo )
$\partial \V_{g,n+1} = -\partial_p R_1$, where $R_1$ denotes the Feynman
graphs with one propagator, and $\partial_p$, the propagator boundary,
that is the set of surfaces obtained by gluing two elementary vertices
via the sewing condition $zw=t$ with $t\in \partial D$, or by gluing
two punctures in the same elementary vertex with the same sewing condition.
These two alternatives are illustrated in Eqn. \twopointone . Since we must
keep track of the orientation of the manifolds where we integrate, the
sign factor relating the vertex boundary to the Feynman diagram boundary
implies that \gettingthere\ becomes
$$(\hbox{I})\,\, =
-\spr_s \ket{\wt\Phi_s}_e \cdot
\int_{\partial_p R_1}\Omega_{\Phi_s B_1\cdots B_n}^{(-1)g,n+1} .\eqn\sofar$$
We have to understand now the behavior of the form $\Omega^{(-1)}$
appearing in the above equation on the set of surfaces $\partial_p R_1$.
To this end we break up the set of surfaces into the its two
components
$$\partial_p R_1 = R_1^{tree} + R_1^{loop}, \eqn\decbound$$
where $R_1^{tree}$ denote the surfaces obtained by gluing two string
vertices and $R_1^{loop}$ denote the surfaces obtained by gluing two
punctures on a single string vertex.
It follows from \decbound\ that Eqn. \sofar\ can be written in the
form
$$(\hbox{I})\,\, =
-\spr_s  \ket{\wt\Phi_s}_e \cdot \biggl(
\,\int_{R_1^{tree}}\Omega_{\Phi_s B_1\cdots B_n}^{(-1)g,n+1}
+\int_{R_1^{loop}}\Omega_{\Phi_s B_1\cdots B_n}^{(-1)g,n+1}
\biggr) .\eqn\sofari$$
We now claim that the following `factorization' equations hold

$$\int_{R_1^{tree}}\Omega_{\Phi_s B_1\cdots B_n}^{(-1)g,n+1}
=\spr_r(-)^{\Phi_r+\Phi_s}
\hskip-10pt\sum\limits_{{g_1+g_2 = g
\atop \{ i_l ,j_k \} ; l,k\geq 0}\atop l+k =n}
\hskip-8pt\sigma (i_l,j_k)
\int_{\V_{g_1,l+2}} {\Omega_{}}_{\Phi_s B_{i_1}\cdots B_{i_l}\wt\Phi_r}
^{(0)g_1,l+2}\,\,\cdot \,\hskip-8pt\int_{\V_{g_2,k+1}}
{\Omega_{}}_{\Phi_r B_{j_1}\cdots B_{j_k}}^{(0)g_2,k+1},\eqn\firstfac$$

\noindent
for the surfaces of the `tree' type, and for the surfaces of the
`loop' type we get

$$\int_{R_1^{loop}}\Omega_{\Phi_s B_1\cdots B_n}^{(-1)g,n+1}
={1\over 2}\spr_r(-)^{\Phi_r+\Phi_s}
\hskip-8pt\int_{\V_{g-1,n+3}}{\Omega_{}}_{\Phi_s\Phi_r\wt\Phi_rB_1\cdots B_n}
^{(0)g-1,n+3} .\eqn\secfac$$

\noindent
Let us assume for the time being the correctness of the factorization
equations above and complete the proof of the main identity.
It follows by linearity that Eqn. \firstfac\ can be written as
$$\int_{R_1^{tree}}\Omega_{\Phi_s B_1\cdots B_n}^{(-1)g,n+1}
=\hskip-8pt\sum\limits_{{g_1+g_2 = g
\atop \{ i_l ,j_k \} ; l,k\geq 0}\atop l+k =n}
\hskip-8pt (-)^{\Phi_s}\sigma (i_l,j_k)
\int_{\V_{g_1,l+2}} {\Omega_{}}_{\Phi_s B_{i_1}\cdots B_{i_l}X}
^{(0)g_1,l+2} , \eqn\doesnotfit$$
where the state $\ket{X}$ is given by
$$\ket{X}=\spr_r(-)^{\Phi_r}\ket{\wt\Phi_r} \cdot\int_{\V_{g_2,k+1}}
{\Omega_{}}_{\Phi_r B_{j_1}\cdots B_{j_k}}^{(0)g_2,k+1}
= \l B_{j_1} \cdots B_{j_k} \r_{g_2},
\eqn\firstfacx$$
where use was made of Eqn.\altformprod .
Using now the definition of the multilinear product in \stringfvertex\
and incorporating the value of the ket $\ket{X}$, Eqn. \doesnotfit\
gives us
$$\int_{R_1^{tree}}\Omega_{\Phi_s B_1\cdots B_n}^{(-1)g,n+1}
=\hskip-6pt\sum\limits_{{g_1+g_2 = g
\atop \{ i_l ,j_k \} ; l,k\geq 0}\atop l+k =n}
\hskip-6pt (-)^{\Phi_s}\sigma (i_l,j_k)
\big\{ \Phi_s,  B_{i_1}\cdots B_{i_l} , \l B_{j_1} \cdots B_{j_k} \r_{g_2}
\big\}_{g_1},\eqn\sixtwelve$$
In a similar way the `loop' factorization equation \secfac\ can be
written as
$$\int_{R_1^{loop}}\Omega_{\Phi_s B_1\cdots B_n}^{(-1)g,n+1}
={1\over 2}\spr_r(-)^{\Phi_r+\Phi_s}
\big\{ \Phi_s, \Phi_r, \wt\Phi_r, B_1,\cdots ,B_n\big\}_{g-1}.\eqn\secfaci$$
Substitute now the results of the last two equations into \sofari\
$$\eqalign{(\hbox{I})\,\, &=
\spr_s (-)^{\Phi_s+1} \ket{\wt\Phi_s}_e \cdot \biggl(
\hskip-6pt\sum\limits_{{g_1+g_2 = g
\atop \{ i_l ,j_k \} ; l,k\geq 0}\atop l+k =n}
\hskip-6pt\sigma (i_l,j_k)
\big\{ \Phi_s,  B_{i_1}\cdots B_{i_l} ,
\l B_{j_1} \cdots B_{j_k} \r_{g_2} \big\}_{g_1} \cr
{}&+ {1\over 2}\spr_r(-)^{\Phi_r}
\big\{ \Phi_s, \Phi_r, \wt\Phi_r, B_1,\cdots ,B_n\big\}_{g-1}
\biggr),\cr} \eqn\sofarii$$
but using the definition of the string product (Eqn.\altformprod ) we find
$$(\hbox{I})\,\, =-\hskip-9pt\sum\limits_{{g_1+g_2 = g
\atop \{ i_l ,j_k \} ; l,k\geq 0}\atop l+k =n}
\hskip-6pt\sigma (i_l,j_k) \l B_{i_1} ,\cdots , B_{i_l} ,
\l B_{j_1}, \cdots , B_{j_k} \r_{g_2} \r_{g_1}-{1\over 2}\spr_r(-)^{\Phi_r}
\l \Phi_r, \wt\Phi_r, B_1,\cdots ,B_n\r_{g-1} .\eqn\sofariii$$
Comparing with Eqn. \fderv\ we have succeeded in showing that
$$\eqalign{
0\,\,&= Q \l B_1 , \cdots , B_n \r_g
+ \sum_{i=1}^n (-)^{(B_1 + \cdots B_{i-1})} \l B_1, \cdots
QB_i, \cdots , B_n \r_g  \cr
{}&+\hskip-6pt\sum\limits_{{g_1+g_2 = g
\atop \{ i_l ,j_k \} ; l,k\geq 0}\atop l+k =n}
\hskip-6pt\sigma (i_l,j_k)
\l B_{i_1} ,\cdots , B_{i_l} ,
\l B_{j_1}, \cdots , B_{j_k} \r_{g_2} \r_{g_1}\cr
{}&+ {1\over 2}\spr_r(-)^{\Phi_r}
\l \Phi_r, \wt\Phi_r, B_1,\cdots ,B_n\r_{g-1} ,\cr}
\eqn\sixfortyone$$
which is the desired result. In order to complete the proof
we must establish the factorization relations that were
assumed in the above derivation.
\medskip
\noindent
$\underline{\hbox{Proving the Factorization Equations}}.\,$
Let us begin with the `tree' type factorization equation.
Consider the form $\Omega^{(-1)}$ for this case
$${\Omega_{}^{}}_{\Phi_s B_1\cdots B_n}^{(-1)g,n+1}
(\wh V_1^\Sigma ,\cdots \wh V_{6g-6+2n+1}^\Sigma )
=N_{g,n+1}\bra{\Sigma} {\bf b}({\bf v}_1^\Sigma )\cdots
{\bf b}({\bf v}_{6g-6+2n+1}^\Sigma )
\ket{\Phi_s}\ket{B_1}\cdots\ket{B_n},\eqn\begfag$$
where the state $\bra{\Sigma}$ corresponds to a Riemann surface
$\Sigma \in R_1^{tree}$. In order to emphasize the fact that the
Schiffer variations are done on the punctures of the surface $\Sigma$
we have included the superscript on the tangent
vectors and on the associated Schiffer variations.

A surface
$\Sigma\in R_1^{tree}$ is built by sewing, so it actually
determines two surfaces $\wh\Sigma_1$ and
$\wh\Sigma_2$ with coordinate curves around their punctures. This means
that there is a phase ambiguity for the local coordinates, in particular
for the local coordinates around the punctures that are to be sewn.
This ambiguity makes the sewing parameter of the glued surface ambiguous.
If we make a choice of marked points
around the punctures of the surfaces $\wh\Sigma_1$ and $\wh\Sigma_2$
we obtain the surfaces $\Sigma_1$ and
$\Sigma_2$ respectively, and we can determine the sewing parameter $t$.
We write
$$\Sigma \in R_1^{tree} \quad \rightarrow\quad
\Sigma = \Sigma_1 \cup_t \Sigma_2 , \quad\hbox{with}\quad
\wh\Sigma_1 \in \V_{g_1,l+2}\,\,\hbox{and}\,\,
\wh\Sigma_2 \in \V_{g_2,k+1},\eqn\settree$$
with $l,k\geq 0$, $l+k=n$, and where $\cup_t$ denotes sewing with
sewing parameter $t=\exp (i\theta )$. Note that the surface $\Sigma_1$
has been assumed to contain the puncture corresponding to the
state $\Phi_s$, the $l$ punctures corresponding to $l$ states
$\ket{B_{i_l}}$, and one extra puncture $r$ which is sewn to the
puncture $r'$ of the surface $\Sigma_2$. This latter surface contains
$k$ additional punctures. The sewing statement implies that the
states representing the surfaces $\Sigma , \Sigma_1$, and $\Sigma_2$
satisfy the relation
$$\bra{\Sigma} = \bra{\Sigma_1} \bra{\Sigma_2} t^{L_0^{(r)}}
\ov t^{\ov L_0^{(r)} }  \ket{R_{rr'}}
=\bra{\Sigma_1} \bra{\Sigma_2} e^{i\theta (L_0^{(r)}- \ov L_0^{(r)} )}
\ket{R_{rr'}}
=\bra{\Sigma_1} \bra{\Sigma_2}
R_{rr'}^\theta \rangle .\eqn\sewtree$$
Recall that the operator formalism construction
of a state associated to a surface requires the complete specification
of coordinates around the punctures. Since we only have (naturally)
coordinates defined up to a phase, to construct
$\bra{\Sigma}$, for example, we specify arbitrarily the marked points,
this choice being irrelevant since the states that are to be contracted
with $\bra{\Sigma}$ satisfy the $L_0-\ov L_0 = 0$ condition.
What we have noted is that the phase ambiguity around the local
coordinates to be sewn must be fixed, in order to be able to define
the sewing parameter. The phases around the other punctures of
the constituent surfaces $\Sigma_1$ and $\Sigma_2$ are irrelevant.
The surface $\Sigma$ with unmarked coordinate curves will be denoted
as $\wh\Sigma$.

We have to consider integration over the set of surfaces in
$R_1^{tree}$. This space can be considered as a base space
$\V_{g_1,l+2} \times \V_{g_2,k+1}$ with a U(1) fiber, representing
the sewing angle. We will split the parameters of integration into
a group of parameters defining the surface $\Sigma_1$, a group of
parameters defining the surface $\Sigma_2$ and the sewing parameter.
Then we will see that the integrand factorizes, and we shall be able
to complete the proof of factorization.

The $U(1)$ fiber is generated by deformations of the surface
$\Sigma$ associated with the sewing parameter. From $\Sigma$ we
determine uniquely two surfaces $\wh\Sigma_1$
and $\wh\Sigma_2$ with coordinate curves. We then make a choice
of marked points around the punctures to be sewn and
obtain the surfaces $\Sigma_1$, $\Sigma_2$ and the sewing parameter $t$.
We let $\wh V^\Sigma_\theta$ denote
the tangent (in $T_{\wh\Sigma}(\wh\P)$) generating the deformation
${\partial\over \partial \theta}$,
and let ${\bf v}^\Sigma_\theta$ denote the corresponding Schiffer
variation. This Schiffer variation is supported on the neighborhood
of the punctures of $\Sigma$, thus in particular, away from the
region of sewing. It is clear from the fact that any tangent can
be represented by a Schiffer variation that ${\bf v}^\Sigma_\theta$
exists, and is defined uniquely up to a Borel vector. This deformation,
however, can be produced directly by changing the coordinates of
the sewing ket by the corresponding phase rotation. We must therefore
have that
$$\bra{\Sigma}{\bf T}({\bf v}^\Sigma_\theta) = \bra{\Sigma_1} \bra{\Sigma_2}
i(L_0-\ov L_0 )^{(r)} |R_{rr'}^\theta\rangle .\eqn\equivtreei$$
Since the connection conditions for the stress energy tensor are
the same as those for the antighost field, the above equation
implies that we must have
$$\bra{\Sigma}{\bf b}({\bf v}^\Sigma_\theta) =\bra{\Sigma_1} \bra{\Sigma_2}
i(b_0-\ov b_0 )^{(r)} \ket{R_{rr'}^\theta}.\eqn\equivxtreei$$

Let us now show how to obtain from a deformation of the surface
$\wh\Sigma_1$ a deformation of the surface $\Sigma$.
Again, from $\Sigma$ we determine uniquely two surfaces $\wh\Sigma_1$
and $\wh\Sigma_2$, and then (by a choice of marked points)
the surfaces $\Sigma_1$, $\Sigma_2$ and the sewing parameter $t$.
Now consider a deformation of the surface $\wh\Sigma_1$, that
is, a tangent vector $\wh V_i^{\wh\Sigma_1} \in T_{\wh\Sigma_1}(\wh\P)$.
We choose a tangent vector $V_i^{\Sigma_1} \in T_{\Sigma_1}(\P)$,
which projects down to $\wh V_i^{\wh\Sigma_1}$ as we forget about the
phases of the local coordinates. We use this tangent vector
$V_i^{\Sigma_1}$ to deform the surface $\Sigma_1$, and then we sew
the deformed $\Sigma_1$ to $\Sigma_2$ with sewing parameter $t$
to get a deformed $\Sigma$ surface. Let us
denote the tangent that generates this deformation of $\Sigma$ by
$\wh V^\Sigma (i,\Sigma_1)$, where the information inside the
parenthesis indicates that this tangent originated from the
$i$-th deformation of $\Sigma_1$. The vector $\wh V^\Sigma (i,\Sigma_1)$
is not uniquely determined, due
to the choices made at several places. The ambiguity, however, is
simple. Different choices would result in the addition to
$\wh V^\Sigma (i,\Sigma_1)$ of a vector along $\wh V^\Sigma_\theta$,
which, as mentioned in the previous paragraph, corresponds to a change
in the sewing parameter.
Denote the associated Schiffer variations by
$$ V_i^{\Sigma_1} \,\leftrightarrow\,\,
{\bf v}_i^{\Sigma_1}\quad\hbox{and}\quad
\wh V^\Sigma (i,\Sigma_1)\,\,\leftrightarrow\, \,
{\bf v}^\Sigma (i,\Sigma_1) .\eqn\translation$$
While the vector
${\bf v}_i^{\Sigma_1}$ is supported on all the punctures of
$\Sigma_1$, including the one to be sewn, the vector
${\bf v}^\Sigma (i,\Sigma_1)$ is supported around the punctures
of $\Sigma$.
The above discussion relating the deformations implies that
$$\bra{\Sigma}{\bf T}({\bf v}^\Sigma (i,\Sigma_1)) =
\bra{\Sigma_1} \bra{\Sigma_2}
{\bf T}({\bf v}_i^{\Sigma_1})\,\ket{R_{rr'}^\theta}.\eqn\equivtree$$
Again, a similar equation must hold for antighosts:
$$\bra{\Sigma}{\bf b}({\bf v}^\Sigma (i,\Sigma_1)) =
\bra{\Sigma_1}\bra{\Sigma_2} {\bf b}({\bf v}_i^{\Sigma_1})\,
\ket{R_{rr'}^\theta}.\eqn\equivxtree$$
The index $i$ runs from one up to $d_1$, where $d_1=2d_{g_1,l+2}$ is the
(real) dimensionality of the set $\V_{g_1,l+2}$. A completely analogous
discussion holds for the component $\Sigma_2$ of the surface. In this
case the index $i$ runs from one up to $d_2$, where $d_2=2d_{g_2,k+1}$
is the (real) dimensionality of $\V_{g_2,k+1}$.

We can now consider the form $\Omega^{(-1)}$ of Eqn. \begfag , with
input tangent vectors the vectors defined by the deformations of
$\wh\Sigma_1$, $\wh\Sigma_2$ and the sewing angle
$$\eqalign{{}&{\Omega_{}^{}}_{\Phi_s B_1\cdots B_n}^{(-1)g,n+1}
\bigl(\,\wh V^\Sigma (1,\Sigma_1),\cdots\wh V^\Sigma (d_1,\Sigma_1)\, ;\,
\wh V^\Sigma (1,\Sigma_2),
\cdots\wh V^\Sigma (d_2,\Sigma_2)\, ;\, \wh V^\Sigma_\theta \,\bigr)\cr
{}&=N_{g,n+1}\bra{\Sigma_1}\bra{\Sigma_2}
\bigl[ {\bf b}({\bf v}^\Sigma (1,\Sigma_1) )\cdots
{\bf b}({\bf v}^\Sigma (d_1,\Sigma_1) ) \bigr] \,
\bigl[ {\bf b}({\bf v}^\Sigma (1,\Sigma_2) )\cdots
{\bf b}({\bf v}^\Sigma (d_2,\Sigma_2) )\bigr] \cr
{}&\quad\cdot {\bf b}({\bf v}^\Sigma_\theta )|R_{rr'}^\theta\rangle
\ket{\Phi_s}\ket{B_1}\cdots\ket{B_n}.\cr}\eqn\xbegfagi$$
Note that the number of antighost insertions associated to each
of the surfaces is even. We now can use Eqns.\equivxtreei\ and \equivxtree\
to rewrite the right hand side as
$$\eqalign{&=(-)^{\Phi_s}N_{g,n+1}\bra{\Sigma_1}{\bf b}({\bf v}_1^{\Sigma_1})
\cdots {\bf b}({\bf v}_{d_1}^{\Sigma_1} ) \ket{\Phi_s}\cr
{}&\,\,\cdot \bra{\Sigma_2}
{\bf b}({\bf v}_1^{\Sigma_2})\cdots {\bf b}({\bf v}_{d_2}^{\Sigma_2} )
(ib_0^{-(r)}|R_{rr'}^\theta\rangle )\ket{B_1}\cdots\ket{B_n},\cr}
\eqn\begfagii$$
where the sign factor appeared by commuting the state $\ket{\Phi_s}$
across an odd number of antighosts. We now rearrange the states to find
$$\eqalign{{} &=(-)^{\Phi_s}\sigma (i_l; j_k)(iN_{g,n+1})
\bra{\Sigma_1}{\bf b}({\bf v}_1^{\Sigma_1})\cdots
{\bf b}({\bf v}_{d_1}^{\Sigma_1} ) \ket{\Phi_s}
\ket{B_{i_1}} \cdots \ket{B_{i_l}} \, \cr
{}&\quad\,\cdot\, \bra{\Sigma_2}
{\bf b}({\bf v}_1^{\Sigma_2})\cdots {\bf b}({\bf v}_{d_2}^{\Sigma_2} )
(b_0^{-(r)}|R_{rr'}^\theta\rangle )
\ket{B_{j_1}}\cdots\ket{B_{j_k}},\cr}\eqn\begfagiii$$
where the sign factor $\sigma$ is the sign involved in
rearranging the sequence $(b_0,B_1, \cdots B_n)$ into the sequence
$(B_{i_1}\cdots B_{i_l};b_0; B_{j_1},\cdots B_{j_k})$.
Let us now expand the sum over states involved in the reflector
to obtain
$$\eqalign{ {} &=\sum_r(-)^{\Phi_r+\Phi_s}\sigma (i_l; j_k)
(2\pi iN_{g,n+1})\bra{\Sigma_1}{\bf b}({\bf v}_1^{\Sigma_1})\cdots
{\bf b}({\bf v}_{d_1}^{\Sigma_1} ) \ket{\Phi_s}
\ket{B_{i_1}} \cdots \ket{B_{i_l}}|\wt\Phi_r\rangle \cr
{}&\quad\quad \cdot\bra{\Sigma_2} {\bf b}({\bf v}_1^{\Sigma_2})\cdots
{\bf b}({\bf v}_{d_2}^{\Sigma_2})\bigl[{1\over 2\pi}\exp (i\theta L_0^-)
\ket{\Phi_r}\bigr] \ket{B_{j_1}}\cdots\ket{B_{j_k}},\cr}\eqn\ybegfagi$$
where in the sum one includes states that
are not annihilated by $L_0^-$. We now use the relation
$$(2\pi i) N_{g,n+1} = N_{g_1,l+2}\, N_{g_2,k+1},\eqn\consisnorm$$
which follows from the definition of $N$ in \normcon\ and
$$d_{g,n+1} = 1+d_{g_1,l+2}+d_{g_2,k+1},\eqn\diffdim$$
which is the familiar statement that the ingredient spaces on the left
and on the right of the geometrical equation differ by complex dimension one.
Equations \ybegfagi\ and \consisnorm\ imply that
$${\Omega_{}^{}}_{\Phi_s B_1\cdots B_n}^{(-1)g,n+1}\bigl(
\,\wh V^\Sigma (1,\Sigma_1),\cdots\wh V^\Sigma (d_1,\Sigma_1)\,;\,
\wh V^\Sigma (1,\Sigma_2),\cdots\wh V^\Sigma (d_2,\Sigma_2)\, ;\,
\wh V^\Sigma_\theta  \,\bigr) $$
$$=\sum_r(-)^{\Phi_r+\Phi_s}\sigma (i_l; j_k)\,
{\Omega_{}^{}}_{\Phi_sB_{i_1}\cdots B_{i_l}\wt\Phi_r}^{(0),g_1,l+2}
(\wh V_1^{\wh\Sigma_1} ,\cdot\cdot\wh V_{d_1}^{\wh\Sigma_1} ) \,\cdot\,
{\Omega_{}^{}}_{[{1\over 2\pi}\exp(i\theta L_0^-)\Phi_r]
B_{j_1}\cdots B_{j_k}}^{(0),g_2,k+1}
(\wh V_1^{\wh\Sigma_2} ,\cdot\cdot\wh V_{d_2}^{\wh\Sigma_2}).\eqn\factforms$$
Note that we have obtained a relation between forms defined on different
spaces. The best way of thinking about the above equation is that
given the sets of tangents in $T_{\wh\Sigma_1}$ and
$T_{\wh\Sigma_2}$ we have shown that the right hand side equals the
form on the left hand side evaluated on associated tangents in $T_{\wh\Sigma}$.
It is necessary that whatever choices we made to get the tangents in
$T_{\wh\Sigma}$ the result of the left hand side is the same. This is
clear, because as we discussed before, the ambiguity amounts to
changing the tangents in $T_{\wh\Sigma}$ by vectors along
$\wh V^\Sigma_\theta$.
Thus, for different choices Eqn.\equivtree\ would read
$$\bra{\Sigma} \bigl[ {\bf T}({\bf v}^\Sigma (i,\Sigma_1))
+ {\bf T}({\bf v}^\Sigma_\theta ) \bigr] =
\bra{\Sigma_1} \bra{\Sigma_2}
{\bf T}({\bf v}_i^{\Sigma_1})\, |R_{rr'}^\theta\rangle .\eqn\eklkree$$
The related equation for the antighosts would read
$$\bra{\Sigma} \bigl[ {\bf b}({\bf v}^\Sigma (i,\Sigma_1))
+ {\bf b}({\bf v}^\Sigma_\theta ) \bigr] =
\bra{\Sigma_1} \bra{\Sigma_2}
{\bf b}({\bf v}_i^{\Sigma_1})\, |R_{rr'}^\theta\rangle ,\eqn\eklkreei$$
and if we used this in relating Eqns.\begfagii\ and \xbegfagi\
we would find that the extra terms vanish due to the presence of
${\bf b}({\bf v}^\Sigma_\theta )$ in \xbegfagi .
We are now essentially done. In integrating over $R_1^{tree}$ we must
consider all possible splittings, as indicated in the geometrical
equation \twopointone . Thus we have
$$\int_{R_1^{tree}} = {1\over 2}\hskip-10pt\sum\limits_{{g_1+g_2 = g
\atop \{ i_l ,j_k \} ; l,k\geq 0}\atop l+k =n}
\int_{\V_{g_1,l+2}}\cdot\int_{\V_{g_2,k+1}}\cdot \int_{S^1} ,\eqn\splitint$$
where $\int_{S^1}$ is the integral over the sewing angle. Since we always
want to consider the state $\Phi_s$ as inserted on the surface which is said
to have genus $g_1$, the terms in the geometrical equation when $\Phi_s$
ends up in the surface of genus $g_2$ are relabeled and simply cancel the
factor of $1/2$ in the above equation.
Integrating equation \factforms\ using \splitint\ we have
$$\int_{R_1^{tree}}\Omega_{\Phi_s B_1\cdots B_n}^{(-1)g,n+1}
=\sum_r(-)^{\Phi_r+\Phi_s}\hskip-12pt\sum\limits_{{g_1+g_2 = g
\atop \{ i_l ,j_k \} ; l,k\geq 0}\atop l+k =n}\hskip-12pt\sigma (i_l,j_k)
\hskip-8pt\int_{\V_{g_1,l+2}}
\hskip-10pt {\Omega_{}}_{\Phi_s B_{i_1}\cdots B_{i_l}\wt\Phi_r}^{(0)g_1,l+2}
\,\cdot\hskip-6pt\int_{\V_{g_2,k+1}}
\hskip-10pt{\Omega_{}}_{[\int {d\theta \over 2\pi}\exp (i\theta L_0^-)
\Phi_r] B_{j_1}\cdots B_{j_k}}^{(0)g_2,k+1},\eqn\xxxstfac$$
where the integral over the sewing parameter went into the state
in the second form. This integral simply projects to the subspace of
states annihilated by $L_0^-$. This reduces the sum over {\it all}
states $r$ to the usual sum $\spr_r$, whereupon the above equation
is precisely the first factorization equation.
\bigskip
We must now deal with the case of the `loop' factorization equation.
Given a surface $\Sigma \in R_1^{loop}$ of genus $g$ and $n$ punctures,
by cutting the surface along the sewing line, and restoring
back the semiinfinite cylinders, we obtain the surface $\wh\Sigma_o$,
where the `$o$' stands for open (the surface was opened up) and the
hat indicates that there is no canonical definition of the phase of the local
coordinates around the two punctures in question. We then make
a choice of the phases by marking arbitrary points in the coordinate
curves associated to the two punctures, and obtain the surface
$\Sigma_o$ (no hat). At this stage the sewing parameter $t$ corresponding
to the surface $\Sigma$ can be determined. We therefore have
$$\bra{\Sigma} = \bra{\Sigma_o} t^{L_0^{(r)}}
\ov t^{\ov L_0^{(r)} }  \ket{R_{rr'}}
=\bra{\Sigma_o} e^{i\theta (L_0^{(r)}- \ov L_0^{(r)} )}
\ket{R_{rr'}}
=\bra{\Sigma_o} R_{rr'}^\theta \rangle .\eqn\sewloop$$
Again, our concern is the form $\Omega^{(-1)}$ that reads
$${\Omega_{}^{}}_{\Phi_s B_1\cdots B_n}^{(-1)g,n+1}
(\wh V_1^\Sigma ,\cdots \wh V_{6g-6+2n+1}^\Sigma )
=N_{g,n+1}\bra{\Sigma} {\bf b}({\bf v}_1^\Sigma )\cdots
{\bf b}({\bf v}_{6g-6+2n+1}^\Sigma )
\ket{\Phi_s}\ket{B_1}\cdots\ket{B_n},\eqn\loopfacss$$
where now $\Sigma \in R_1^{loop}$. This time the $6g-6+2n+1$
dimensional region of integration, corresponding to that number of
antighost insertions in the above, must be related to an
integration over $\V_{g-1,n+3}$, which is of dimension
$6g-6+2n$, plus an integration over the sewing angle.
Just as we did in the `tree' case, we now have that
$$\eqalign{
{}&\bra{\Sigma}{\bf T}({\bf v}^\Sigma_\theta) =
\bra{\Sigma_o} i(L_0-\ov L_0 )^{(r)} |R_{rr'}^\theta\rangle ,\cr
{}&\bra{\Sigma}{\bf b}({\bf v}^\Sigma_\theta) =
\bra{\Sigma_o} i(b_0-\ov b_0 )^{(r)} |R_{rr'}^\theta\rangle ,\cr}
\eqn\equivloops$$
where ${\bf v}^\Sigma_\theta$ is the Schiffer variation corresponding
to the tangent vector $\wh V^\Sigma_\theta$ that generates the
deformation $\partial /\partial \theta$. Also (in a way analogous
to the way we discussed the other deformations in the tree case)
for a given $\Sigma$ once we fix $\Sigma_o$ and the sewing parameter,
a tangent $\wh V_i^{\wh\Sigma_o}$ can be used to choose a tangent
$V_i^{\Sigma_o}$, and this tangent is used to deform the surface
$\Sigma_o$, which is then sewn back to give a deformed $\Sigma$,
thus defining a tangent $\wh V^\Sigma (i,\Sigma_o)$. This tangent
is ambiguous up to a vector along the tangent $\wh V^\Sigma_\theta$.
We therefore have, denoting the Schiffer variations by
$$V_i^{\Sigma_o} \,\leftrightarrow\,\,
{\bf v}_i^{\Sigma_o}\quad\hbox{and}\quad
\wh V^\Sigma (i,\Sigma_o)\,\,\leftrightarrow\, \,
{\bf v}^\Sigma (i,\Sigma_o) ,\eqn\translllation$$
that
$$\eqalign{
{}&\bra{\Sigma}{\bf T}({\bf v}^\Sigma (i,\Sigma_o)) =
\bra{\Sigma_o} {\bf T}({\bf v}_i^{\Sigma_o})\, |R_{rr'}^\theta \rangle , \cr
{}&\bra{\Sigma}{\bf b}({\bf v}^\Sigma (i,\Sigma_o)) =
\bra{\Sigma_o}{\bf b}({\bf v}_i^{\Sigma_o})\,
|R_{rr'}^\theta\rangle .\cr}\eqn\equivloopm$$
Letting $d_o \equiv 2d_{g-1,n+3} =6g-6+2n$,
we then have that Eqn.\loopfacss\
can be evaluated for the tangents discussed above giving
$$\eqalign{{}&{\Omega_{}^{}}_{\Phi_s B_1\cdots B_n}^{(-1)g,n+1}\bigl(
\wh V^\Sigma (1,\Sigma_o),\cdots\wh V^\Sigma (d_o,\Sigma_o)\, ;\,
\wh V^\Sigma_\theta \bigr)\cr
{}&=N_{g,n+1}\bra{\Sigma_o}\bigl[ {\bf b}({\bf v}^\Sigma (1,\Sigma_o) )\cdots
{\bf b}({\bf v}^\Sigma (d_o,\Sigma_o) ) \bigr]
{\bf b}({\bf v}^\Sigma_\theta )|R_{rr'}^\theta\rangle
\ket{\Phi_s}\ket{B_1}\cdots \ket{B_n}.\cr}\eqn\xbegloopi$$
We now can use Eqns. \equivloops\ and \equivloopm\
to rewrite the right hand side as
$$=\sum_r(-)^{\Phi_s}(2\pi i N_{g,n+1})
\bra{\Sigma_o}{\bf b}({\bf v}_1^{\Sigma_o})\cdots
{\bf b}({\bf v}_{d_o}^{\Sigma_o} ) \ket{\Phi_s}
\bigl[ b_0^{-(r)}{1\over 2\pi}\exp (i\theta L_0^-)\ket{\Phi_r}\bigr]
\ket{\Phi_r^c} \ket{B_1}\cdots\ket{B_n},\eqn\elevennine$$
where the sign factor appeared by moving the state $\ket{\Phi_s}$
through $b_0^-$. Using the form notation (and the same observations
as we made before with the normalization factor) we find
$${\Omega_{}^{}}_{\Phi_s B_1\cdots B_n}^{(-1)g,n+1} \bigl( \wh V^\Sigma
(1,\Sigma_o),\cdots\wh V^\Sigma (d_o,\Sigma_o)\, ;\,\wh V^\Sigma_\theta\bigr)$$
$$= \sum_r(-)^{\Phi_r+\Phi_s}\,\,
{\Omega_{}^{}}_{\Phi_s [{1\over 2\pi}\exp (i\theta L_0^-)\Phi_r]
\wt\Phi_r B_1\cdots B_n}^{(0)g-1,n+3}
\bigl( \wh V^{\Sigma_o}_1 ,\cdots \wh V^{\Sigma_o}_{d_o} \bigr).
\eqn\elevenfourteen$$
The integral over $R_1^{loop}$ can be written as
$$\int_{R_1^{loop}} = {1\over 2} \int_{\V_{g-1,n+3}} \cdot \int_{S^1}
,\eqn\integloop$$
and using this on \elevenfourteen\ we find
$$\int_{R_1^{loop}}
{\Omega_{}^{}}_{\Phi_s B_1\cdots B_n}^{(-1)g,n+1}
={1\over 2}\sum_r(-)^{\Phi_r+\Phi_s}
\hskip-6pt\int_{\V_{g-1,n+3}}\hskip-6pt {\Omega_{}^{}}_{\Phi_s [\int
{d\theta \over 2\pi}\exp (i\theta L_0^-)\Phi_r]
\wt\Phi_r B_1\cdots B_n}^{(0)g-1,n+3}, \eqn\eleventwentythree$$
and upon doing the integral over $\theta$ the sum over states reduces
to the sum over states satisfying the $L_0^-$ constraint, and the
above equation becomes precisely the `loop' factorization equation
we wanted to establish. All in all this concludes our proof of the
main identity.

\REF\klebanov{I. Klebanov, `Ward Identities in Two-Dimensional
String Theory', PUPT-1302, December 1991}

\chapter{Algebraic Structures on the BRST Cohomology}

In this section we wish to connect our study of the algebraic
structure of both the classical and quantum closed string field
theory to the algebraic structures derived in the context of
two-dimensional closed string theory. It was observed by
Witten and the author [\wittenzwiebach ] that the tree-level Ward
identities of the theory could be neatly summarized as a homotopy
Lie algebra on the BRST cohomology, with products $m_2,m_3,m_4\cdots..$.
E. Verlinde [\everlinde ] showed that the complete higher genus
Ward identities could be obtained from a generating function
satisfying a Batalin-Vilkovisky type master equation (for applications
of the Ward identities see also Ref. [\klebanov ]). In both works
the possible relation to similar structures in closed string field
theory was pointed out. In this section we derive the connection,
by showing how to obtain from the basic identity relating the
products in string field theory the structures on cohomology.

We have shown in \S6.4 that the string vertices have `stubs' of length $\pi$
around each puncture. These stubs prevent the appearance of curves
that are shorter than $2\pi$ when vertices are sewn together.
Accordingly the string vertices $\V_{g,n}$ include all surfaces
whose internal foliations are all of height smaller than $2\pi$. We observed
that the length of the stubs can be changed at will, as long as
they are kept longer than $\pi$. For stubs of length $l$ the
string vertices $\V_{g,n}(l)$ include all surfaces whose foliations are all
of height smaller than $2l$. The important thing to realize is that
as we let the parameter $l$ grow every set $\V_{g,n}$ becomes larger
and larger. This happens because  $\V_{g,n}(l) \subset \V_{g,n}(l')$
for $l < l'$, which in turn, is manifest from the
above description of the $\V$'s.
Thus, the sets $\ov \M_{g,n} - \V_{g,n}(l)$,
as $l$ increases, are becoming smaller and smaller neighborhoods of
the compactification divisor of $\ov \M_{g,n}$ (the set of noded
surfaces in $\ov\M_{g,n}$). As $l\rightarrow \infty$ these sets
go simply into the noded surfaces. Equivalently, as $l\rightarrow \infty$
$\V_{g,n} \rightarrow \M_{g,n}$ (the moduli space without compactification,
or nodes).
One may think, at first glance, that the growth of {\it all}
vertices is incompatible with the fact that the Feynman graphs produce
smaller and smaller neighborhoods of the compactification divisor, but
this is clearly wrong, as the above discussion makes manifest. Indeed,
think of the limit $l\rightarrow \infty$, when the propagator
simply joins surfaces through a node.
The nodes of $\ov\M_{g,n}$ actually correspond to the sets
$\ov\M_{g_1,n_1} \times \ov\M_{g_2,n_2}$ with
$g_1+g_2 =g;\, n_1+n_2 = n+2$, plus the set $\ov\M_{g-1,n+2}$;
namely, the nodes include full copies of the lower dimensional
compactified moduli spaces. Therefore, given that the propagator
just joins surfaces through nodes, the lower vertices $\V_{g,n}$ must
all coincide with the spaces $\M_{g,n}$ in order for the Feynman
graphs to produce the whole set of noded surfaces.

Consider the structure of the multilinear functions $\{ \cdots \}_g$,
constructed explicitly in \S7. They give, for a given set of of $n$ states
$B_i$, the integral over $\V_{g,n}$, with the Polyakov measure, of the
correlation function of those states. Thus, they simply give us
the part of the string amplitude corresponding to the subset
$\V_{g,n}$ of $\M_{g,n}$. Therefore as $l\rightarrow \infty$,
and as a consequence $\V_{g,n}(l)\rightarrow \M_{g,n}$, the multilinear
product goes into the full string amplitude for the states $B_i$,
associated to the corresponding genus (Of course, this assumes that
for the given choice of external momenta, or parameters, the amplitude
is finite, since otherwise the infinite part would indeed come from
the neighborhoods of the noded surfaces, and the limits would not
make sense).

Consider the following sum of multilinear functions in which each term has
a BRST operator acting on a state
$$\big\{ QB_1 , B_2 , \cdots , B_n \big\}_g +
(-)^{B_1}\big\{ B_1 , QB_2 , \cdots , B_n \big\}_g + \cdots
+(-)^{B_1+\cdots B_{n-1}}\big\{ B_1, \cdots , Q B_n \big\}_g ,
\eqn\summultf$$
and the multilinear functions now correspond to integrals over
the full moduli space $\M_{g,n}$. This can be rewritten as
$$\eqalign{
\sum_{i=1}^n (-)^{\sum_{k=0}^{i-1}B_k}\big\{ B_1 , \cdots ,
QB_i , \cdots , B_n \big\}_g
&= \int_{\M_{g,n}} {\Omega_{}^{}}_{(\sum Q)B_1\cdots B_n}^{(0)g,n}
=\int_{\M_{g,n}} \hbox{d}{\Omega_{}^{}}_{B_1\cdots B_n}^{(-1)g,n}, \cr
{}&=\int_{\partial\M_{g,n}}{\Omega_{}^{}}_{B_1\cdots B_n}^{(-1)g,n} ,
\cr}\eqn\brstdec$$
where we used the definition of the
multilinear functions, the BRST property of the forms, and
Stokes theorem. We now argue that the final expression should
be taken to be zero
$$\sum_{i=1}^n (-)^{\sum_{k=0}^{i-1}B_k}\big\{ B_1 , \cdots ,
QB_i , \cdots , B_n \big\}_g = 0 . \eqn\brstdeco$$
Basically, what we are saying is that if the
original integrand is well defined in $\ov \M_{g,n}$ (rather than
just $\M_{g,n}$), that is, the integrand does not diverge at the
degenerate surfaces, then the original integral could have been
done over $\ov\M_{g,n}$, and since this space has no boundary we do
not get boundary terms. This is the essence of BRST decoupling, it is
clearly a formal statement since one must check nice behavior near
degeneration (Making sure this holds can require analytic continuation
on external parameters, and typically at the quantum level, correcting
the two-dimensional conformal field theory). We will assume Eqn.
\brstdeco\ holds. The usual statement of BRST decoupling is that
$$\big\{ \Phi_t , B_2,B_3, \cdots B_n \big\}_g = 0 ,\eqn\ubrstdec$$
where $\Phi_t$ denotes a BRST trivial state, namely,
$\ket{\Phi_t} = Q \ket{\alpha}$, and all the $\ket{B_i}$ states,
with $i\geq 2$, are annihilated by the BRST operator. This equation
follows from \brstdeco\ if we take $B_1 = \alpha$. Moreover, for
arbitrary states
$\ket{\alpha}$ and $\ket{\beta}$, and states $\ket{B_i}$ annihilated
by $Q$ for $i\geq 3$, we also have
$$\big\{ Q\alpha ,\beta ,B_3,\cdots ,B_n \big\}_g
+(-)^\alpha\big\{ \alpha ,Q\beta ,B_3,\cdots B_n \big\}_g=0,\eqn\hbrstdec$$
which follows from \brstdeco\ by taking $B_1 = \alpha$ and
$B_2 = \beta$. The two equations above will be of utility to us later.

We now consider choosing a suitable basis of states. The states
$\ket{\Phi_s}$ of the conformal field theory, satisfying the
$b_0^-$ and $L_0^-$ constraints will be decomposed
into physical states $\Phi_p$, or semi-relative cohomology classes
(states annihilated by $Q$, that are not $Q$ on something), trivial
states $\ket{\Phi_t}$ (states that are of the type $Q$ on something)
and unphysical states $\ket{\Phi_u}$ (states that are not annihilated
by $Q$). The physical states are ambiguous up to trivial states, and
the unphysical states are ambiguous up to the addition of physical
and trivial states. We choose a basis of states such that the
unphysical states, are all orthogonal to each other and orthogonal
to the physical states. Moreover, since the trivial states must
be all orthogonal to each other and to the physical states; upon
choosing this basis we have that the states conjugate to physical
states must be physical states and that the trivial and unphysical
states are conjugates to each other. A detailed discussion on how
to arrive to such basis has been given in Appendix C of the second
paper in Ref.[\sen ]. Given an unphysical state
$\ket{\Phi_{u_i}}$ denote the associated conjugated state by
$\ket{\Phi_{{t'}_i}}$, that is
$$\ket{\Phi_{u_i}^c} = \ket{\Phi_{{t'}_i}}, \eqn\setupbasis$$
we then define
$$\ket{\wt\Phi_{u_i}} =
b_0^-\ket{\Phi_{u_i}^c} = b_0^-\ket{\Phi_{{t'}_i}}
\equiv \ket{\Phi_{t_i}}.\eqn\setiing$$
Therefore we have that
$$\ket{\wt\Phi_{u_i}} = \ket{\Phi_{t_i}}, \quad\hbox{and}\quad
\ket{\wt\Phi_{t_i}} = -\ket{\Phi_{u_i}}, \eqn\mainbase$$
where the second relation is readily established.\foot{The
first relation implies $c_0^-\ket{\Phi_t} = \ket{\Phi_u^c}$ and
this together with \otherorder\ implies that
$\bra{\Phi_u}c_0^-\ket{\Phi_t} = (-)^{\Phi_u}$.  The second
relation implies $\ket{\Phi_t^c}=-c_0^-\ket{\Phi_u}$, which,
using \qonref , gives us $\bra{\Phi_t^c}=(-)^{\Phi_u}\bra{\Phi_u}c_0^-$.
This result, together with $\bra{\Phi_t^c}\Phi_t\rangle =1$, agrees
with the result derived from the first relation.}
Using this basis of states, an arbitrary string
product can be written in the form (using the second equation
in \altformprod )
$$\eqalign{
\l B_1 , B_2, \cdots , B_n \r_g &= \spr_s (-)^{\Phi_s} \ket{\wt\Phi_s}
\cdot \big\{ \Phi_s , B_1 ,\cdots , B_n \big\}_g \cr
&= \sum_p (-)^{\Phi_p} \ket{\wt\Phi_p}
\cdot \big\{ \Phi_p , B_1 ,\cdots , B_n \big\}_g \cr
&+ \sum_i (-)^{\Phi_{u_i}} \biggl(  \ket{\Phi_{t_i}}
\cdot \big\{ \Phi_{u_i} , B_1 ,\cdots , B_n \big\}_g  + \ket{\Phi_{u_i}}
\cdot \big\{ \Phi_{t_i} , B_1 ,\cdots , B_n \big\}_g \biggr) . \cr}
\eqn\expprod$$
where two minus signs (the first from the opposite statistics
of $\Phi_{u_i}$ and $\Phi_{t_i}$, and the second from \mainbase )
cancelled in the last term of the right hand side.

We can now begin our reduction of the main identity \bidentity\ to physical
states. Accordingly we take all of the $B$ states to be physical and
use the above to obtain
$$\eqalign{
0 &= \sum_i (-)^{\Phi_{u_i}} Q \ket{\Phi_{u_i}}
\cdot \big\{ \Phi_{t_i} , B_1 ,\cdots , B_n \big\}_g  \cr
{}&+ \hskip-10pt\sum\limits_{g_1,g_2 \atop \{ i_l ,j_k \} l,k }
\sigma (i_l,j_k) \biggl(
\sum_p (-)^{\Phi_p}\l B_{i_1} \cdots B_{i_l} , \wt\Phi_p \r_{g_1}
\cdot \big\{ \Phi_p , B_{j_1} ,\cdots , B_{j_k} \big\}_{g_2} \cr
{}&\quad\quad\quad\quad\quad\quad
+\sum_i (-)^{\Phi_{u_i}}\l B_{i_1} \cdots B_{i_l} , \Phi_{t_i} \r_{g_1}
\cdot \big\{ \Phi_{u_i} , B_{j_1} ,\cdots , B_{j_k} \big\}_{g_2} \cr
{}&\quad\quad\quad\quad\quad\quad
+\sum_i (-)^{\Phi_{u_i}}\l B_{i_1} \cdots B_{i_l} , \Phi_{u_i} \r_{g_1}
\cdot \big\{ \Phi_{t_i} , B_{j_1} ,\cdots , B_{j_k} \big\}_{g_2} \biggr) \cr
{}&+{1\over 2}\spr_s (-)^{\Phi_s} \biggl(
\sum_p (-)^{\Phi_p} \ket{\wt\Phi_p}
\cdot \big\{ \Phi_p , \Phi_s , \wt\Phi_s ,  B_1 ,\cdots , B_n \big\}_{g-1}\cr
&\quad\quad\quad\quad\quad + \sum_i (-)^{\Phi_{u_i}} \ket{\Phi_{t_i}}
\cdot \big\{ \Phi_{u_i},\Phi_s,\wt\Phi_s , B_1 ,\cdots , B_n \big\}_{g-1}\cr
&\quad\quad\quad\quad\quad + \sum_i (-)^{\Phi_{u_i}} \ket{\Phi_{u_i}}
\cdot \big\{ \Phi_{t_i},\Phi_s , \wt\Phi_s , B_1 ,\cdots , B_n \big\}_{g-1}
\biggr) . \cr}\eqn\expandinden$$
This equation is really three equations, that is, the coefficients
of the physical , unphysical and trivial states should vanish. Let us
pay attention now only to physical states. From the above equation we
then find that for each physical state labeled by $p'$ we must have
$$\eqalign{
0 &= \hskip-10pt\sum\limits_{g_1,g_2 \atop \{i_l ,j_k\} l,k }
\hskip-5pt\sigma (i_l,j_k) \biggl(
\sum_p (-)^{\Phi_p}
\big\{ \Phi_{p'} ,  B_{i_1} \cdots B_{i_l} , \wt\Phi_p \big\}_{g_1}
\cdot \big\{ \Phi_p , B_{j_1} ,\cdots , B_{j_k} \big\}_{g_2} \cr
{}&\quad\quad\quad\quad\quad\quad
+\sum_i (-)^{\Phi_{u_i}}
\big\{ \Phi_{p'}, B_{i_1} \cdots B_{i_l} , \Phi_{t_i} \big\}_{g_1}
\cdot \big\{ \Phi_{u_i} , B_{j_1} ,\cdots , B_{j_k} \big\}_{g_2} \cr
{}&\quad\quad\quad\quad\quad\quad
+\sum_i (-)^{\Phi_{u_i}}
\big\{ \Phi_{p'}, B_{i_1} \cdots B_{i_l} , \Phi_{u_i} \big\}_{g_1}
\cdot \big\{ \Phi_{t_i} , B_{j_1} ,\cdots , B_{j_k} \big\}_{g_2} \biggr) \cr
+&\quad {1\over 2}\spr_s (-)^{\Phi_s}
\big\{ \Phi_{p'},\Phi_s,\wt\Phi_s,B_1,\cdots ,B_n \big\}_{g-1}.
\cr}\eqn\onphys$$
Let us now consider the case when the multilinear functions use
the full moduli spaces $\M_{g,n}$, and become the string amplitudes.
BRST decoupling of trivial states in Eqn.\ubrstdec\ implies that the
second and third term in the above equation vanish. Therefore, if we
work at genus zero, when the last term is not present, we have
$$0= \hskip-10pt\sum\limits_{p, \{i_l ,j_k\} l\geq 1,k\geq 2 \atop l+k=n}
\hskip-5pt\sigma (i_l,j_k) (-)^{\Phi_p}
\big\{ \Phi_{p'} ,  B_{i_1} \cdots B_{i_l} , \wt\Phi_p \big\}_0
\cdot \big\{ \Phi_p , B_{j_1} ,\cdots , B_{j_k} \big\}_0 ,\eqn\wzward$$
which is the form of the Ward identities derived in Ref. [\wittenzwiebach ]
in the context of 2D string theory. They involve only physical states.
For the case of higher genus we must still do some work since the sum
over states $s$ in the quantum term of \onphys\ involves physical, as well
as unphysical and trivial states. Consider this term and expand the
sum over states as usual
$$\eqalign{{1\over 2}\spr_s (-)^{\Phi_s}
\big\{ \Phi_{p'},\Phi_s,\wt\Phi_s,B_1,\cdots ,B_n \big\}_{g-1}
&= {1\over 2}\sum_p (-)^{\Phi_p}
\big\{ \Phi_{p'},\Phi_p,\wt\Phi_p,B_1,\cdots ,B_n \big\}_{g-1} \cr
&+{1\over 2}\sum_i (-)^{\Phi_{t_i}}
\big\{ \Phi_{p'},\Phi_{t_i}, -\Phi_{u_i} ,B_1,\cdots ,B_n \big\}_{g-1} \cr
&+{1\over 2}\sum_i (-)^{\Phi_{u_i}}
\big\{ \Phi_{p'},\Phi_{u_i}, \Phi_{t_i} ,B_1,\cdots ,B_n \big\}_{g-1}. \cr}
\eqn\analyzeloop$$
Since the statistics of $\Phi_{u_i}$ and $\Phi_{t_i}$ are opposite to
each other the last two terms add (rather than cancel out!)
$$\eqalign{{1\over 2}\spr_s (-)^{\Phi_s}
\big\{ \Phi_{p'},\Phi_s,\wt\Phi_s,B_1,\cdots ,B_n \big\}_{g-1}
&= {1\over 2}\sum_p (-)^{\Phi_p}
\big\{ \Phi_{p'},\Phi_p,\wt\Phi_p,B_1,\cdots ,B_n \big\}_{g-1} \cr
&+\sum_i (-)^{\Phi_{u_i}}
\big\{ \Phi_{p'},\Phi_{u_i}, \wt\Phi_{u_i} ,B_1,\cdots ,B_n \big\}_{g-1}. \cr}
\eqn\analyzeop$$
This is not surprising since we have not yet used the BRST properties of
trivial and unphysical states. We will now show that indeed
$$\sum_i (-)^{\Phi_{u_i}} \big\{ \Phi_{p'},\Phi_{u_i}, \wt\Phi_{u_i} ,B_1,
\cdots ,B_n \big\}_{g-1} = 0, \eqn\itworksr$$
by use of the quartet mechanism of Kugo and Ojima [\kugoojima ].
The basic idea in this mechanism is that an unphysical state determines
three other states, and the four together form a quartet. The unphysical
state, acted by the BRST operator gives us a trivial state. These
two states form a BRST doublet. Now, the unphysical state must
have a tilde conjugate state, this conjugate state must be trivial,
and therefore together with the state it arises from by action of
$Q$ forms another doublet. These two doublets form a quartet.
This may be sketched in the following diagram
$$\matrix{\Phi_u & \buildrel \wt {} \over \longrightarrow & Q\alpha_u \cr
\mapdown{Q} & {} &\mapup{Q} \cr
(-)^{\Phi_u}Q\Phi_u & \buildrel \wt {} \over \longleftarrow & \alpha_u \cr}
\eqn\koquartet$$
As we see, the unphysical field, via tilde conjugation gives us
the trivial state $Q\alpha_u$, then clearly $\alpha_u$ is unphysical.
The nice property is that the tilde-conjugate of the unphysical state
$\alpha_u$ is the trivial state $Q\Phi_u$, with the sign given correctly
in the diagram. As equations, the above diagram says that
$$ \ket{\wt\Phi_u} = Q\ket{\alpha_u} \quad \rightarrow \quad
\ket{\wt\alpha_u} = (-)^{\Phi_u} Q \ket{\Phi_u} . \eqn\koquarteti$$
Note that the states $\Phi_u$ and $\alpha_u$ have the same statistics.
\foot{A derivation of the signs is sketched for the convenience
of the interested reader. The first relation in \koquarteti\ gives
$\ket{\Phi_u^c} = c_0^- Q \ket{\alpha_u}$ and this together with
$\bra{\Phi_u}\Phi_u^c\rangle = (-)^{\Phi_u}$ gives us
$\bra{\Phi_u}c_0^-Q\ket{\alpha_u} = (-)^{\Phi_u}$. On the other
hand the second relation in \koquarteti\ gives
$\ket{\alpha_u^c} = (-)^{\Phi_u}c_0^- Q \ket{\Phi_u}$, and using
\magiccon\ and \qonref , one finds that
$\bra{\alpha_u^c}= (-)^{\Phi_u + 1} \bra{\Phi_u}Qc_0^-$, which implies
$\bra{\alpha_u^c}\alpha_u\rangle =(-)^{\Phi_u+1}
\bra{\Phi_u}Qc_0^-\ket{\alpha_u} = 1$. This last inner product agrees
with the one following from the first relation in \koquarteti\
after using $\bra{\Phi_u} = \bra{\Phi_u}c_0^-b_0^-$ to interchange
the relative positions of $Q$ and $c_0^-$.}
Now, going back to Eqn. \itworksr , we have to sum over unphysical
states. We do this sum by pairing the unphysical states, and we will
sum over the possible pairs $(\Phi_u , \alpha_u )$. We therefore get
$$\sum_{pairs} (-)^{\Phi_{u}} \biggl(
\big\{ \Phi_{p'},\Phi_{u}, Q\alpha_u ,B_1, \cdots ,B_n \big\}_{g-1}
+\big\{ \Phi_{p'},\alpha_{u}, (-)^{\alpha_u} Q\Phi_u ,B_1, \cdots ,
B_n \big\}_{g-1} \biggr). \eqn\decsumkoq$$
It now follows from \hbrstdec\ and the graded-commutativity of the
multilinear functions that
$$\eqalign{
\big\{ \Phi_{p'},\alpha_{u},(-)^{\alpha_u}Q\Phi_u,B_1,\cdots ,
B_n \big\}_{g-1}
&=-\big\{ \Phi_{p'},Q\alpha_{u},\Phi_u,B_1,\cdots ,B_n\big\}_{g-1}\cr
&=-\big\{ \Phi_{p'},\Phi_u,Q\alpha_{u},B_1,\cdots ,B_n\big\}_{g-1},\cr}
\eqn\itxrksr$$
and back in \decsumkoq\ we get the desired cancellation. It should
be noted that for unphysical states whose unphysical partner is
the same state ($\alpha_u \sim \Phi_u$), as may happen for states
of ghost number two, the single contribution to the sum is vanishes
by the same argument. Therefore we have verified that the
unphysical and trivial states decouple from the sum and we finally
have
$$\eqalign{
0 &= \hskip-10pt\sum\limits_{g_1,g_2 \atop \{i_l ,j_k\} l,k }
\hskip-5pt\sigma (i_l,j_k) \sum_p (-)^{\Phi_p}
\big\{ \Phi_{p'} ,  B_{i_1} \cdots B_{i_l} , \wt\Phi_p \big\}_{g_1}
\cdot \big\{ \Phi_p , B_{j_1} ,\cdots , B_{j_k} \big\}_{g_2} \cr
+&{1\over 2}\sum_p (-)^{\Phi_p}
\big\{ \Phi_{p'},\Phi_p,\wt\Phi_p,B_1,\cdots ,B_n \big\}_{g-1}.
\cr}\eqn\onphyxxs$$
This equation is the reduction of the main identity to physical states,
and indeed all sums over states run only over the physical states of
the theory (the semi-relative BRST cohomology classes). If we now
define the new physical string product
$$\l B_1 , \cdots B_n \r_g \equiv
\sum_p (-)^{\Phi_p} \ket{\wt\Phi_p}
\cdot \,\big\{ \Phi_p , B_1 ,\cdots , B_n \big\}_g , \eqn\physprod$$
where the sum only involves physical states, the previous
identity reads
$$
0= \hskip-10pt\sum\limits_{g_1,g_2 \atop \{i_l ,j_k\} l,k }
\hskip-5pt\sigma (i_l,j_k)
\big\{ \Phi_{p'},B_{i_1}\cdots B_{i_l},\l B_{j_1},\cdots ,
B_{j_k}\r_{g_2}\big\}_{g_1}
+{1\over 2}\sum_p (-)^{\Phi_p}
\big\{ \Phi_{p'},\Phi_p,\wt\Phi_p,B_1,\cdots ,B_n \big\}_{g-1}.
\eqn\onphyxxis$$
As in Eqn.\condsum\ which applies to the starting point for our present
derivation of \onphyxxis , we have that the sum over splittings runs over
all $l,k\geq 0$ with $l+k = n\geq 0$, but with $l\geq 1$, when
$g_1=0$ and $k\geq 2$ when $g_2=0$. These latter restrictions
are not necessary, however, since the $l=0$ terms vanish due to
$\{ \Phi_{p'},\l B_{j_1}\cdots B_{j_k}\r_{g_2}\}_0$
$=\langle \Phi_{p'} , Q\l B_{j_1}\cdots B_{j_k}\r_{g_2} \rangle = 0$,
and the $k=0$ and $k=1$ terms vanish because $\l\cdot\r_0\equiv 0$, and
$\l B \r_0 \equiv QB =0$, respectively.

Let us now recover the master equation for physical states.
We define now the {\it on-shell string field} $\ket{\Psi_o}$ which
is simply the string field with the sum over states in the Hilbert
space of the conformal field theory restricted to the states in
the semi-relative cohomology. Thus in defining fields and antifields
we simply have
$$\ket{\Psi_o}=\ket{\Psi_{o-}} +\ket{\Psi_{o+}},\quad
\ket{\Psi_{o-}}=\sum_{G(\Phi_p)\leq 2}\ket{\Phi_p}\,\psi^p,\quad
\ket{\Psi_{o+}}=\sum_{G(\Phi_p)\leq 2}\ket{\wt\Phi_p}\,\psi_p^*,\eqn\onshfd$$
We take for the master action, the same master action given in
Eqn.\thefullaction , but where the string field is now the on-shell field
defined above and the multilinear functions use the full moduli spaces.
This will be called the {\it on-shell master action}. Since we only have
physical fields, there is no genus zero quadratic term in the on-shell master
action. Every term in the on-shell action is the string amplitude
for the corresponding states. We will now argue that this on-shell master
action satisfies the master equation.

Our first step is to get Eqn. \onphyxxis\ in a more convenient form.
We take $B_i= \Psi_o$ for all $i$, and moreover
$\Phi_{p'} = \Psi_o$.  Then with
$l=n_1-1,\,  k=n_2$, and $l+k = n-1$, the above equation gives us
$$0= \hskip-10pt\sum\limits_{{g_1,g_2 \atop n_1\geq 1, n_2\geq 0}
\atop n_1+n_2 = n\geq 1}{(n-1)!\over (n_1-1)!n_2!}
\hskip-5pt\
\big\{ \Psi_o^{n_1} ,\l \Psi_o^{n_2}\r_{g_2}\big\}_{g_1}
+{1\over 2}\sum_p (-)^{\Phi_p}
\big\{ \Phi_p,\wt\Phi_p, \Psi_o^n \big\}_{g-1}.
\eqn\onphyxijs$$
Since this equation is entirely analogous to equation \mbsidentityii\ we have
that the above implies
$$0= \hskip-10pt\sum\limits_{{g_1,g_2 \atop n_1\geq 1, n_2\geq 0}
\atop n_1+n_2 = n\geq 1}{n!\over n_1!n_2!}
\hskip-5pt\
\big\{ \Psi_o^{n_1} ,\l \Psi_o^{n_2}\r_{g_2}\big\}_{g_1}
+{1\over 2}\sum_p (-)^{\Phi_p}
\big\{ \Phi_p,\wt\Phi_p, \Psi_o^n \big\}_{g-1}.
\eqn\onnunus$$
which is the analog of Eqn.\mbsidentityiii . Since this equation was the
basic ingredient in the verification of the master equation in the
general case, we will now just go over the steps involved and see
that everything goes through in the present conditions. The derivation
begins in Eqn.\bvmexp , and wherever we see $\spr_s$ we must remember
the sum is only over physical states. The derivation goes through until
Eqn.\wweell , where it is clear that we can indeed replace the restricted
sum over physical states by the complete sum $\spr_s$. Then we can simply use
the derivation up to Eqn.\firstpart . The analysis of the second term of the
BV equation simply goes through, using the restricted sums. This
shows that indeed the BV master equation is satisfied by the
on-shell master action and concludes our derivation.

In Ref.[\wittenzwiebach] the action reproducing the three point couplings
of the physical states represents (the restriction to ghost number two
states of) the cubic term in the on-shell master action for two-dimensional
closed string field theory. The physical states in two-dimensional
closed string theory can be described in terms
of differential forms on a three dimensional space and their couplings in
terms of gauge theory in that space. The significance of the on-shell
master action suggests that it should be of interest to derive its
complete form for two-dimensional closed string theory.
\medskip
\ack
I am indebted to E. Witten for discussions that contributed much to the
understanding of the algebraic structure of the theory. I am grateful to
A. Giveon and J. Stasheff for their useful comments on the paper. I wish to
thank P. Nelson for discussions on the operator formalism, to K. Ranganathan
for discussions on Schiffer variations, and to M. Wolf for discussions on
the minimal area problem. Useful conversations with M. Kontsevich,
A. S. Schwarz and W. Thurston are gratefully acknowledged.
\refout
\figout
\end